\documentclass{article}

\usepackage[final,nonatbib]{neurips_2022}

\usepackage[utf8]{inputenc}
\usepackage[T1]{fontenc}
\usepackage{hyperref}
\usepackage{url}
\usepackage{booktabs}
\usepackage{longtable}
\usepackage{array}
\usepackage{subcaption}
\usepackage{amsmath,amssymb,amsfonts}
\usepackage{nicefrac}
\usepackage{microtype}
\usepackage{graphicx}
\usepackage{float}
\usepackage{enumitem}
\usepackage{standalone}
\usepackage{eso-pic}
\usepackage{xcolor}
\usepackage{framed}

\newcounter{algocnt}
\newenvironment{myalgo}[1]{%
  \refstepcounter{algocnt}%
  \par\medskip\noindent%
  \begin{framed}%
  \noindent\small\textbf{Algorithm~\thealgocnt\,:}\enspace\textit{#1}\par\smallskip%
  \noindent\rule{\linewidth}{0.4pt}\par\smallskip%
}{%
  \end{framed}\par\medskip%
}
\newenvironment{grayleftbar}{%
  \MakeFramed{\advance\hsize-\width\FrameRestore}}%
{\endMakeFramed}

\newcommand{\IN}{\hspace*{1.5em}}
\newcommand{\IIN}{\hspace*{3.0em}}
\newcommand{\IIIN}{\hspace*{4.5em}}
\newcommand{\kw}[1]{\textbf{#1}}

\newcommand{\tabref}[1]{Table~\ref{#1}}

\makeatletter
\renewcommand{\@noticestring}{}
\makeatother

\hypersetup{
  colorlinks=true,
  linkcolor=blue!50!black,
  urlcolor=blue!50!black,
  citecolor=blue!50!black
}

\title{From Image to Music Language: A Two-Stage Structure Decoding Approach for Complex Polyphonic OMR\thanks{Online live demo: \url{https://huggingface.co/spaces/k-l-lambda/starry}.}}
\author{
Nan XU \\
FindLab
\And
Shiheng LI \\
FindLab
\And
Shengchao HOU \\
FindLab
}
\date{}

\begin{document}

\maketitle

\raggedbottom

\begin{abstract}
We propose a new approach for a practical two-stage Optical Music Recognition (OMR) pipeline, with a particular focus on its second stage. Given symbol and event candidates from the visual pipeline, we decode them into an editable, verifiable, and exportable score structure. We focus on complex polyphonic staff notation, especially piano scores, where voice separation and intra-measure timing are the main bottlenecks. Our approach formulates second-stage decoding as a structure decoding problem and uses topology recognition with probability-guided search (BeadSolver) as its core method. We also describe a data strategy that combines procedural generation with recognition-feedback annotations. The result is a practical decoding component for real OMR systems and a path to accumulate structured score data for future end-to-end, multimodal, and RL-style methods.
\end{abstract}

\section{Introduction}

\subsection{Background and Motivation}

The recent success of large language models is rooted in the availability of massive human-written corpora. Natural language already exists at enormous scale in digital form, so large-scale pretraining becomes feasible and scaling laws start to matter. Music also has a long-term cultural archive. From Renaissance to contemporary, a vast amount of musical knowledge has been written down in staff notation. In principle, this should be a rich source of training material for symbolic music and notation understanding.

In practice, that potential is still far from fully unlocked. Much of the world's notated music is not available as high-quality structured digital data. It exists as scans, printed books, photographed pages, and publisher PDFs, rather than editable symbolic representations. Some symbolic corpora and engraved datasets do exist, but their scale and structural completeness are still limited. If the music domain is to benefit from similar data scaling, a key prerequisite is reliable conversion from score images to structured, machine-usable formats.

This is the practical motivation of our work. We propose a production-grade OMR method for complex polyphonic staff notation, especially piano scores. The goal is not only to detect local symbols, but to produce output that is editable, verifiable, and exportable. The system serves two purposes. It solves hard cases in practice, and it helps turn historical and modern sheet music collections into structured data that future generative models can learn from.

This perspective also shapes the technical focus of the paper. For printed music, the main bottleneck is often not whether a notehead, stem, or rest was detected at all. It is how the detected objects should be assembled into a coherent musical structure. Piano notation makes this especially clear. Voices can overlap at nearly the same horizontal position, appear only locally, cross staff boundaries, or depend on tuplets and grace-note timing. A system that stops at candidate detection is still far from a usable score representation.

Current OMR work can be broadly grouped into two main routes. The first is the two-stage route. Stage one recognizes symbols from the score image. Stage two assembles them into structured notation and outputs a serialized description, such as MusicXML, LilyPond, ABC Notation, or an internal notation format. The second is the end-to-end route, where a system maps an image directly to a serialized representation, such as MusicXML or a notation token sequence. In the long run, this is the more attractive direction. It is analogous to image-to-LaTeX transcription for mathematical expressions in document understanding. Today, however, it is still constrained by the cost of producing large amounts of reliable structured supervision.

Our work addresses that gap. We aim to make the two-stage route as strong as possible for complex multi-voice scores. We do this through careful decoding design and a data strategy that combines model-based inference, structural rules, and automatically generated training samples. The system is not positioned against future LLM- or RL-based OMR methods. Instead, it is meant to prepare the ground for them by turning difficult recognition cases into reusable structured data. This paper therefore focuses on the part of the two-stage pipeline that most directly determines usability: decoding a set of event candidates into a coherent score structure.

\subsection{Problem Definition: Structure Decoding and Regulation}

The system discussed in this paper is built on a complete two-stage OMR pipeline. Its visual stage processes scanned pages, camera photos of printed pages, and screen-captured score images. It predicts page layout, staff regions, and a set of semantic candidates. These predictions are then assembled into measure-level event candidate sets. At that point, the system already has useful information: approximate geometry, staff identity, local visual features, and candidate predictions for some duration- and notation-related attributes. Even so, the score structure is still not uniquely determined. For difficult measures, usability depends on the temporal relations between events, which events are simultaneous, how events split into voices, and which local duration choices make the measure consistent.

For that reason, we refer to the full second-stage process as \emph{regulation}, and we formulate this process as a structure decoding problem. The input is a set of candidate events with geometry and local attribute tendencies. The output is an editable measure structure with at least three parts: a set of voice chains ordered in time, an absolute tick position for each event in the measure, and corrected duration- and notation-related attributes. Once the output is stable, it can be converted into an internal score representation, MusicXML, LilyPond, or other downstream formats. It can also enter a correction workflow, where humans or agents refine the structure and feed repaired results back into the dataset.

The main difficulty is that structure cannot be recovered reliably from local rules alone. For a given candidate set, several local connections may look plausible. The usable solution is usually the one that satisfies multiple global constraints at once. Horizontal position and time should be roughly monotonic. Duration accumulation inside a voice should be self-consistent. Different voices must proceed in parallel, but they should not intertwine in impossible ways. Whole-measure rests, grace notes, and tuplets also change the assumptions behind ordinary duration inference. For these reasons, pure rule systems hit a ceiling quickly. Pure local classification is also not enough to guarantee an editable and exportable result.

Within this broader regulation process, we use \emph{topology recognition} to refer to its core topology-dependent solving component: the method that predicts how candidate events in a measure are organized into voice chains, and how those chains together form a temporally coherent piece of notation. The BeadSolver proposed in this paper follows this view. Instead of treating detector output as final labels, it treats the output as a probabilistic candidate set. It then uses search with structural evaluation to select a better topological solution among multiple possibilities.

\subsection{Why Voice Separation Is the Difficult Part}

The hardest part of structure recovery is usually voice separation. In piano music this is not an edge case. It is the normal case. Many measures do not contain a clear visual cue that tells the system which note should continue which chain. Instead, notation often compresses several voices into nearby positions. It relies on stem direction, beam organization, local context, and rhythmic logic to convey structure.

\begin{figure}[t]
\centering
\begin{minipage}{0.48\textwidth}
  \centering
  \includegraphics[width=\linewidth]{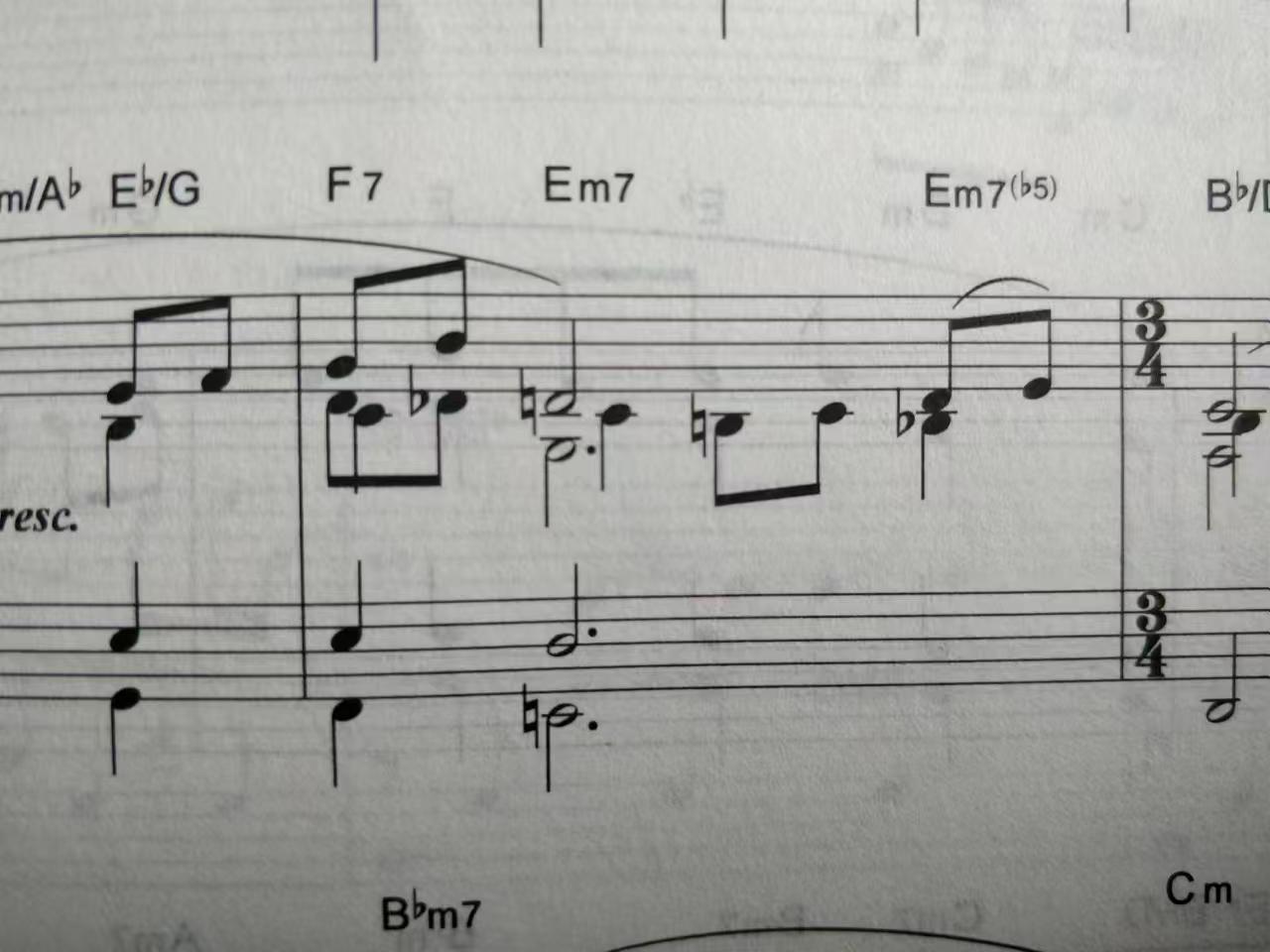}
  \vspace{1mm}
  {\small \textbf{(a)} Multi-voice overlap at adjacent $x$.}
\end{minipage}\hfill
\begin{minipage}{0.48\textwidth}
  \centering
  \includegraphics[width=\linewidth]{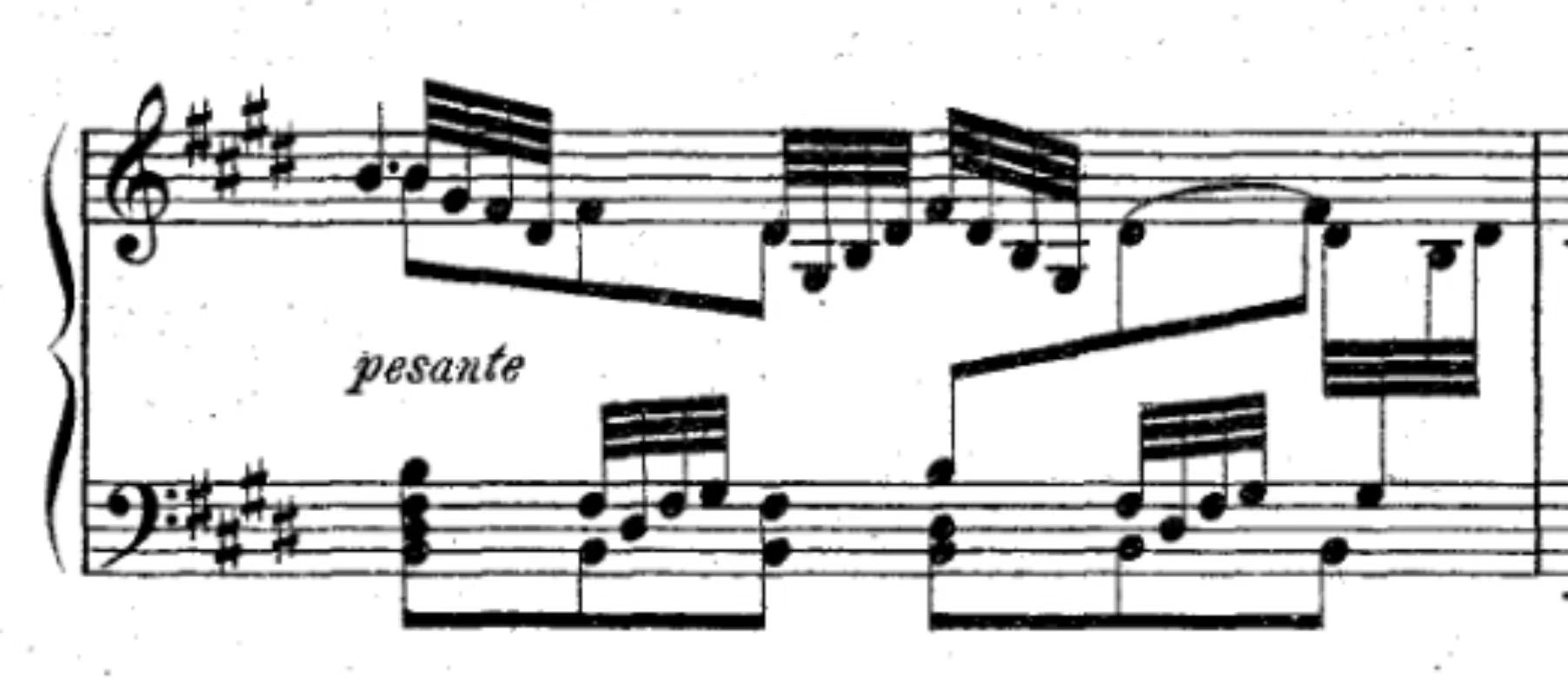}
  \vspace{1mm}
  {\small \textbf{(b)} Cross-staff writing.}
\end{minipage}

\vspace{1mm}
\begin{minipage}{0.72\textwidth}
  \centering
  \includegraphics[width=\linewidth]{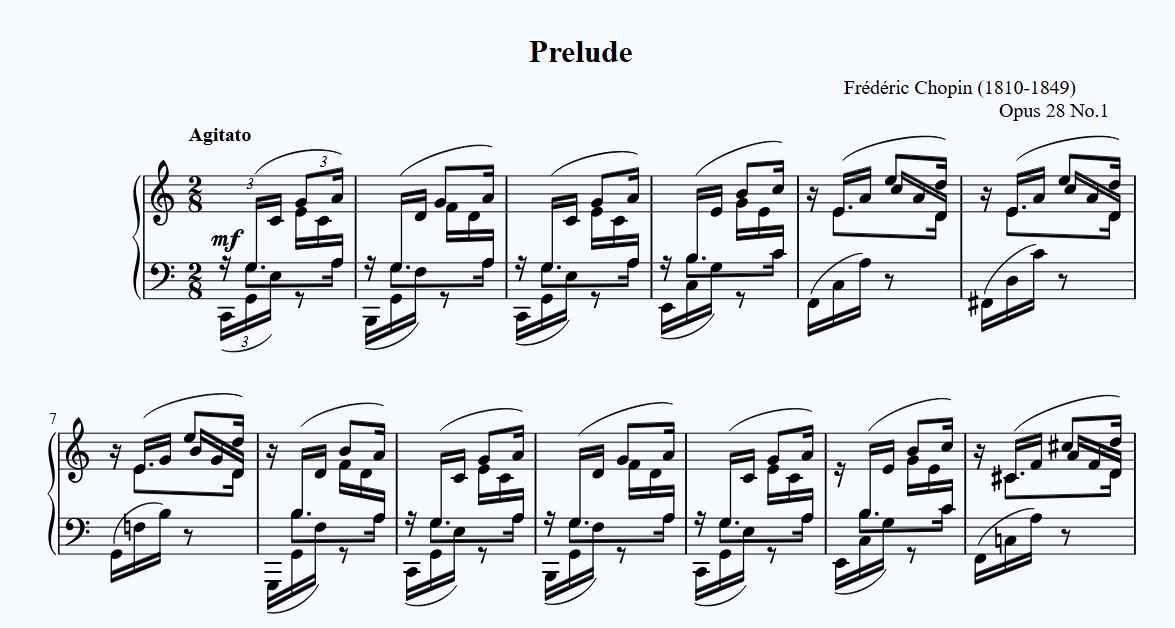}
  \vspace{1mm}
  {\small \textbf{(c)} Tuplet and time-warp complexity.}
\end{minipage}

\caption{Examples of structural ambiguity in complex piano notation. (a) Multiple voices can overlap at nearly the same horizontal position, so local geometry alone is not enough to decide voice continuation. Note that this measure is in a 4/4 time signature. (b) Cross-staff writing breaks simple staff-based assignment. Note that all those 32th notes make up a single voice. (c) Tuplets introduce non-uniform time scaling, which couples duration choices and voice structure; also, rests are merged across voices, which deepens the ambiguity.}
\label{fig:fig1}
\end{figure}

Figure~\ref{fig:fig1} shows common patterns that make voice separation and rhythm reconstruction difficult in practice: multi-voice overlap at similar horizontal positions, cross-staff writing, and tuplet-related timing complexity (often coupled with rest-merge ambiguities across voices).

This difficulty has at least four recurring sources. First, multiple voices may overlap at the same horizontal position. As a result, ``which events happen together'' and ``which events belong to the same voice'' become two different questions. Second, partial voices may appear only locally, which makes simple global ordering rules unreliable. Third, cross-staff writing breaks any one-to-one mapping between staff boundaries and voice boundaries. Fourth, tuplets and other time-warp effects introduce non-uniform temporal scaling, so local duration decisions must be interpreted in a wider context. This is why we handle voice separation and time organization in the same solving framework: in the cases that matter most, they are tightly coupled.

\subsection{Our Contributions}

Against that background, this paper describes a structure decoding method that already serves a full OMR pipeline in practice. The main contributions are as follows.

\begin{enumerate}
\item We present a production-grade two-stage OMR approach that reliably maps score images to an editable, verifiable, and exportable serialized representation (e.g., MusicXML, LilyPond, or an internal format). The key is to bridge the long-standing engineering gap between candidate detection and a usable structural output.
\item We propose probability-guided tree search for structure decoding inside second-stage regulation. Model probabilities provide the search guidance, while usability-oriented structural costs enforce global consistency and help select a usable solution among multiple candidates. This substantially raises the practical ceiling of OMR algorithms on complex polyphonic scores.
\item We introduce a mixed-radix \emph{vtick} encoding for musical time, so that tick prediction follows common rhythmic subdivision patterns rather than generic scalar regression. This gives the model a representation that is closer to the discrete structure of notation and to common MIDI tick conventions.
\item We propose a practical data-generation strategy to compensate for the shortage of real sheet-music training samples for topology recognition. By training an autoregressive symbolic generator on a small seed corpus and then sampling measure-level structures at scale, we can construct structurally faithful supervision without requiring large manually annotated OMR corpora from the outset.
\item We integrate the second-stage decoding pipeline into an annotation and data feedback loop. Automatic solving, quality evaluation, and human/agent correction produce high-quality structured labels (voice, tick, duration, and measure structure) that can be fed back for training and can serve as a foundation for future end-to-end, multimodal, and RL-style methods.
\end{enumerate}

\section{Related Work}

We focus on work that is most relevant to the second half of a two-stage OMR pipeline: turning detected objects into a coherent musical structure. The goal is not a full historical survey, but a clear technical lineage for ``candidates $\to$ voices/time/duration'', together with the data representations and evaluation settings that make this stage comparable.

\subsection{OMR Pipelines and Task Decomposition}
Survey papers commonly describe OMR as a pipeline that separates visual symbol recognition from downstream semantic reconstruction \cite{rebelo2012omr,calvo2020understanding}. This framing is useful for our setting because it highlights an engineering bottleneck: many real failures are not missing noteheads, but incorrect relationships between events, such as voice continuation, simultaneity, and duration normalization. More generally, recent work on document recognition has argued that such tasks should often be understood as transcription from a document into a structured record, rather than as isolated visual detection alone \cite{meyer2025structured}. This broader perspective is also compatible with our view of second-stage OMR as explicit structure decoding. Our work follows this decomposition, but treats the second stage as a standalone decoding problem that deserves explicit modeling.

\subsection{Datasets, Representations, and Evaluation}
Most public datasets emphasize front-end tasks such as detection, classification, or segmentation of music symbols. DeepScores and its successors are representative printed-notation benchmarks in this direction \cite{tuggener2018deepscores,tuggener2021deepscoresv2}. In contrast, datasets such as MUSCIMA++ explicitly annotate object relations for handwritten music, which makes graph-structured supervision and evaluation possible \cite{hajic2017muscima}. Work on OMR evaluation has also stressed the need for shared datasets, terminology, and objective metrics for comparing systems \cite{bellini2007assessing}. More recent work has gone further and proposed unified intermediate representations, notably MTN (Music Tree Notation), to support more comparable metrics across OMR systems \cite{torras2023unified}. These efforts motivate our focus on measure-level structural quality (voices and ticks) rather than only per-symbol accuracy.

\subsection{Notation Assembly and Graph Structure}
Between low-level detection and final export, many systems perform an assembly step that groups primitives (noteheads, stems, beams, barlines) into higher-level musical objects and relations. A common formulation is to represent notation as a graph and recover pairwise relations between detected symbols, but earlier work often assumed perfect or gold-standard detection input. More recent work has emphasized robustness to imperfect detection and more realistic end-to-end evaluation of graph-like notation assembly \cite{yang2024complete}. We share this goal, but focus on a specific core subproblem: recovering voice and time organization at the measure level. More broadly, our aim is to move toward a more complete OMR solution by bridging the gap from images to a music-language representation that supports both reliable playback and re-engraving.

\subsection{Structure Engines and Rule-Based Reconstruction}
There is also a long line of engineering-oriented systems that treat structure reconstruction as a dedicated engine, often driven by rules and constraint checking \cite{bainbridge2003engine}. This view is aligned with our motivation: structure recovery is not a thin post-processing step. Our difference is in how uncertainty is handled. Rather than committing to one deterministic parse, we combine a learned policy with search and structural evaluation, so the system can explore multiple plausible hypotheses and choose the one that yields the most coherent measure.

\subsection{End-to-End Transcription}
End-to-end models that map images directly to a serialized score representation are an attractive long-term direction. Early work often focused on monophonic scores using sequence models and CTC-style objectives \cite{calvo2018monophonic,liu2020crnn}, while more recent work has moved toward full-page pianoform notation \cite{riosvila2023piano}. In complex polyphony, however, end-to-end training is strongly constrained by data availability and annotation cost. Our method targets the practical regime where dense structure labels are scarce: we keep a two-stage design and use probability-guided decoding to recover structure from imperfect candidates, while also creating a path to accumulate better structured data for future end-to-end models.

\section{System Overview}

\begin{figure}[htbp]
\centering
\includestandalone[mode=image,width=0.98\textwidth]{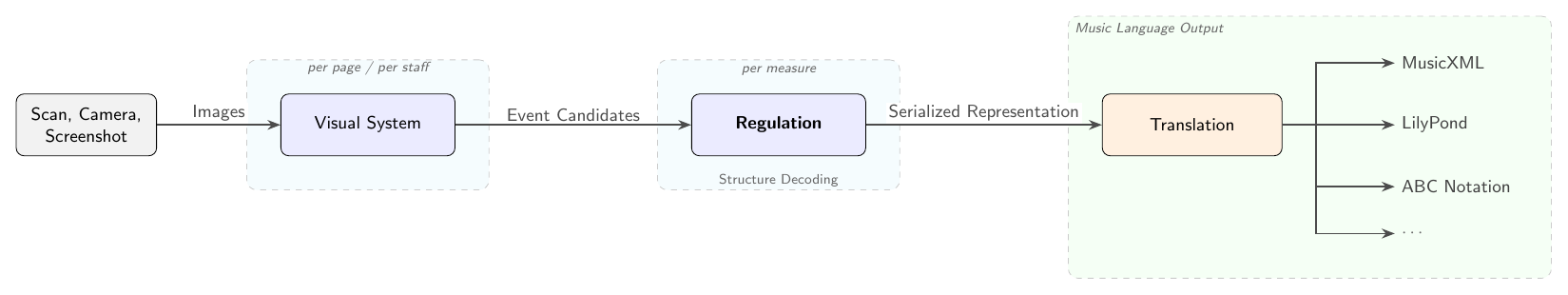}
\caption{A compact overview of the Starry OMR pipeline.}
\label{fig:starry-overview}
\end{figure}

This section summarizes our OMR pipeline at a system level. The goal is to clarify what information is produced by the visual pipeline, what is deferred to the decoding stage, and what the final structured output looks like.
Figure~\ref{fig:starry-overview} gives a compact overview of the full data flow, from input images through visual candidate generation and measure-level regulation to downstream music-language formats.

\subsection{From Pixels to Symbol Candidates}

We focus on printed music, not handwritten scores. The input may be a scanned page, a camera photo of a printed page, or a photo of a score displayed on an electronic screen. These sources differ in noise and geometry. Photos may include perspective distortion, page curvature, uneven lighting, blur, or moir\'e patterns. The system is therefore designed to produce robust local evidence first, and to postpone global structural decisions.

\begin{figure}[htbp]
\centering
\includegraphics[width=0.92\textwidth]{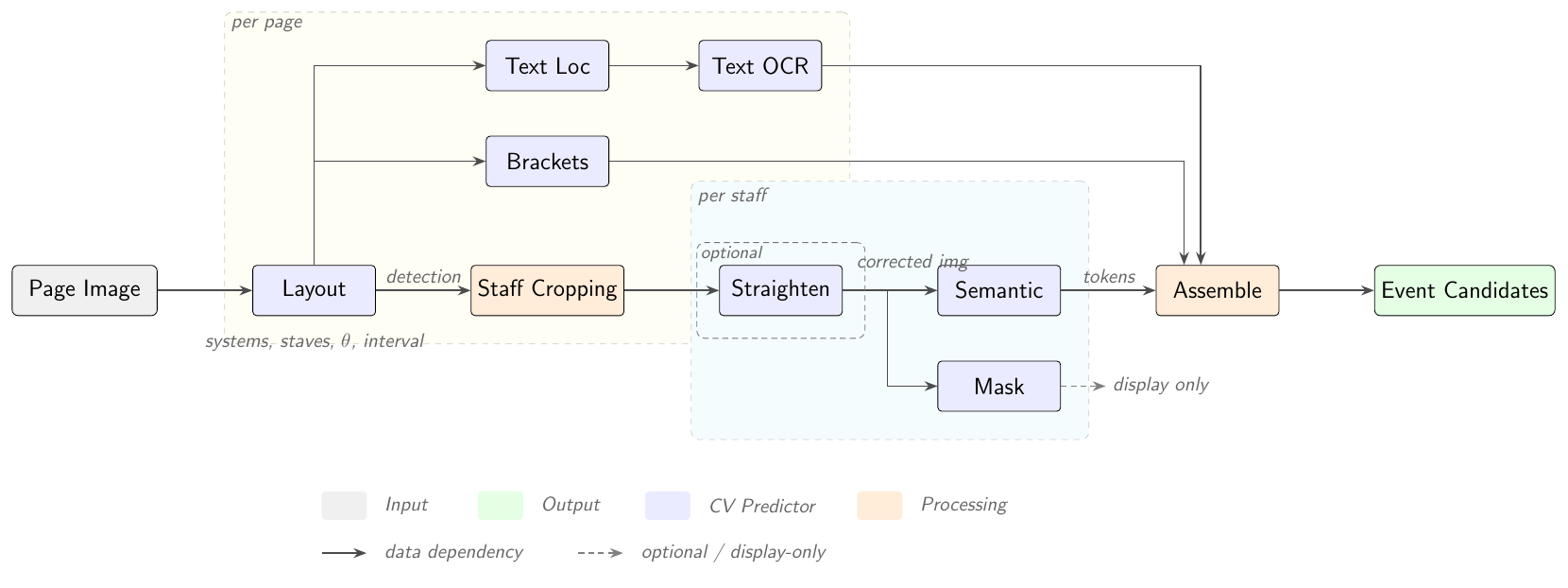}
\caption{Visual pipeline for candidate generation. The page-level stage detects systems and staves, then staff-level predictors produce corrected crops, semantic cues, and auxiliary display outputs. Assembly merges these local predictions into measure-level event candidates for downstream decoding.}
\label{fig:cv-pipeline}
\end{figure}

The visual pipeline is composed of specialized predictors for layout and staff processing, semantic detection, and text recognition. Figure~\ref{fig:cv-pipeline} shows the dependency structure of this pipeline. At a high level, layout analysis uses a non-mutually-exclusive semantic segmentation model to identify systems and staves and to establish a consistent coordinate system. A geometric correction stage can then optionally straighten staff images for photos with deformation; this gauge stage is a regression model with two outputs: one channel predicts the local semantic $y$ coordinate of each pixel within the current staff, using the third staff line as zero, and the other predicts the local slope of vertical lines. In the saved RGB representation, this slope channel is stored with 16-bit precision to reduce floating-point loss. During rectification, the warped local staff image is treated as a texture on a polygon mesh. The gauge outputs are integrated into a smooth deformation field, which determines the mesh vertex coordinates; rendering this mesh then produces the straightened staff image. A mask stage uses a simple semantic segmentation model with two output channels (foreground and background), although in practice we usually only use the foreground channel.
Unlike the traditional detection-based setup, our semantic model is also a multi-channel semantic segmentation network, where channels are not mutually exclusive. Each channel predicts a specific kind of notation evidence (e.g., clefs, time signatures, noteheads, stems, rests, and barlines) as a dense heatmap, and multiple channels may fire at the same location.
In post-processing, we extract concrete geometric primitives from these heatmaps using OpenCV contour detection. Text detection and OCR cover titles, composers, tempo markings, lyrics, fingering, and related annotations. For full scores, an additional staff-layout component can detect left-side brackets that define staff grouping.

An important design choice is that this stage aims for high recall and weak commitment. In particular, we do not require the CV stage to directly output long-range structures like slurs or beams as full curves or connections. Instead, such structures are decomposed into local anchors or endpoints whenever possible. This reduces the difficulty of the visual task and fits naturally with downstream symbolic representations, where long-range notation is often expressed by local marks attached to discrete events.

\begin{figure}[htbp]
\centering
\begin{minipage}{0.9\textwidth}
  \centering
  \includegraphics[width=0.47\linewidth]{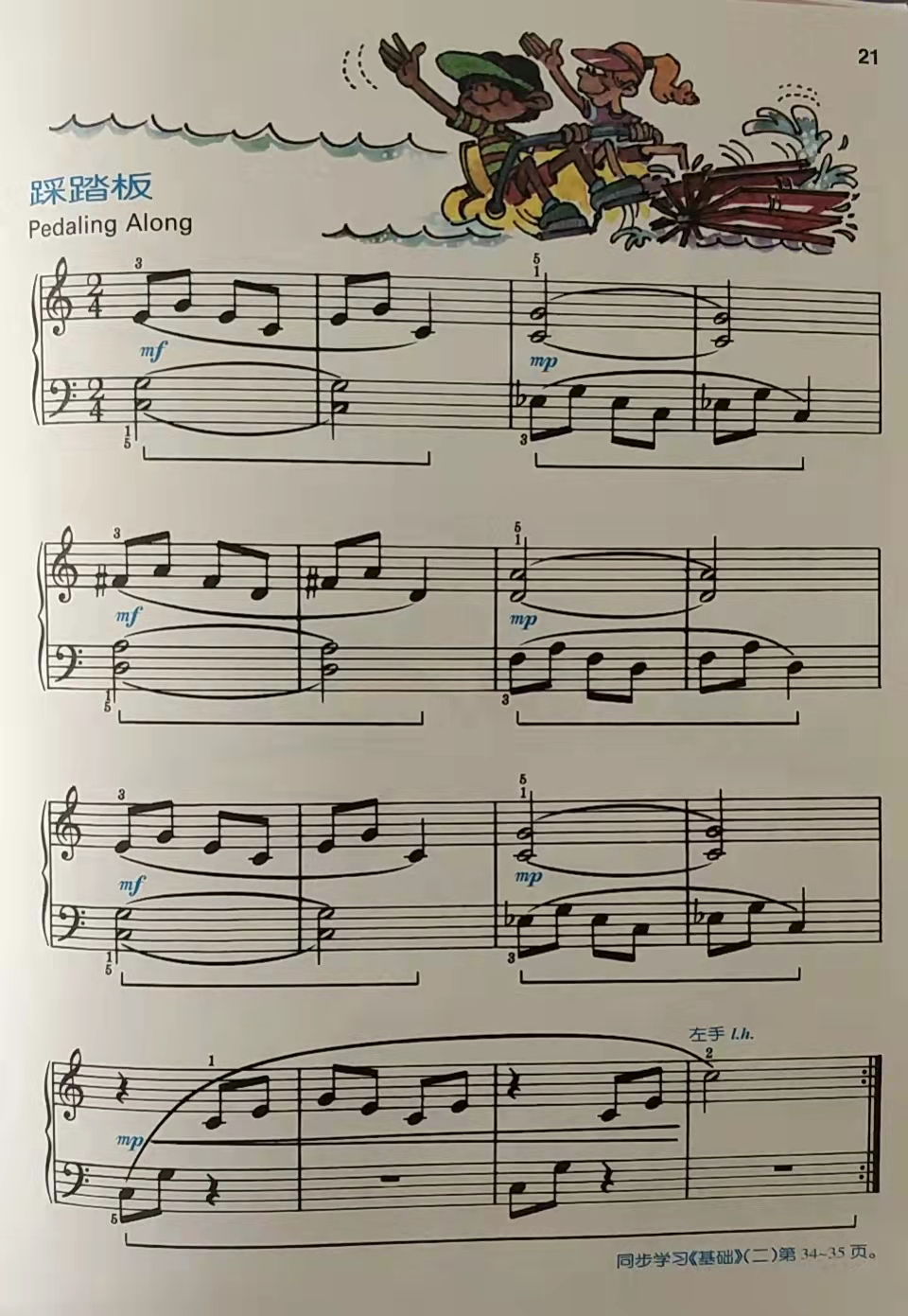}\hfill
  \includegraphics[width=0.47\linewidth]{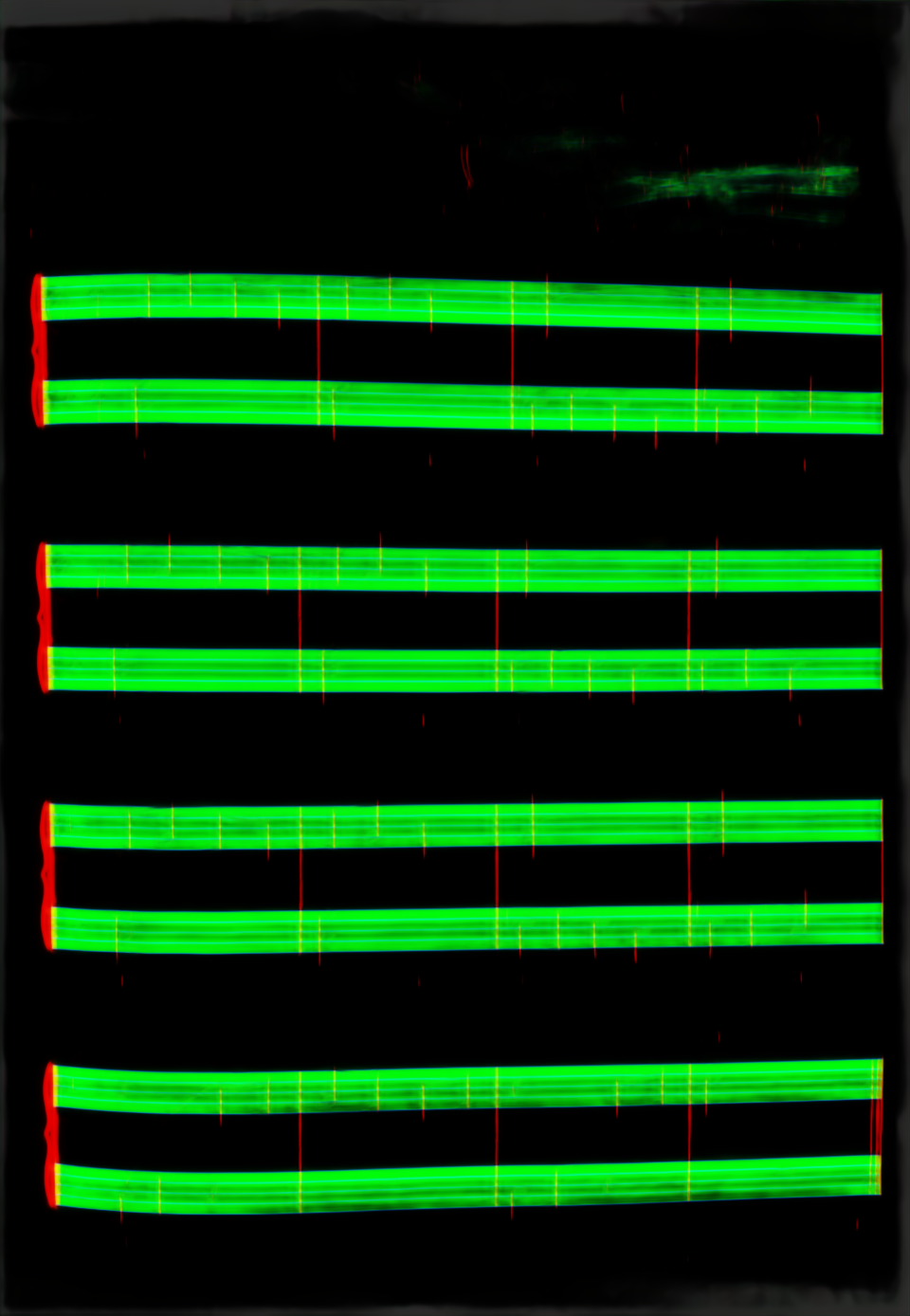}
  \vspace{1mm}
  
  {\small \textbf{(a)} Layout: page input and page-level prediction map. The output is a 3-channel heatmap shown as RGB: red = vertical lines (barlines and system boundaries), green = staff regions, blue = horizontal staff lines.}
\end{minipage}

\vspace{2mm}
\begin{minipage}{0.9\textwidth}
  \centering
  \includegraphics[width=0.31\linewidth]{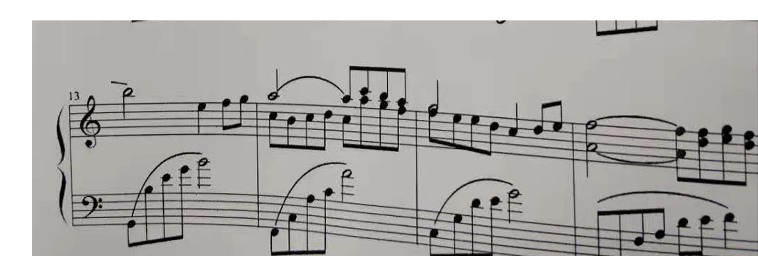}\hfill
  \includegraphics[width=0.31\linewidth]{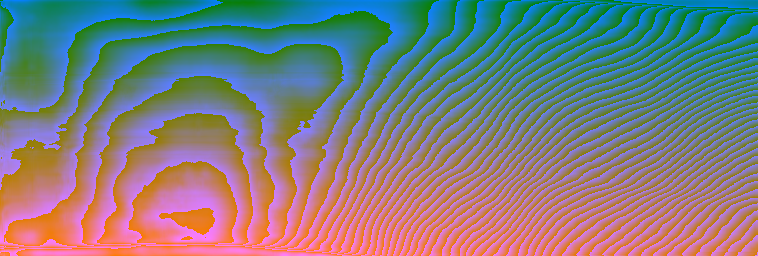}\hfill
  \includegraphics[width=0.31\linewidth]{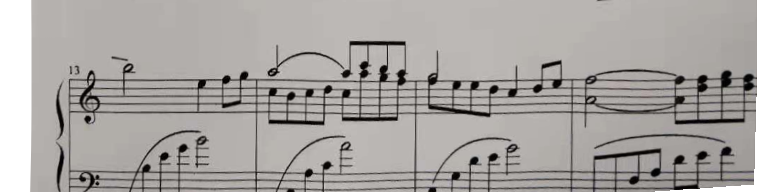}
  \vspace{1mm}
  
  {\small \textbf{(b)} Gauge: warped staff crop, model output, and rectified rendering. The middle image encodes the two predicted gauge channels in RGB: red and blue store the integer and fractional parts of vertical displacement $Y$, while green stores the local slope term $K$ used for horizontal remapping.}
\end{minipage}

\vspace{2mm}
\begin{minipage}{0.9\textwidth}
  \centering
  \includegraphics[width=0.47\linewidth]{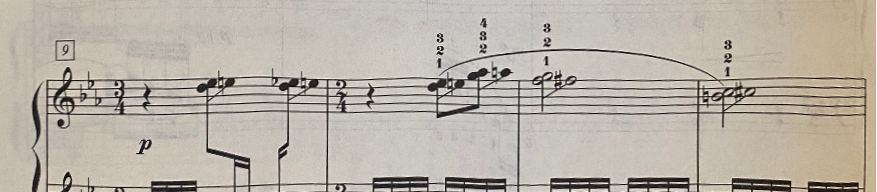}\hfill
  \includegraphics[width=0.47\linewidth]{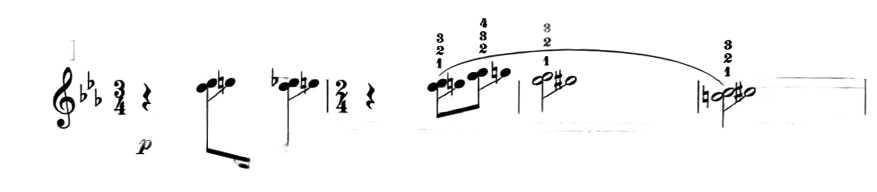}
  \vspace{1mm}
  
  {\small \textbf{(c)} Mask: staff crop and mask prediction. The mask model focuses on the current staff content, mainly for noise removal and foreground separation.}
\end{minipage}

\vspace{2mm}
\begin{minipage}{0.9\textwidth}
  \centering
  \includegraphics[width=0.47\linewidth]{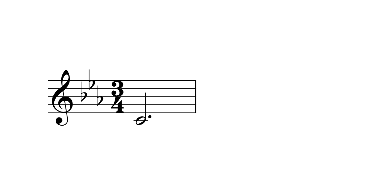}\hfill
  \includegraphics[width=0.47\linewidth]{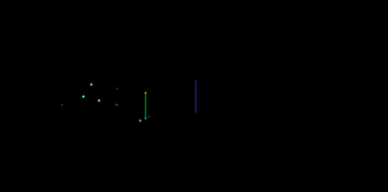}
  \vspace{1mm}
  
  {\small \textbf{(d)} Semantic: staff crop and symbol semantic prediction.\\ Each semantic is represented by a separate heatmap channel.}
\end{minipage}
\caption{Examples of visual predictors used before symbolic assembly. Panels (a), (c), and (d) show input-output pairs. Panel (b) shows the gauge input, the intermediate model output, and the final rectified rendering.}
\label{fig:cv-examples}
\end{figure}

The role of the assembly stage is to turn these local cues into measure-level candidates that the decoding stage can operate on. Figure~\ref{fig:cv-examples} gives concrete examples of the main predictors used before assembly. Layout establishes a page-to-system-to-staff hierarchy through a non-exclusive segmentation output. The gauge model is a regression network that predicts a semantic $y$ field and a local vertical-line slope field, which together support staff rectification for photos. The mask model is a two-channel foreground/background segmentation model used mainly for foreground extraction. The semantic model predicts localized notation evidence rather than complete long-range structures. Assembly then groups primitive detections (e.g., noteheads and stems) into higher-level musical objects and produces a candidate set for each measure.
Concretely, chord candidates are constructed by grouping noteheads that share the same stem evidence (for instance, a consistent stem tip or a shared stem $x$ position), and then collecting all noteheads that fall into the resulting stem-aligned region. Local attributes such as duration class, dots, and beam state are attached by matching nearby semantic cues (flags, dots, beam anchors) around the stem and noteheads. Rest candidates are built directly from rest detections, optionally augmented by nearby dots. The candidates include these event-like objects (chords and rests) as well as boundary nodes used by the topology model (e.g., BOS/EOS). Each candidate is associated with a staff identity and geometry features such as horizontal position and vertical extent. Local attribute hints are also attached when available, such as duration category, dots, beam or stem information, grace flags, and time-warp cues.

\begin{figure}[t]
\centering
\includegraphics[width=0.9\linewidth]{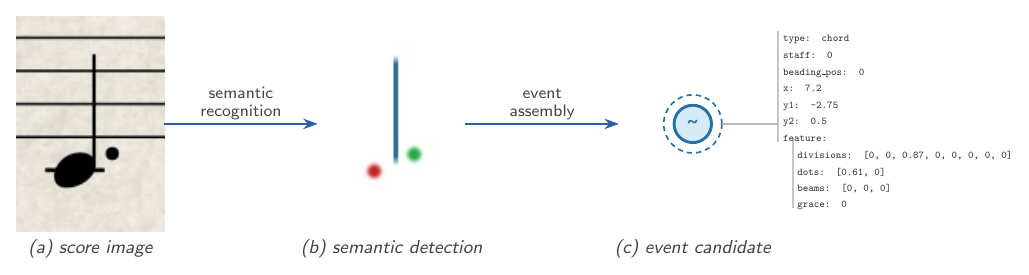}
\caption{Simplified pipeline from semantic recognition to event-candidate assembly. Dense semantic evidence is first predicted on the score image, then grouped into measure-level event candidates with attached geometry and local attribute hints. These event attributes come from geometric measurements detected during assembly together with the confidence values of the semantic points.}
\label{fig:assembly-flow}
\end{figure}

Figure~\ref{fig:assembly-flow} gives a simplified view of the transition from semantic recognition to event-candidate assembly.

\subsection{From Candidates to Editable Structure}

\begin{figure}[t]
\centering
\begin{minipage}{0.48\linewidth}
  \centering
  \includegraphics[width=\linewidth]{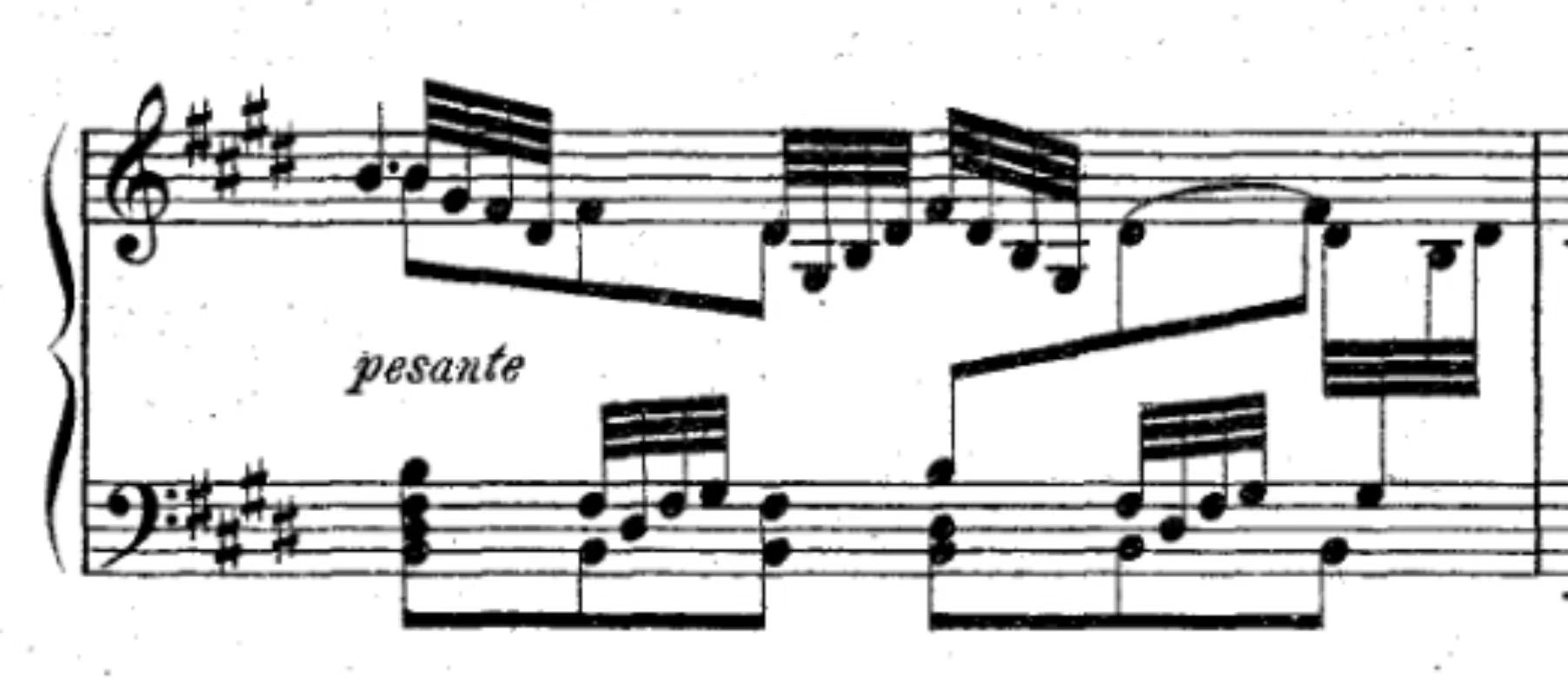}
  \vspace{1mm}

  {\small \textbf{(a)} Raw measure with severe voice ambiguity.}
\end{minipage}\hfill
\begin{minipage}{0.48\linewidth}
  \centering
  \includegraphics[width=\linewidth]{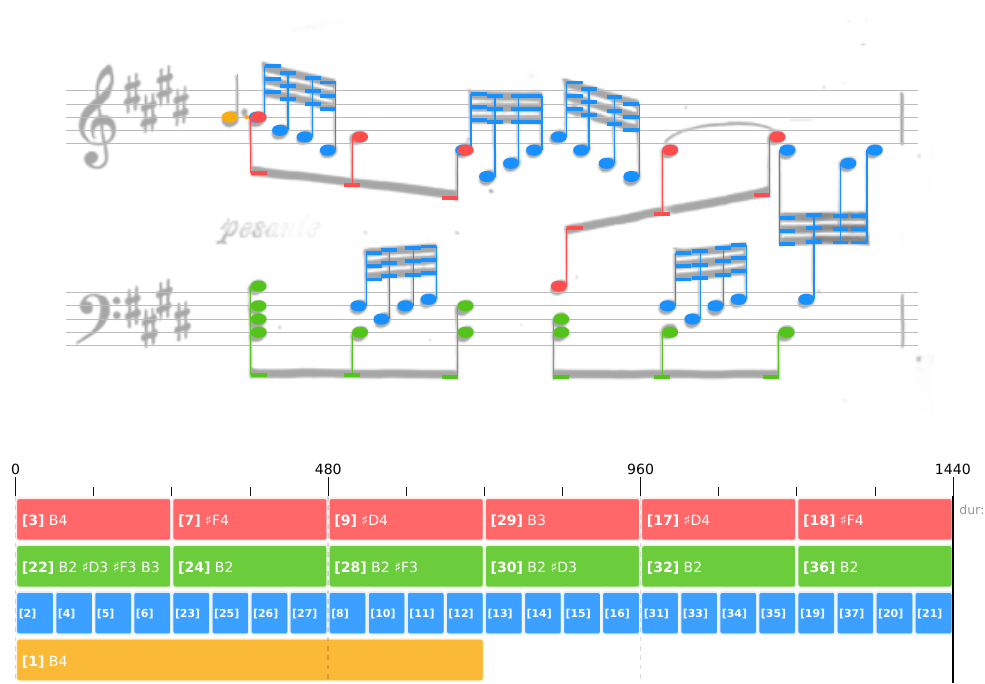}
  \vspace{1mm}

  {\small \textbf{(b)} Target voice structure after regulation.}
\end{minipage}
\caption{Regulation target on a difficult piano measure. (a) The image shows the original notation, where voice continuation and cross-staff assignment are ambiguous. (b) Each color denotes one voice in the regulated result. The strip below is a per-voice timeline view whose horizontal axis is the measure tick coordinate; each small block corresponds to one event shown above. This makes the recovered voice structure and time allocation visible at the same time.}
\label{fig:voice-target}
\end{figure}

The second stage takes the assembled candidate set and recovers an editable structure. We refer to this full second-stage process as \emph{regulation}. It fills the gap between local recognition output and a usable score representation. In engineering terms, it bridges the distance between local detections such as clefs, chords, and rests, and a globally usable structure defined by voices, tick positions, and corrected duration attributes. Within regulation, \emph{topology recognition} is the topology-dependent method that predicts and solves the voice- and time-organization of the candidate events. The core question is not whether a notehead exists, but how events should be organized into voices and time. Figure~\ref{fig:voice-target} illustrates the target of this step: the input is still a visually ambiguous measure, while the desired output is a voice-separated and temporally coherent event structure. In the solved view, each color denotes one voice, and the strip below shows the same result in a per-voice time-axis view.

Before regulation starts, the page has already been organized into a hierarchy of systems, staves, bars, and measures. We use the measure as the basic unit of decoding. A measure keeps the page and system context that is needed for interpretation, while preserving the staff identity of each detected object.

Within each measure, we distinguish between two kinds of discrete terms. \textbf{Context terms} provide local or global notation context but are not themselves scheduled as time events, such as clefs, key signatures, time signatures, and octave-shift markings. They help explain pitch coordinates, duration semantics, and local layout context. \textbf{Event terms} are the objects that participate in voice and time reconstruction, such as chords, rests, and explicit boundary nodes such as BOS and EOS. BOS is the start node of a voice chain in the topology and is anchored at the beginning of the measure. EOS is the corresponding end node and is anchored at the terminal bar position of the measure. These event terms are the direct objects operated on by the topology-recognition model and the broader regulation procedure.

At the representation level, each candidate event is encoded as an EventElement with a discrete type such as CHORD, REST, BOS, or EOS, together with geometry fields such as $x$, $y_1$, $y_2$, and $\mathit{pivotX}$. Each event also carries a local feature vector and soft attribute hints, including fields such as division, dots, beam, stem direction, grace, and other notation cues. The topology-recognition model uses these local signals as conditional evidence, while the full regulation process is free to revise them when a different global structure yields a more coherent measure.

\subsection{Outputs and Downstream Use}

The output of regulation is first designed to support conversion into multiple score formats. Once a measure has stable voices, tick positions, and rhythmic attributes, it can be exported to MusicXML, LilyPond, or an internal structured representation, depending on the downstream task. This conversion support is essential because the same regulated result must serve notation editors, rendering tools, playback systems, and internal data pipelines.

At the measure level, the core output fields include voice sequences, where each voice is an ordered list of event IDs; absolute tick positions inside the measure; and corrected rhythmic and notational attributes such as division, dots, time-warp, beam, stem, and grace-related fields. These fields make the result not only serializable, but also inspectable and editable.

Once the structure is in this form, it supports several downstream uses: validation, error correction, playback, and export. It is also the interface for annotation feedback. Difficult measures can be inspected and corrected by humans or agents, then written back as high-quality structured labels for later training and evaluation. In that sense, editability is not only a user-facing goal, but also the mechanism by which the system accumulates better data over time.

\section{Method}

Western staff notation is a formal symbolic language, but its structure sits somewhere between mathematics and natural language. It is not as strict as a mathematical formula, where every token usually has a single well-defined place in a rigid syntax. At the same time, it is far less free than natural language, because voice continuation, rhythmic grouping, pitch position, and bar-level timing still obey strong structural constraints. This intermediate character shapes our method.

Our design tries to use the strengths of both rule-based reasoning and deep learning. The role of rules is not to enumerate all musical meanings by hand, but to carry the parts of the problem that benefit from explicit deduction and validation. The role of deep learning is to absorb the ambiguity that is hard to formalize locally, especially when visual evidence is noisy, incomplete, or genuinely multi-modal. In other words, we try to let the model handle recognition where the task is loose and many-sided, while letting the solver enforce structure where the task becomes checkable.

This leads to two concrete principles. First, we make maximal use of the loose correspondence between horizontal layout and musical time. The $x$ coordinate is not a reliable clock, but it still carries enough weak timing information to constrain the search when combined with rhythmic hypotheses. Second, we try to decouple two kinds of information that are entangled in the image: pitch and voice. Vertical position is directly informative about pitch, but much less reliable as a voice label in dense polyphonic writing. Our topology-recognition method is designed around this separation. It uses learned predictions to propose local continuations, then uses search and structural evaluation to recover a measure-level organization in which time, pitch, and voice can be assigned more cleanly.

\subsection{Key Idea: Probability-Guided Chained Decision Search}

\begin{figure}[htbp]
\centering
\begin{minipage}[t]{0.48\linewidth}
  \centering
  \includegraphics[width=\linewidth]{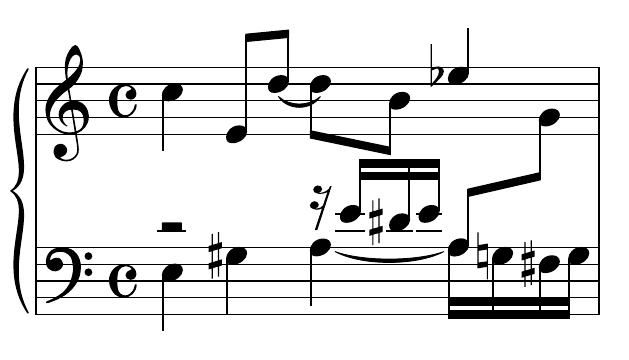}

  {\small \textbf{(a)}}
\end{minipage}\hfill
\begin{minipage}[t]{0.48\linewidth}
  \centering
  \includegraphics[width=\linewidth]{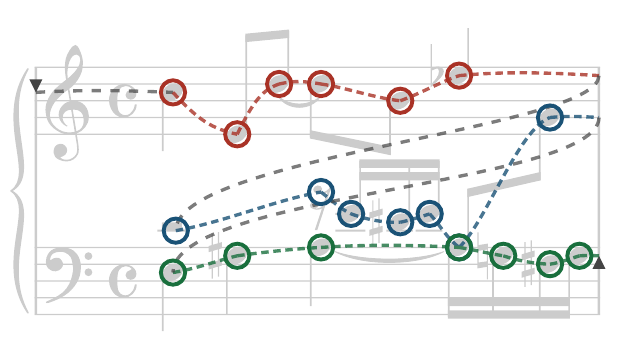}

  {\small \textbf{(b)}}
\end{minipage}
\caption{Measure regulation as chained structure recovery. (a) A polyphonic measure example. (b) Connect all events in a chain so that we can model the regulation problem as a MDP.}
\label{fig:bead-idea}
\end{figure}

Figure~\ref{fig:bead-idea} illustrates the intuition behind the name \emph{bead}. We treat the event candidates in one measure as beads that must be threaded into a single directed chain. In panel (b), the colored paths are drawn as one continuous threaded structure rather than as an unordered set of pairwise relations. This is the representation that BeadSolver operates on.

The modeling choice was motivated by the structure of polyphonic measures themselves. Within a measure, voices do not have a fixed global order, and they may interleave across the same time span. In principle, the event graph could therefore be constructed from any starting corner and in many different traversal orders, for example by following a reliability-first edge policy. But this freedom also creates heavy path redundancy: the same final structure can be reached through roughly $O(n!)$ equivalent construction paths. Allowing the order of voices to vary while keeping the node order inside each voice fixed is a reasonable compromise. It preserves the ambiguity that matters and removes most of the combinatorial waste. From there, the search can be viewed as a prefix-expansion problem, which is where the Monte Carlo tree search intuition enters: the solver expands a prefix state, evaluates its candidate continuations, and repeatedly balances probability-guided exploration with structural pruning.

Let the candidate event set of a measure be
\[
E = \{e_0, e_1, \dots, e_n, e_{n+1}\},
\]
where $e_0=\mathrm{BOS}$ and $e_{n+1}=\mathrm{EOS}$ are explicit boundary nodes. We define a search state at step $t$ as
\[
s_t = (P_t, R_t, \rho_t),
\]
where $P_t$ is the current chain prefix, $R_t = E \setminus P_t$ is the remaining candidate set, and $\rho_t$ collects the rhythmic attributes already assigned along the prefix. In the tree search, each branch corresponds to a distinct sequence of committed assignments; in our implementation this is realized by applying these assignments in-place on a shared cluster state with rollback on backtracking (and only snapshotting the best-so-far state), rather than deep-copying $\rho_t$ at every expansion. An action is a three-part local decision
\[
a_t = (u_t, d_t, o_t),
\]
obtained by taking the argmax from the current posterior distributions over the next-node, division, and dots choices:
\[
u_t = \arg\max \hat{p}^{(u)}_t,\qquad
d_t = \arg\max \hat{p}^{(d)}_t,\qquad
o_t = \arg\max \hat{p}^{(o)}_t.
\]
Here $u_t \in R_t \cup \{\mathrm{EOS}\}$ is the next node, $d_t$ is the division class, and $o_t$ is the dot count. Together, they determine the note duration in ticks,
\[
\mathrm{duration}(d_t, o_t) = 1920 \cdot 2^{-d_t} \cdot \bigl(2 - 2^{-o_t}\bigr),
\]
where $o_t=0,1,2$ corresponds to undotted, dotted, and double-dotted values. For example, a dotted eighth note corresponds to $d_t=3$ and $o_t=1$, giving $1920 \cdot 2^{-3} \cdot (2-2^{-1}) = 360$ ticks. The transition is the chained update
\[
P_{t+1} = P_t \oplus (u_t, d_t, o_t),
\]
where the prefix is initialized with \textsc{BOS}. Each action appends $u_t$ to the end of $P_t$; if $u_t=\mathrm{EOS}$, the current voice is closed, \textsc{BOS} is removed from the prefix, and \textsc{BOS} is appended again at the end to mark the start of the next voice. In this sense, \textsc{BOS} appears only once in $P_t$ at any time.
and the decoding objective is
\[
\hat{\zeta}
=
\arg\min_{\zeta \in \mathcal{T}(E)}
\mathcal{L}(\zeta),
\label{eq:topology-search}
\]
where $\zeta$ is a complete decoding trajectory for a measure and $\mathcal{L}$ is the global structural loss defined over the whole measure.

A simplified illustration of the overall tree-search workflow is shown in Figure~\ref{fig:tree-search-flow}.

\begin{center}
\captionsetup{type=figure}
\captionof{figure}{A simplified illustration of the overall tree-search workflow in principle. In the Pass step, the bars above the event nodes represent the predicted probabilities of the next-node choices.}%
\label{fig:tree-search-flow}
\end{center}
{\centering
\includegraphics[width=0.72\linewidth]{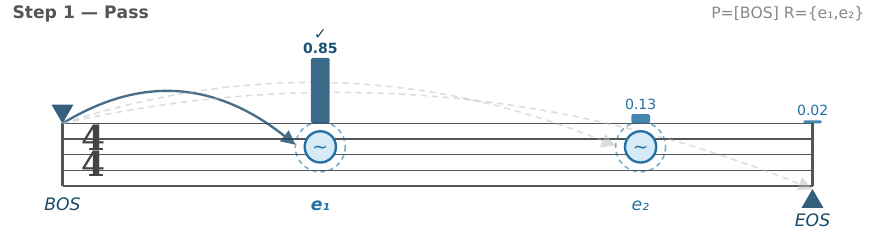}\par\vspace{1pt}%
\includegraphics[width=0.72\linewidth]{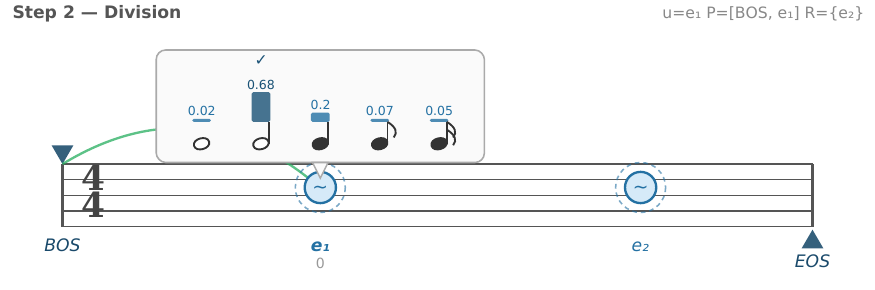}\par\vspace{1pt}%
\includegraphics[width=0.72\linewidth]{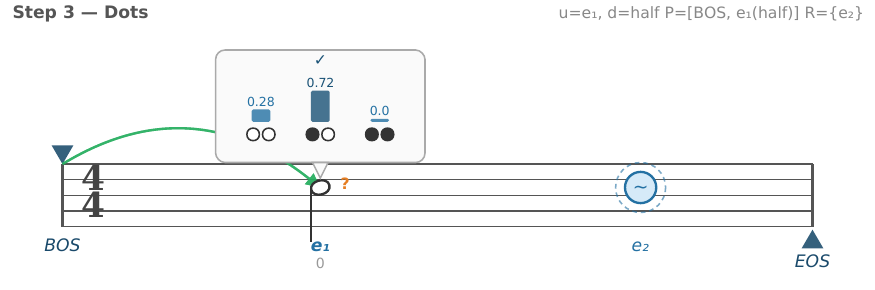}\par\vspace{1pt}%
\includegraphics[width=0.72\linewidth]{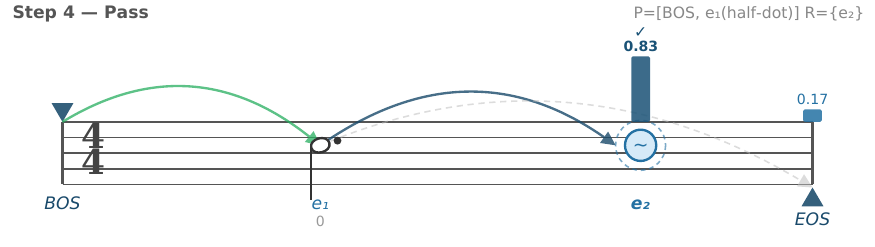}\par\vspace{1pt}%
\includegraphics[width=0.72\linewidth]{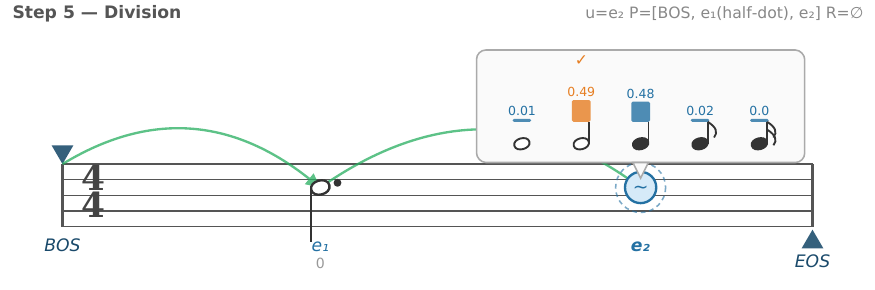}\par\vspace{1pt}%
\includegraphics[width=0.72\linewidth]{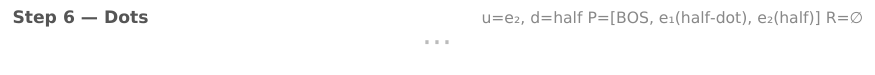}\par\vspace{1pt}%
\includegraphics[width=0.72\linewidth]{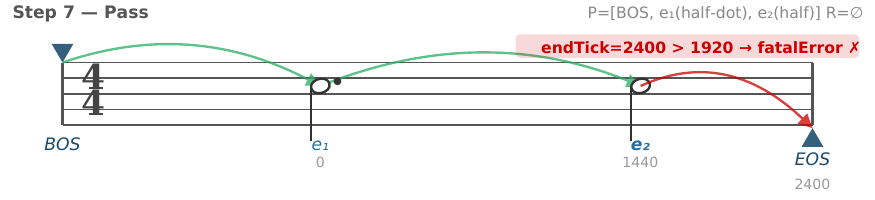}\par\vspace{1pt}%
\includegraphics[width=0.72\linewidth]{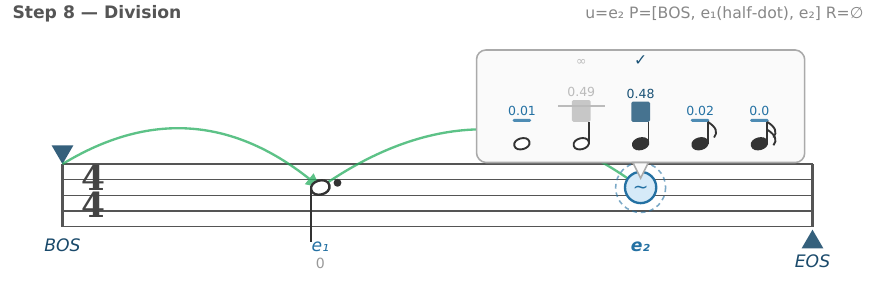}\par\vspace{1pt}%
\includegraphics[width=0.72\linewidth]{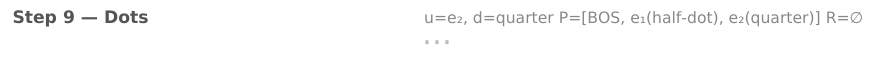}\par\vspace{1pt}%
\includegraphics[width=0.72\linewidth]{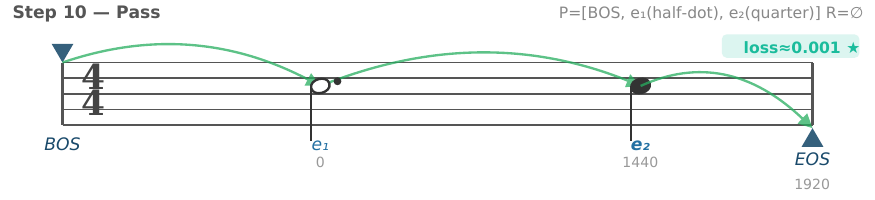}\par\medskip
}

In this view, topology recognition becomes a Markov decision process (MDP). At each step the solver maintains a state $s_t$ that summarizes the current prefix $P_t$, the remaining candidates $R_t$, and the rhythmic attributes already committed along the prefix. An action $a_t$ extends the prefix by selecting the next node $u_t$ (or a boundary marker) and assigning a small set of local rhythmic attributes. We represent voice switches explicitly in the chain, which allows the solver to interleave voices while keeping the order inside each voice meaningful.

The objective selects the trajectory with the best global structure under $\mathcal{L}$. While the loss can combine several consistency checks, the core idea is geometric: musical time should not be wildly inconsistent with horizontal layout. We treat $x$ as a weak time signal and penalize structures that force time to twist back and forth along the $x$ axis. This is especially useful in dense polyphony, where purely local choices have many plausible alternatives, but only a few lead to a globally coherent measure.

\begin{figure}[htb]
\centering
\begin{minipage}[b]{0.52\linewidth}
  \centering
  \includegraphics[width=\linewidth]{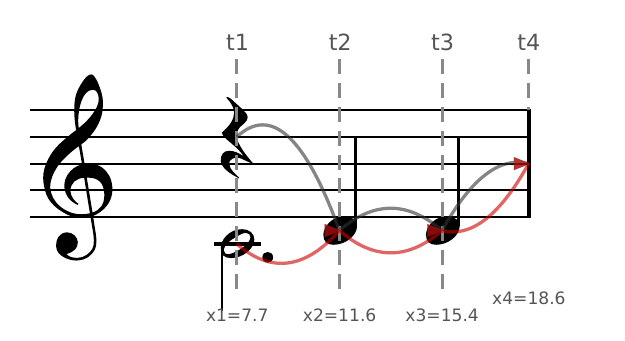}
\end{minipage}\hfill
\begin{minipage}[b]{0.44\linewidth}
  \centering
  \includegraphics[width=\linewidth]{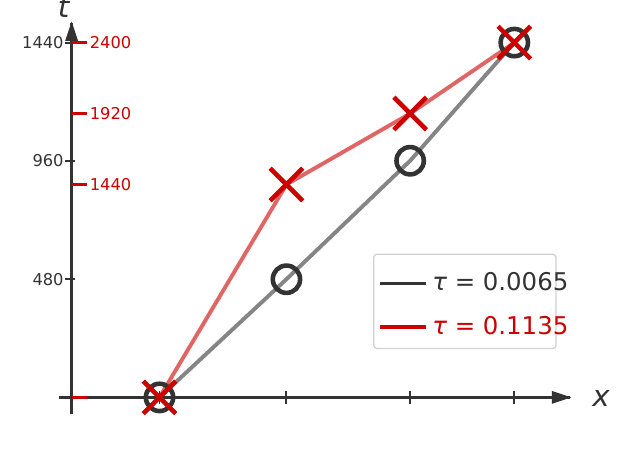}
\end{minipage}
\caption{%
  \textbf{$x$--tick geometry consistency.}
  \emph{Left:} an ambiguous measure; there are 2 potential topology candidates for regulation.
  \emph{Right:} events plotted in the $(x, t)$ plane, where $t$ is the cumulative tick position obtained by accumulating event durations along the voice chain; each group is normalised by its total measure duration.
  A well-regulated voice (black circles) distributes its events nearly uniformly along both axes.
  An ill assignment (red crosses) results in a slope that deviates from 1.
  We designed a metric named \emph{tick twist} ($\tau$) to compute the peak error value from the most abnormal local segment in a topology structure.
}
\label{fig:xtick}
\end{figure}

Figure~\ref{fig:xtick} illustrates this $x$--tick geometry intuition. Two candidate topologies can look locally reasonable, yet behave very differently when their events are plotted in the $(x,t)$ plane. A coherent voice tends to advance in both $x$ and tick at a roughly stable rate, while an ill assignment introduces an abnormal segment that deviates sharply. Our loss captures this through a metric such as \emph{tick twist} ($\tau$), which highlights the most inconsistent local segment and provides a strong pruning signal during search. By design, abnormal but still monotone slopes receive penalties in the range $[0,1)$, while backward-going segments receive penalties greater than $1$ and are treated as fatal during search; the precise definition is given in Appendix~B.

Finally, the local predictions produced by the visual stage are treated as probability distributions rather than fixed labels. For example, duration-related fields (division and dots) and other rhythmic cues are carried as soft evidence attached to each candidate event. The solver can therefore correct local mistakes when a different global structure yields a better-consistent measure, instead of being forced to accept the first-stage decision verbatim.

\subsection{Role of the Model}

The tree search solver relies on a learned model. We refer to this model as \emph{BeadPicker}. It serves as a reusable scoring function that can be queried repeatedly during decoding, under different chain prefixes.

Formally, given the current decoding state $s_t$ and the measure-level candidate set $E$, BeadPicker defines a conditional prediction function
\[
f_\theta(s_t, E)
=
\bigl(\mathbf{p}^{(u)}_t,\; \mathbf{p}^{(d)}_t,\; \mathbf{p}^{(o)}_t,\; \mathbf{z}_t\bigr),
\]
where $\mathbf{p}^{(u)}_t = \{p_i\}$ is the posterior distribution over next-node choices used in Section~4.3, $\mathbf{p}^{(d)}_t$ is the posterior over division classes, $\mathbf{p}^{(o)}_t$ is the posterior over dot counts, and $\mathbf{z}_t$ collects the remaining outputs used mainly by evaluation, such as tick, beam, and completion-related fields.

In the MDP view, BeadPicker plays two roles.
\begin{itemize}
  \item \textbf{Policy (action guidance).} Given the current prefix, the model predicts a probability distribution over which remaining event should be selected next (including the option to emit a boundary marker to end the current voice). It also predicts predispositions for the local rhythmic decisions that are explicitly part of the action space, notably the division class and dot count. Together, these outputs guide the tree search to expand promising branches first.
  \item \textbf{Value reference (evaluation signals).} The model also outputs fields that are mainly consumed by the evaluation function, rather than being treated as actions. The most important one is an absolute tick for each event, which directly enters the $x$--tick consistency checks. In addition, a small set of completion-related hints (e.g., grace/full-measure/fake flags) help detect unfinished or degenerate hypotheses, and beam-related predictions can help penalize broken grouping patterns.
\end{itemize}

\paragraph{Inputs.}
Each query operates on a measure-level candidate set produced by the first stage. For every candidate event, the input includes its discrete type (chord, rest, boundary), staff identity, geometry fields (notably horizontal position $x$ and vertical extents), and a local feature vector extracted from the visual predictors. The query also includes a prefix encoding that marks which elements are already fixed in the current chain. This prefix context is the main mechanism that turns a single encoder-style model into an approximate autoregressive decoder when called repeatedly during search.

\paragraph{Outputs.}
For each candidate event, BeadPicker outputs a successor score and a set of attribute predispositions. During decoding, the successor head is interpreted as an action distribution for the \textsc{Pass} step. Predispositions for division and dots are used to guide the corresponding local action choices. Tick and evaluation-related fields (including beam) are primarily used inside global evaluation, where they shape losses such as tick twist, enforce basic consistency (e.g., beam grouping), and help detect unfinished or invalid hypotheses.

\paragraph{Successor supervision.}
The successor label is typically one-hot: for a given prefix, exactly one next event is marked as positive. At voice boundaries, we use a multi-hot label that marks the valid next-voice start candidates simultaneously. This avoids baking an arbitrary ordering between voices into the model, while still training it to predict plausible continuations under the chain representation.

Tick prediction is also useful beyond regulation search. For formats that do not require explicit voice assignments (e.g., MIDI), once each event is placed on a tick timeline, recognition is essentially complete. In such cases, the model can be run in a single forward pass to produce a reasonable tick estimate for each event and export a playable sequence. Moreover, the predicted tick of the EOS node provides an estimate of the measure length in ticks, which can serve as an implicit time-signature cue. In practice, we use it as an important reference signal to sanity-check the meter inferred by the vision stage and to flag potential time-signature errors early.

\subsection{Decoding and Evaluation}

The goal of decoding is to turn a set of measure-level candidates into a usable structure: a small number of voice chains, with plausible tick positions and rhythmic attributes, and with enough consistency that the result can be exported and edited. The full algorithmic description is provided in Appendix~B. Here we focus on the intuition behind the search loop and how evaluation keeps it grounded.

\paragraph{Branch selection and pruning.}
The search tree is expanded in a probability-guided way~\cite{coulom2007mcts,browne2012mcts}. For a node with posterior branch probabilities $\{p_i\}$, we pick the next branch by an adjusted score that balances exploitation and exploration,
\[
  \hat{p}_i = \frac{p_i}{\mathrm{visits}(c_i) + 1},
\]
where $c_i$ denotes the $i$-th child node. This is a simplified PUCT-like adjustment in the spirit of AlphaGo-style search~\cite{silver2017go}: high-probability branches are explored early, but every branch is eventually revisited less and less often as its visit count grows.

At the same time, we do not maintain the standard MCTS $Q$ value. For score topology, evaluator-acceptable solutions are extremely sparse in the full search space, and the final outcome is relatively sensitive to each local choice. A locally correct step usually does not contribute in an accumulative way along the trajectory, as it often does in board games; a branch may look plausible for many steps and fail abruptly because of a later structural contradiction, while an unlikely local choice may become correct only under a very specific continuation. For this reason, we rely on posterior guidance plus explicit evaluation, rather than on a stable per-node value estimate.

Pruning is then handled separately. We accumulate a log-probability penalty along the path, which we call \emph{pretentiousness}, and cut off paths whose cumulative penalty exceeds a tolerance. This behaves like a lazy beam threshold inside tree search: branches that are globally disfavored by the model are removed early, while evaluation can still rescue a locally unlikely step if it leads to a substantially better structure.

\paragraph{Evaluation.}
Evaluation turns a partial chain into a scalar loss that reflects both structural usability and global consistency. We first propagate tick positions along each voice using the chosen division/dots and the measure context. The loss then penalizes problems such as remaining uncovered events, excessive voice fragmentation, large gaps inside a voice, and violations of the $x$--tick geometry constraint (e.g., high tick twist). Finally, we include a term that reflects how far the chosen path deviates from the model's distribution, so the solver prefers structures that are not only consistent, but also plausible under learned priors. Some failures are treated as fatal and trigger immediate pruning, such as severe $x$--tick reversals or structurally degenerate partial solutions.

\paragraph{Stopping.}
Search stops when the budget is exhausted, the tree is fully explored, or a sufficiently low-loss solution is found.

\paragraph{Outputs.}
The best chain is converted back into a regulated measure structure. Concretely, we extract voice sequences by splitting the chain at voice boundaries, finalize per-event tick positions, and commit the rhythmic attributes needed for duration (division, dots, and time-warp when present), together with auxiliary fields such as beam and grace flags. The result is stored as an editable measure-level representation: a list of events with geometry and semantic attributes, a small set of voice tracks defined as ordered event IDs, and a measure duration in ticks.

This representation is designed to support multiple downstream views without re-solving the structure. For example, it can be converted into per-voice timelines where events are ordered by tick and gaps are made explicit as rests or spaces, which is essential for rendering and playback. When the recovered ticks disagree with nominal durations (e.g., due to tuplets or local overlaps), the same structure can carry a minimal local time-warp correction so that the timeline remains consistent. These views then serve export (MusicXML/LilyPond/MIDI) as well as interactive inspection and correction.

\subsection{Train a Topology Model}

By introducing tree search, we have turned a single-shot topology prediction problem into a multi-step, conditional prediction problem. Instead of asking a model to output a complete graph in one pass, we query it repeatedly under different prefixes and let it answer simpler questions: which event should come next, and which local rhythmic options are most plausible for that choice.

The candidate set within one measure is also small enough to support global reasoning. A sheet music measure typically contains far fewer than 100 event terms after assembly. We therefore use a Transformer-based model~\cite{vaswani2017attention} with global self-attention over the whole candidate set, and we condition it on the current chain prefix. This gives the model full context for each decision while keeping the interface compatible with search.

\subsubsection{Architecture}

\begin{figure}[htbp]
\centering
\includegraphics[width=0.98\linewidth]{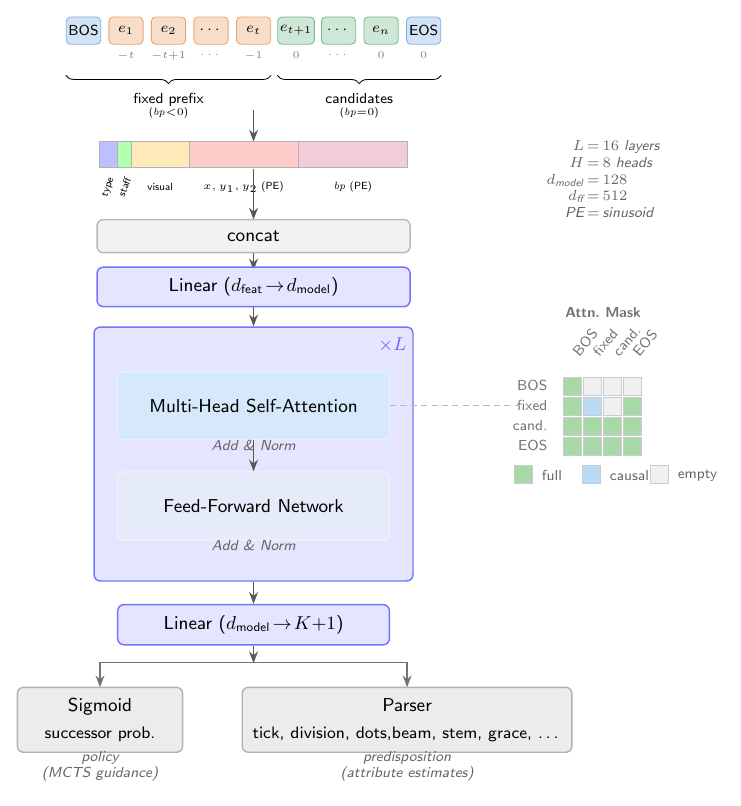}
\caption{BeadPicker architecture for topology recognition. The model reads measure-level event candidates together with geometry, local attribute hints, and chain-prefix context, and predicts successor probabilities and related fields used by the solver. In the figure, the values shown under each $e_i$ indicate its bead position. In practice, fixed-prefix elements are sorted by bead position (not by element index), and the two orders do not necessarily coincide; the diagram uses a simplified example for readability.}
\label{fig:beadpicker-arch}
\end{figure}

BeadPicker is an encoder-style Transformer that reads the full set of measure candidates at once and predicts several output heads jointly (Figure~\ref{fig:beadpicker-arch}). The main head is the successor distribution used by the \textsc{Pass} step. Additional heads provide the attribute predispositions used by the solver, including tick estimates, division and dots preferences, and lightweight structural hints such as beam state and completion-related flags.

\paragraph{Bead Position Encoding}

We treat an unfinished measure during search as a partially realized sequence, in the same spirit as autoregressive decoding. We therefore assign each element a position encoding called \emph{bead position}. All unfixed candidates share position $0$, while the fixed prefix is encoded by negative offsets that backtrack from the current tip (more recent prefix events are closer to $0$). This gives the model explicit information about which part of the measure has already been committed.

\paragraph{Partial Causal Masking}

For a fixed event whose voice assignment has been decided, its tick value is determined by the events to its left within that voice. We therefore treat the fixed prefix as a causal chain and apply a causal mask after sorting fixed elements by horizontal position $x$. This prevents a fixed event from attending to later fixed events, which aligns the model with the left-to-right tick accumulation used by decoding. There is one exception: the EOS boundary node remains globally visible. This lets every element perceive the measure span and boundary conditions.

We sort by $x$ rather than by bead position. Bead position reflects the current chain order, but a voice can contain implicit rests (gaps) that are not represented as explicit events. These ``bubbles'' break the assumption that consecutive chain elements carry all necessary left context. Sorting fixed elements by $x$ restores a layout-consistent notion of ``past'' that remains informative.
The causal mask is therefore not isolated by voice; fixed events share the same $x$-ordered causal view.

In contrast, unfixed candidates still need global context to judge their potential role in the final structure, so they keep full attention over the non-padding elements.

\paragraph{Loss Functions}

BeadPicker is trained with a multi-task objective that matches its role in search. The successor head is trained with a binary cross-entropy loss, using one-hot labels for ordinary steps and multi-hot labels at voice boundaries to avoid learning an arbitrary voice ordering. This successor loss is weighted higher than the attribute heads, since choosing the next event is the most consequential decision for the search trajectory.

The remaining heads supervise the attribute predispositions used by decoding and evaluation. Division, dots, beam state, and stem direction are trained as discrete classifications; grace, time-warp, full-measure, and fake flags are trained as binary predictions. Tick supervision uses both an absolute tick target and an auxiliary relative-tick consistency term defined over event pairs, which encourages the model to preserve local time differences even when the absolute origin is ambiguous. Together, these losses teach the model to provide usable priors, while still allowing the solver to revise local decisions under global constraints.

\paragraph{Tick Prediction as a Vector (\emph{vtick})}

Predicting tick as a single scalar regression target is workable, but it is not an ideal fit for musical time. In music, tick values are derived from beats and are best understood as a family of fractions rather than as generic real numbers. Two values can be numerically close but have very different musical meaning and prior likelihood. For example, $1/2$ naturally suggests a regular duple subdivision, while a near value like $11/23$ is more likely to appear as an irregular slice inside a cadenza-like passage. To reflect this structured, non-uniform distribution, we also support a vectorized tick target: the tick value is converted into a factorized multi-class code under a fixed base that covers typical measure lengths. The model predicts this code with a multi-label loss and the code is decoded back into an integer tick.

\noindent\textbf{Example.}
Let the base be $1920 = 2^7 \cdot 3 \cdot 5$. We represent an integer tick $t$ by digits in the mixed radix $(2,2,2,2,2,2,2,3,5)$ (i.e., the prime factors, from most significant to least). These radices reflect common rhythmic subdivision patterns in music, and are also consistent with the common MIDI tick convention.
For $t=1234$, the digits are
\[
  (1,0,1,0,0,1,0,0,4),
\]
because $1234 = 1\cdot 960 + 1\cdot 240 + 1\cdot 30 + 4\cdot 1$.
The corresponding vector code concatenates one-hot encodings of each digit with the zero state dropped, giving
\[
  \mathrm{vtick}(1234) =
  [1 \;|\; 0 \;|\; 1 \;|\; 0 \;|\; 0 \;|\; 1 \;|\; 0 \;|\; 0,0 \;|\; 0,0,0,1].
\]

This representation gives the model an easier learning problem: it can learn the ``digits'' of a tick value as separate sub-decisions, rather than fitting a single regression head across a wide range. In practice, it improves numerical stability and encourages exact tick outputs, which directly benefits the downstream evaluation terms that depend on tick geometry.

\subsubsection{Training Data Pipeline}

In the vision stage, we already rely on a full score-synthesis pipeline to produce diverse training images. Topology recognition has a different data requirement. Here the priority is \emph{structural fidelity} rather than visual appearance: we need many measure-level samples whose voice structure and rhythmic semantics follow the real distribution of written music. Acquiring such corpora at scale is also a long-term goal of OMR itself. However, our supervision only needs independent, single-measure samples, which makes the problem easier. With a small seed corpus of real scores, we can train an autoregressive music-language model and then use large-scale random sampling to obtain many non-duplicate measure structures that are close to the target distribution.

\paragraph{Music DSL \emph{Paraff}.}
To make this generation and conversion pipeline practical, we devised a compact language named Paraff. Paraff is inspired by LilyPond and follows an interleaved, measure-level description style, similar to LilyPond's parallel notation mode. Its syntax is intentionally simplified: tokens are grouped into a finite set of part-of-speech categories, and allowed local continuations are defined by a part-of-speech transition matrix. During sampling, this lets us apply mask-guided decoding that rules out illegal next tokens, so random exploration does not break grammar and the generated measures remain parseable and structurally valid.
To encourage the model to generate complex measures, we also prepend a small set of rule-derived prompt tokens during training to indicate desired structural tendencies (e.g., voice count and rhythmic diversity). At generation time, these prompt tokens condition sampling and provide a simple handle to steer complexity. See details in Appendix~D.

\begin{figure}[htbp]
\centering
\includegraphics[width=0.98\linewidth]{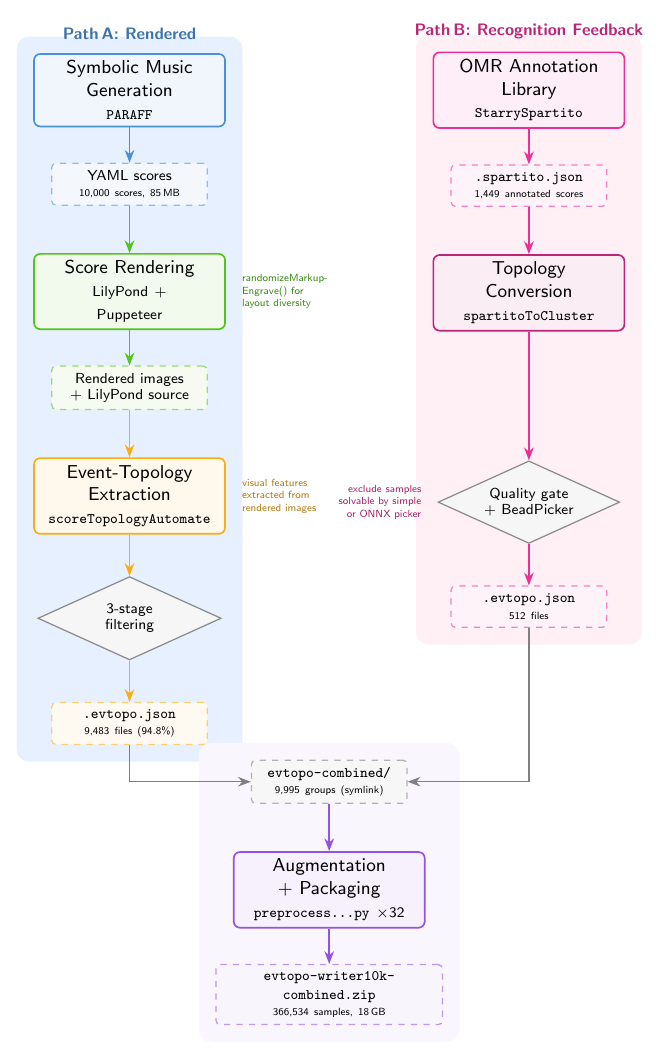}
\caption{Training-data pipeline for topology recognition. Structured symbolic music is rendered into score form, converted into event-topology supervision, then augmented and packaged into measure-level training samples.}
\label{fig:topology-data-pipeline}
\end{figure}

\paragraph{From symbolic music to topology samples.}
Figure~\ref{fig:topology-data-pipeline} summarizes the pipeline. We start from structured symbolic music that already has clean voice and duration semantics. We render it into staff notation to obtain realistic spatial layouts and visual cues, then convert the rendered result into the same measure-level event candidates used by the solver. Because the symbolic source provides the ground-truth structure, we can automatically derive the successor labels and attribute targets needed to train BeadPicker. Finally, we apply augmentation and packaging steps to match the diversity of real-world layouts.
To preserve geometric realism, symbol positions come from engraving: we convert Paraff to LilyPond and let LilyPond lay out the measure, then use the Lotus\footnote{\href{https://github.com/k-l-lambda/lotus}{https://github.com/k-l-lambda/lotus}} SVG pipeline to parse the engraved SVG and recover per-glyph geometry in staff-space units. Candidate events inherit their $x$ positions and extents from these measured glyph locations, so the resulting topology samples follow real engraving constraints.
The ``randomized layout'' in this stage refers to randomized engraving conditions rather than random geometry noise: we vary staff size and page dimensions (which affect line breaking and horizontal spacing), toggle paper options such as raggedness, and optionally switch to a compact layout preset. This yields diverse but still physically plausible coordinate distributions for training.

\paragraph{Feedback and dataset mixing.}
The synthetic pipeline covers a wide range of rhythmic and polyphonic patterns, but it cannot represent every artifact found in photos and historical prints. For this reason, we also include measures corrected through interactive workflows as high-quality samples. Over time, this creates a practical closed loop: the deployed two-stage system produces candidates, the solver proposes a structure, hard cases are corrected, and the corrected structures are recycled as training data.

\section{Experiments}

The main body of experiments in this paper was completed during 2021--2023, as part of building and iterating the production OMR system. Besides the BeadSolver method described above, we had already implemented two alternative second-stage regulation strategies in earlier iterations: a fully rule-based greedy regulator, and a \emph{linear-equations} regulator that combines strong rules with lightweight learned cues. We report these as baselines to contextualize the practical value of probability-guided search.

\subsection{Baselines}
\paragraph{Greedy regulation (rules only).}
This baseline groups events into roughly x-aligned columns by horizontal overlap, assigns ticks by sweeping columns left-to-right and advancing time to the earliest unfinished end time, then forms voices by chaining events forward using the assigned ticks. This algorithm works well for monophonic and single voice chords scores. It is fast and deterministic, but it cannot backtrack.

\paragraph{Linear-equations regulation (strong rules + weak learning).}
This baseline estimates horizontal and vertical link evidence between events and searches over graph-building actions under a fixed budget. Each candidate topology is then validated by solving a small linear system that infers per-event time-warp factors, so infeasible structures can be pruned early.
The key constraint exploited by this check is the \emph{isochrony} implied by voice-aligned structure: events that are aligned across voices should agree on their accumulated time, which yields a compact set of linear equalities.
The learned part in this baseline only provides a weak local prior: it suggests both which event-to-event continuations look more plausible and which events are more likely to align vertically in time, in a single shot for the whole measure.

\subsection{Datasets}
We organize experiments around two questions: (1) whether the topology model can be trained on difficult polyphonic patterns under limited real supervision, and (2) whether probability-guided search improves regulation quality relative to rule-based baselines.

\paragraph{Training set: \texttt{evtopo-writer10k-combined}.}
The main training set is a large, measure-level event-topology corpus built by mixing two sources (see \tabref{tab:train-stats}). The first source is rendered data: Paraff-generated symbolic music is converted to LilyPond, engraved with randomized layout, and then converted into measure-level event candidates with ground-truth structure. The second source is recognition-feedback data extracted from corrected scores, which provides hard cases that match the real distribution more closely. To emphasize high-quality corrected samples, we use three sampling weights (1, 4, 100), which yields 918,930 effective weighted samples.
We do not currently report a dedicated cross-staff prevalence statistic for the training corpus. The generator and the feedback portion both include cross-staff patterns, but their relative frequency may differ from the independent LilyPond test set; we plan to include such stratified statistics in a future dataset release.

\paragraph{Three-stage quality filtering.}
Because the rendered source is produced by a long toolchain, we apply quality checks at three points and drop any score that fails. The principle is to remove samples that are structurally ill-defined, numerically invalid, or obviously inconsistent in space-time ordering, rather than to overfit to a particular style.
First, at the symbolic stage, we only keep measures that are parseable and structurally complete, and we reject degenerate cases such as empty declared staves or unclosed rhythmic constructs (e.g., broken beaming or invalid tuplet closure).
Second, at the engraving stage, we treat rendering failures and known unstable engraving patterns as hard filters (e.g., severe stem or beam anomalies reported by the engraver log), and we also drop samples that exceed a runtime budget.
Third, after converting the rendered result into event-topology clusters, we keep only clusters that pass basic structural validity: they must contain real note/rest events, all coordinates and ticks must be finite, and the implied time order should not strongly contradict horizontal layout (a strong ``time goes forward but moves left'' pattern). During packaging we further remove clusters that are too small to train on, fall outside target label ranges, or lack a valid supervision order.

\begin{table}[t]
\centering
\small
\setlength{\tabcolsep}{5pt}
\begin{tabular}{lr}
\toprule
\textbf{Training data statistics} & \textbf{Value} \\
\midrule
Groups (scores) & 9,995 \\
Augmented measure samples & 366,534 \\
Augmentation & 32$\times$ \\
Rendered-source groups & 9,483 \\
Feedback-source groups & 512 \\
Sampling weights & 1 / 4 / 100 \\
Effective weighted samples & 918,930 \\
\bottomrule
\end{tabular}
\caption[Training Dataset Statistics]{Training Dataset Statistics. Key properties of \texttt{evtopo-writer10k-combined}.}
\label{tab:train-stats}
\end{table}

\paragraph{Weighted sampling.}
The recognition-feedback portion is small by count but disproportionately important: it contains hard measures that match production error patterns. We therefore use weighted sampling so that corrected examples are seen more often during training, without losing the breadth of the rendered corpus. Table~\ref{tab:train-weights} summarizes the three weight tiers and their effective contribution.

\begin{table}[t]
\centering
\small
\setlength{\tabcolsep}{4.5pt}
\begin{tabular}{lrrrr}
\toprule
\textbf{Category} & \textbf{Samples} & \textbf{\%} & \textbf{Weight} & \textbf{Effective \%} \\
\midrule
Unannotated & 295,554 & 80.6 & 1 & 32.2 \\
Verified (not patched) & 67,444 & 18.4 & 4 & 29.4 \\
Patched & 3,536 & 1.0 & 100 & 38.5 \\
\bottomrule
\end{tabular}
\caption[Training Sample Weights]{Training Sample Weights. Example weighting used for training-time sampling.}
\label{tab:train-weights}
\end{table}

\paragraph{Split protocol.}
We split by score group to avoid leakage between training and validation. Following \tabref{tab:split}, we use 9,697 groups (356,431 samples) for training and 198 groups (6,392 samples) for validation; 100 groups are left unused by this split scheme.

\begin{table}[t]
\centering
\small
\setlength{\tabcolsep}{5pt}
\begin{tabular}{lrr}
\toprule
\textbf{Split} & \textbf{Groups} & \textbf{Samples} \\
\midrule
Train & 9,697 & 356,431 \\
Validation & 198 & 6,392 \\
Unused & 100 & 3,711 \\
\bottomrule
\end{tabular}
\caption[Train/Validation Split]{Train/Validation Split. Group-level split for \texttt{evtopo-writer10k-combined}.}
\label{tab:split}
\end{table}

\paragraph{Group size distribution.}
Most groups have a similar number of measures, but anthologies and collections create a long tail. Table~\ref{tab:group-size} provides a compact summary.

\begin{table}[t]
\centering
\small
\setlength{\tabcolsep}{4.5pt}
\begin{tabular}{lrrrrrr}
\toprule
Percentile & P1 & P5 & P50 & P75 & P95 & P99 \\
\midrule
Samples per group & 11 & 24 & 31 & 35 & 39 & 188 \\
\bottomrule
\end{tabular}
\caption[Group Size Distribution]{Group Size Distribution. Distribution of sample counts per score group.}
\label{tab:group-size}
\end{table}

\paragraph{Independent test set: \texttt{lilypond-topology}.}

For evaluation, we use an independent LilyPond-based test set summarized in \texttt{data/test-overview.tex}. It contains 107 scores and 5,317 measure-level samples. For the direct spartito-versus-ground-truth comparison reported below, this corresponds to 88,754 scored events after excluding boundary tokens and unmatched extras from the denominator. The collection contains a large fraction of genuinely polyphonic writing: 96.3\% of scores contain at least two voices, 48.6\% include cross-staff writing, 37.4\% include grace notes, and 29.0\% include tuplets.

The composer distribution is intentionally diverse but not uniform: Bach, Chopin, and Czerny account for about 79\% of measures, while the remaining composers contribute smaller but stylistically distinct subsets (see \tabref{tab:test-composers}). The set spans common meters, dominated by 4/4 and 3/4, with a meaningful share of 2/4 and compound meters (see \tabref{tab:test-meter}). It is also mostly in major keys. The five most common key signatures are C major (23), G major (17), F major (13), E major (11), and B-flat major (10); 6 scores contain modulations. Complexity also varies by composer: Chopin contributes a higher share of tuplets and grace notes, while Bach exhibits more cross-staff writing.

The original LilyPond sources of this dataset mainly come from MutopiaProject. Its structured measure-level annotations are produced through LilyPond score rendering, SVG geometry analysis, and cross-checking by multiple structure-solving algorithms. The final correctness of every measure in the test set was verified manually.

\begin{table*}[t]
\centering
\small
\begin{subtable}[t]{0.26\linewidth}
  \centering
  \setlength{\tabcolsep}{5pt}
  \begin{tabular}{lr}
  \toprule
  \textbf{Property} & \textbf{Value} \\
  \midrule
  Groups (scores) & 107 \\
  Measure samples & 5,317 \\
  Scored events & 88,754 \\
  Multi-voice scores & 96.3\% \\
  Cross-staff scores & 48.6\% \\
  Grace-note scores & 37.4\% \\
  Tuplet scores & 29.0\% \\
  \bottomrule
  \end{tabular}
  \caption{Overview statistics.}
  \label{tab:test-stats}
\end{subtable}
\hfill
\begin{subtable}[t]{0.44\linewidth}
  \centering
  \setlength{\tabcolsep}{4.5pt}
  \begin{tabular}{lrrr}
  \toprule
  \textbf{Composer} & \textbf{Scores} & \textbf{Measures} & \textbf{\%} \\
  \midrule
  Bach & 39 & 1,321 & 24.8 \\
  Chopin & 24 & 1,504 & 28.3 \\
  Czerny & 33 & 1,388 & 26.1 \\
  Beethoven & 2 & 172 & 3.2 \\
  Liszt & 2 & 401 & 7.5 \\
  Mozart & 2 & 219 & 4.1 \\
  Others & 2 & 129 & 2.4 \\
  Debussy & 1 & 72 & 1.4 \\
  Rachmaninoff & 1 & 62 & 1.2 \\
  Kuhlau & 1 & 49 & 0.9 \\
  \bottomrule
  \end{tabular}
  \caption{Composer distribution.}
  \label{tab:test-composers}
\end{subtable}
\hfill
\begin{subtable}[t]{0.26\linewidth}
  \centering
  \setlength{\tabcolsep}{5pt}
  \begin{tabular}{lrr}
  \toprule
  \textbf{Time sig.} & \textbf{Scores} & \textbf{\%} \\
  \midrule
  4/4 & 34 & 31.8 \\
  3/4 & 33 & 30.8 \\
  2/4 & 17 & 15.9 \\
  3/8 & 9 & 8.4 \\
  2/2 & 8 & 7.5 \\
  6/8 & 4 & 3.7 \\
  12/8 & 2 & 1.9 \\
  9/8 & 2 & 1.9 \\
  Other & 2 & 1.9 \\
  \bottomrule
  \end{tabular}
  \caption{Meter distribution; 4 scores contain changes.}
  \label{tab:test-meter}
\end{subtable}
\caption{Test set overview (\texttt{lilypond-topology}).
(a) Key statistics of the 107-score test set.
(b) Composer breakdown; Bach, Chopin, and Czerny account for $\sim$79\% of measures.
(c) Time-signature distribution.}
\label{tab:test-overview}
\end{table*}

\paragraph{Train-test independence.}
The rendered-source portion of training provides large-scale coverage with realistic spatial layouts, while recognition-feedback samples reflect production error patterns. The test set is independent from both sources and is drawn from a separate LilyPond corpus, which makes it a meaningful generalization check (summarized in \tabref{tab:train-test-indep}).

\begin{table}[t]
\centering
\small
\setlength{\tabcolsep}{5pt}
\begin{tabular}{lcc}
\toprule
 & \textbf{Training} & \textbf{Test} \\
\midrule
Dataset & \texttt{writer10k-combined} & \texttt{lilypond-topology} \\
Primary source & Paraff render + feedback & Independent LilyPond corpus \\
Groups (scores) & 9,995 & 107 \\
Measure samples & 366,534 & 5,317 \\
Weighting & 1 / 4 / 100 & uniform \\
Purpose & model training & generalization evaluation \\
\bottomrule
\end{tabular}
\caption[Train-Test Independence]{Train-Test Independence. Summary of train vs test construction.}
\label{tab:train-test-indep}
\end{table}

\subsection{Metrics}
We evaluate regulation at the measure level. Each method receives the same candidate event set from the visual stage and must output a structured result: voice chains, tick positions, and duration-related attributes.

\paragraph{Event-level comparison against ground truth.}
We match events by stable event ID and use the ground-truth data as the denominator. The main event-level view is therefore a direct field-error comparison rather than a detector-style precision/recall score. We report per-field error rates for \texttt{tick}, \texttt{division}, \texttt{dots}, \texttt{beam}, and \texttt{timeWarp}, plus an \emph{any-field} error rate that marks an event wrong if any of the six core fields differ. Tick RMSE is reported separately in raw tick units. Following the comparison script, tick matching uses a tolerance of $\pm 1$ tick for the event-level error rate, while RMSE uses the raw difference.

\paragraph{Measure-level exactness and internal diagnostics.}
At the measure level, we report three exact-match tiers against ground truth: \emph{perfect}, \emph{voice match}, and \emph{tick exact}. A measure is \emph{perfect} when the voice partition matches ground truth and all core event fields match. \emph{Voice match} ignores attribute differences and only requires the canonicalized voice sets to agree. \emph{Tick exact} requires a voice match and exact tick values for all events. In addition, we report internal structural diagnostics from the evaluator in Appendix~B, including average quality score, tick-twist, surplus time, and broken-beam rate. These diagnostics do not compare against ground truth directly; they measure the self-consistency of the produced result.

\subsection{Configurations}
\paragraph{Training configuration (BeadPicker).}
BeadPicker is a 16-layer Transformer encoder ($d_\text{model}{=}128$, 8 heads, $d_\text{inner}{=}512$) with dropout 0.1, totaling about 3.24M parameters. We initialize all weight matrices with Xavier-uniform initialization and a depth-scaled gain ($g = 32^{-1/2}$) to keep the 16-layer stack stable at the start of training. Training uses multi-task supervision with a dominant weight on the successor head; we mix large-scale rendered samples with recognition-feedback samples and apply moderate augmentation to remain robust to layout variation and imperfect candidates. The full hyperparameter list is provided in Appendix~\ref{sec:impl}.

\paragraph{Decoding configuration (BeadSolver).}
We evaluate two decoding budgets. \emph{Single-pass} BeadSolver uses a fixed search budget and is meant to reflect the core capability of the model-guided solver under a practical compute constraint. \emph{Multi-pass} BeadSolver runs multiple passes with escalating budgets and returns the best solution found across passes; it is used to probe the upper bound of the method when more compute is available. In our implementation, the passes share the same model and evaluation function, and only differ in search quota; the multi-pass setting can also include a lightweight pre-pass that assigns measure-level predispositions (e.g., estimated measure duration) before search, which improves robustness on ambiguous measures.

\subsection{Results}
Our main comparison evaluates on the independent LilyPond test set against ground-truth topology and reports both event-level and measure-level metrics (see \tabref{tab:main-results}).

\begin{table*}[t]
\centering
\small
\setlength{\tabcolsep}{5pt}
\begin{tabular}{lcccc}
\toprule
 & \textbf{Greedy} & \textbf{Equations} & \textbf{BeadSolver} & \textbf{BeadSolver-Multipass} \\
\midrule
\multicolumn{5}{l}{\textit{Event-level error vs ground truth (lower is better)}} \\
Any-field error (\%) & 30.61 & 15.84 & 5.57 & \textbf{3.18} \\
Tick RMSE & 472.5 & 236.6 & 201.9 & \textbf{42.9} \\
Tick error (\%) & 27.92 & 13.17 & 4.27 & \textbf{1.64} \\
Division error (\%) & 1.61 & 1.60 & \textbf{0.53} & 0.54 \\
Dots error (\%) & 0.27 & 0.26 & \textbf{0.12} & 0.13 \\
Beam error (\%) & 2.51 & 2.51 & \textbf{1.03} & 1.09 \\
Time-warp error (\%) & 13.20 & 7.00 & 1.31 & \textbf{1.26} \\
\midrule
\multicolumn{5}{l}{\textit{Measure-level exactness and internal quality}} \\
Perfect (\%) & 50.54 & 65.94 & 83.04 & \textbf{86.83} \\
Voice match (\%) & 55.09 & 70.96 & 88.08 & \textbf{92.42} \\
Tick exact (\%) & 66.75 & 75.55 & 91.18 & \textbf{95.69} \\
Avg quality score & 0.634 & 0.817 & 0.905 & \textbf{0.957} \\
Avg tick-twist & 0.110 & 0.196 & \textbf{0.049} & 0.106 \\
Broken beam (\%) & 19.62 & 6.96 & 2.13 & \textbf{2.07} \\
\midrule
Wall time (107 scores) & $\sim$18s & $\sim$84min & $\sim$24min & $\sim$33min \\
\bottomrule
\end{tabular}
\caption[Main Results on LilyPond Test Set]{Main Results on the LilyPond Test Set. Comparison on \texttt{lilypond-topology}. BeadSolver uses probability-guided tree search with BeadPicker; Multipass runs three passes with escalating budgets. Event-level values are computed by direct comparison against the ground-truth spartito over all 88,754 scored events; measure-level tiers are computed over all 5,317 measures. In this evaluation, BeadPicker is executed as an ONNX model on a CPU backend (Intel Xeon E5-2680 v4, 14 cores / 28 threads).}
\label{tab:main-results}
\end{table*}

Overall, greedy regulation struggles mainly on tick placement and voice organization: 30.65\% of scored events contain at least one field error, and only 50.71\% of measures are fully correct. The linear-equations baseline closes much of this gap, but it remains brittle on ambiguous timing and time-warp structure. BeadSolver improves substantially once learned local guidance is combined with search, reducing any-field event error to 5.91\% and raising the perfect-measure rate to 83.00\%.

Multipass decoding improves the tail further. It reaches 86.80\% perfect measures and lowers tick RMSE from 202.4 to 41.2. One detail is worth noting: the internal evaluator terms do not always move in the same direction as the ground-truth comparison, because on difficult cases the extra exploration pushes the solver more strongly toward structural correctness, which can in turn cause slight rebounds in tick-twist and some event-duration-related attributes. This may reflect a small trade-off for structural improvement, and it also highlights the gap between the current evaluator and a perfect ground-truth judge.

\subsection{Ablations and Qualitative Examples}
We report three ablations that remove one component at a time.

\paragraph{Replace learned picker by rule-based picker.}
We replace BeadPicker with a purely rule-based picker and keep the same search procedure. This isolates the effect of learned guidance in the solver. Without the learned picker, search alone provides only limited gains over greedy and leaves many measures unusable (see \tabref{tab:rulebased}).

\begin{table}[t]
\centering
\small
\setlength{\tabcolsep}{4.5pt}
\begin{tabular}{lccc}
\toprule
 & \textbf{Greedy} & \textbf{Rule-based} & \textbf{BeadSolver} \\
\midrule
\multicolumn{4}{@{}l}{\textit{Event-level error vs ground truth}} \\
Any-field error (\%) & 30.61 & 29.56 & \textbf{5.57} \\
Tick RMSE & 472.5 & 457.9 & \textbf{201.9} \\
\midrule
\multicolumn{4}{@{}l}{\textit{Measure-level exactness and diagnostics}} \\
Perfect (\%) & 50.54 & 56.72 & \textbf{83.04} \\
Voice match (\%) & 55.09 & 58.70 & \textbf{88.08} \\
Tick exact (\%) & 66.75 & 63.34 & \textbf{91.18} \\
Avg quality score & 0.634 & 0.637 & \textbf{0.905} \\
Avg tick-twist & 0.110 & 0.072 & \textbf{0.049} \\
Avg surplus time & \textbf{0.1} & 88.9 & 4.4 \\
Broken beam (\%) & 19.62 & 15.65 & \textbf{2.13} \\
\bottomrule
\end{tabular}
\caption[No Learned Picker]{Learned picker vs rule-based picker. Greedy is the no-search baseline, Rule-based uses the same tree-search framework with a heuristic picker, and BeadSolver uses the learned BeadPicker. The learned picker is essential: search alone improves beam continuity and coarse quality only slightly, but it does not recover the large gains in tick placement and voice structure.}
\label{tab:rulebased}
\end{table}

The design principle of the rule-based picker is to use as much accessible information as possible to construct explicit rules that reflect local topology prediction. It builds successor preferences from coarse horizontal grouping and left-to-right continuation, estimates ticks by linear interpolation between nearby anchors, passes through upstream feature vectors for division, dots, beam, and stem direction, and uses threshold rules for special cases such as grace notes, whole-measure rests, and likely duplicate events. In other words, it tries to describe the local next-step policy by hand, while leaving global consistency to the same search and evaluator used by BeadSolver.

The results in \tabref{tab:rulebased} show that this is far from sufficient. Although the search framework is unchanged, replacing the learned picker by rules drops the perfect-measure rate from 83.00\% to 56.87\%, voice match from 88.02\% to 58.68\%, and raises tick RMSE from 202.4 to 458.0. This large degradation demonstrates that the learned picker is not an optional refinement but a necessary component. More broadly, it reflects the difficulty of describing real staff-topology behavior with explicit rules alone: local continuation depends not only on horizontal proximity, but also on context-dependent duration semantics, voice exchange, cross-staff motion, tuplets, grace timing, and many engraving-dependent exceptions.

At the same time, rule-based search is still slightly better than greedy decoding in some respects. Compared with greedy, it improves the perfect-measure rate from 50.71\% to 56.87\%, reduces any-field event error from 30.65\% to 29.61\%, and lowers broken-beam and tick-twist diagnostics. This indicates that explicit search plus structural checking can repair some limited issues.

\paragraph{Replace \emph{vtick} by scalar MSE tick loss.}
We remove the mixed-radix vector code for tick prediction and train with the scalar tick head only. This increases tick-related errors and degrades duration consistency (see \tabref{tab:vtick}). This ablation is evaluated only on model-side metrics from training and one-pass forward prediction. We do not run the full search loop here, because floating-point tick regression changes the downstream rule evaluation itself and would make the search-stage comparison harder to interpret.

\begin{table}[t]
\centering
\small
\setlength{\tabcolsep}{4.5pt}
\begin{tabular}{lrr}
\toprule
 & \textbf{With \emph{vtick}} & \textbf{Without \emph{vtick}} \\
\midrule
tick term & \textbf{2.475} & 8.188 \\
tick\_fixed term & \textbf{0.528} & 6.117 \\
rel\_tick term & \textbf{3.618} & 8.126 \\
duration term & \textbf{1.255} & 9.476 \\
\bottomrule
\end{tabular}
\caption[No \emph{vtick}]{\emph{vtick} ablation results. Diagnostic terms reported by the topology model (lower is better): \texttt{tick} is RMSE of absolute tick prediction; \texttt{tick\_fixed} is the same RMSE computed on fixed-prefix elements; \texttt{rel\_tick} is RMSE of predicted pairwise tick differences; \texttt{duration} is RMSE of the \texttt{EOS} tick (i.e. the entire measure duration). All values are measured in the internal tick unit, where a quarter note equals 480 ticks. These numbers are computed from a one-pass forward (glimpse) evaluation on all measures in \texttt{lilypond-topology}. In the full BeadSolver loop, an incorrect search trajectory can temporarily worsen the structure and lead to higher tick RMSE than this glimpse metric.}
\label{tab:vtick}
\end{table}

\paragraph{Skip BeadSolver pre-pass.}
In Multipass decoding, we remove the lightweight pre-pass that assigns measure-level predispositions before search. This ablation mainly hurts measures with whole-measure rests and ambiguous duration cues. The largest effect appears on tick-related metrics, because the pre-pass gives the same BeadPicker a cleaner initial estimate of measure duration and fake-event semantics before the expensive search begins (see \tabref{tab:prepass}).

\begin{table}[t]
\centering
\small
\setlength{\tabcolsep}{4.5pt}
\begin{tabular}{lrr}
\toprule
 & \textbf{Without pre-pass} & \textbf{With pre-pass} \\
\midrule
\multicolumn{3}{@{}l}{\textit{Event-level error vs ground truth}} \\
Any-field error (\%) & 5.91 & \textbf{3.18} \\
Tick RMSE & 206.5 & \textbf{42.9} \\
Tick error (\%) & 4.57 & \textbf{1.64} \\
Time-warp error (\%) & 2.55 & \textbf{1.26} \\
Has-time-warp error (\%) & 2.32 & \textbf{0.55} \\
\midrule
\multicolumn{3}{@{}l}{\textit{Measure-level exactness and diagnostics}} \\
Perfect (\%) & 85.46 & \textbf{86.83} \\
Voice match (\%) & 90.20 & \textbf{92.42} \\
Tick exact (\%) & 93.61 & \textbf{95.69} \\
Avg quality score & 0.933 & \textbf{0.957} \\
Tick-twist $\geq 1$ (\%) & \textbf{0.08} & 1.03 \\
Broken beam (\%) & 2.41 & \textbf{2.07} \\
\bottomrule
\end{tabular}
\caption[No Pre-Pass]{Pre-pass ablation results. The pre-pass is not a separate model; it is a lightweight glimpse with the same BeadPicker before search. It materially improves ground-truth-aligned accuracy, especially on tick-related metrics, except one internal diagnostic (\texttt{tickTwist} $\ge 1$) becomes worse.}
\label{tab:prepass}
\end{table}

Finally, Appendix~C provides a qualitative failure-to-success progression across regulation strategies. These examples highlight typical error modes: greedy failures that can be repaired by equations, equations failures that require probability-guided search, and remaining hard cases where Multipass search or manual correction is needed.

\section{Discussion}

\subsection{Methodological Trade-Offs}

\paragraph{Why search matters}
The flexibility of polyphonic staff notation makes reading and decoding difficult in a very concrete way: quick visual intuition is often not enough to decide what is correct. Many measures admit several locally plausible interpretations, and the right one only becomes clear after checking global constraints such as voice continuity, measure duration, and rhythmic consistency. In this sense, structure decoding is closer to logical analysis than to pattern matching.

This is similar to board games. A one-pass model can be surprisingly capable, but its ceiling is limited when the decision requires exploring alternatives. When combined with Monte Carlo Tree Search, the same model can reach a much higher level because it can branch, test, and back up evidence instead of committing too early \cite{silver2017go}. Our solver follows the same principle: learned predictions provide strong local guidance, while rule-based exploration and verification are used to resolve ambiguity.

For this reason, we believe that rule-driven exploration will remain a necessary component of an ideal OMR system, whether the system is two-stage or end-to-end. Even with stronger models, complex notation will continue to require explicit reasoning over constraints, especially in the long tail where training data is sparse.

\paragraph{Why not reinforcement learning}
We did not directly formulate the problem as reinforcement learning for two main reasons. First, score regulation still lacks a universal rule standard for evaluating solutions. We have designed several evaluation functions, but each comes with exception cases; unlike board games, there is no single explicit rule judgment that remains reliable across the full data distribution. Second, replacing these rules with a learned reward model would return us to the same data bottleneck that the current method is designed to alleviate. In that sense, our present search-based formulation is a practical compromise: it uses learned local priors where supervision is available, while keeping global decision-making anchored by explicit constraints.

\subsection{Issues remaining}

Despite these benefits, our current system still has some limitations.

\paragraph{Weak global evaluation.}
Search only helps if we can reliably rank candidate structures. In our implementation, the ``value'' of a partial or complete solution is dominated by hand-designed terms with empirical weights. Some terms do incorporate model outputs, but the overall evaluator is still relatively weak. In practice, we have seen exception cases where the correct solution receives a worse score because it violates a heuristic assumption (e.g., informal rhythmic usage, slight engraving misalignment, or crowded layout patterns that break a simple space-time prior). As a result, a non-trivial fraction of manual annotations are caused by the evaluator underscoring the correct topology. A more robust learned evaluator, or a tighter verification layer with more sophisticated heuristics, is an important next step.

\paragraph{Compute cost on very large measures.}
Our measure-level decoding is efficient for typical piano measures (tens of events), but it can become expensive for extreme cases with unusually many events. In the worst case, the search space is combinatorial: if $n$ candidates can be ordered and assigned in many interchangeable ways, the number of distinct partial structures can grow on the order of $\alpha n^2$, where $\alpha$ denotes the fraction of events that actually induce branching choices in search\footnote{A more precise formula is: $n \cdot (\frac{\alpha n}{2} + 1)$.}. Empirically, $\alpha$ is roughly $0.1$. At the same time, each Pass expansion invokes a global-attention Transformer over the full candidate set, whose cost is roughly $O(n^2)$ due to self-attention. Taken together, the worst-case time complexity can be viewed as $O(n^4)$, which quickly exceeds a practical budget. This is why very large measures are a bottleneck in production. Mitigations include tighter candidate pruning, coarse-to-fine search, and models with more efficient attention or factorized representations.

Here, ``large measures'' refers specifically to the size of a single part in staff notation, that is, one instrument track. In full ensemble scores this issue usually does not grow in the same way, because most non-keyboard instruments have limited chordal capacity and their notation is not structurally coupled across parts.

\paragraph{Dependence on visually sensitive semantics.}
Finally, the second-stage solver assumes that certain visually sensitive cues are already correct, because they define the structural frame of a measure. Examples include barlines (which define measure boundaries), staff assignment, and time-signature hints that define expected measure duration (though already partly mitigated by the design of the pre-pass glimpse). When these upstream predictions are wrong (or when the page is degraded and the cues are ambiguous), regulation may fail in ways that cannot be repaired by search alone, since the candidate set is missing the correct boundary conditions. Improving the robustness of these visual semantics, and exposing uncertainty in them to the decoder, remain important engineering priorities.

\subsection{Starry as an Agent-Assisted Annotation Scaffold}
Multimodal LLMs are another plausible direction for OMR in the long run, similar in spirit to how current LLM systems are used for data-to-formula recognition. However, given the current scale of music-structure supervision, general-purpose LLMs are still underfit in this domain. To reduce hallucinations and enforce correctness, we instead treat the existing OMR system and its rule-based evaluator as a scaffold for an annotation agent.
As an early experiment toward scaling structured supervision, we explored using the Starry OMR system as a scaffold for an annotation agent. The basic idea is to turn ``regulation failures'' into small, self-contained tasks at the measure level. For each issue measure, the agent requests a cropped score image together with a compact serialization of the event candidates and the current (failed) regulation state. It then proposes a fix in a structured patch format that edits ticks, duration attributes, time-warp flags, and voice chains. A quality evaluator checks the proposed fix immediately and returns diagnostic metrics; if the measure still fails, the agent iterates.

\begin{figure}[htb]
\centering
\includegraphics[width=0.92\linewidth]{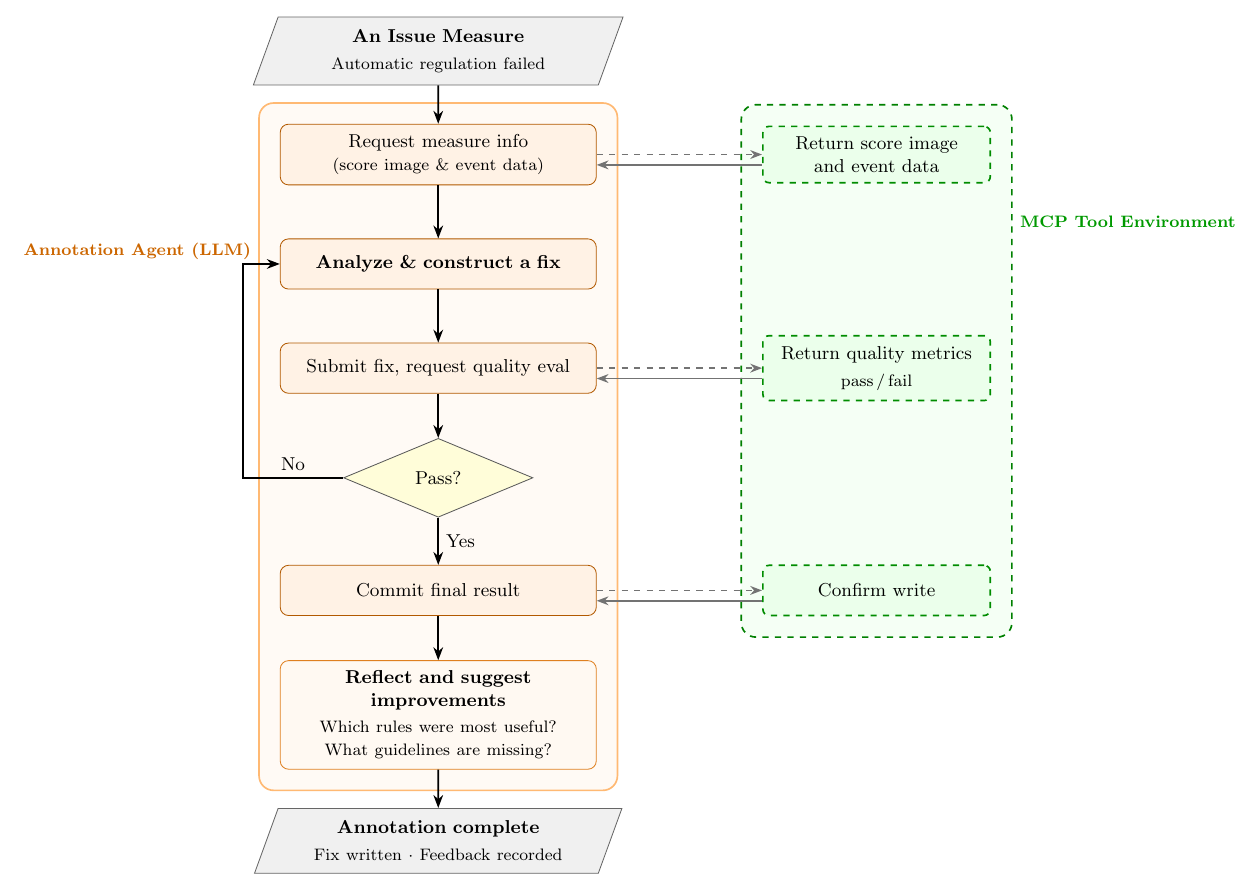}
\caption{A preliminary agent-assisted annotation loop for issue measures. The agent requests measure-level context (image crop and event data), proposes a structured fix, and uses automatic quality evaluation as a guardrail.}
\label{fig:annotation-agent}
\end{figure}

This loop is designed to be conservative. The agent is forced to output explicit structured edits rather than free-form text, and every fix is validated by the same structural checks used elsewhere in the pipeline. When a fix passes, the result is committed and the session can also record short feedback about which rules or guidelines were helpful. Figure~\ref{fig:annotation-agent} sketches the workflow. While this is only a preliminary trial, it illustrates a practical path to grow a high-fidelity structure corpus under limited human time: the production system proposes solutions, and an agent (with optional human oversight) focuses on the tail where decoding is hardest.

Appendix~F provides a complete rollout example of this agent-assisted annotation loop.

\section{Conclusion and Future Work}

We presented a production-grade two-stage OMR pipeline for complex piano scores, with an emphasis on the second stage: turning imperfect symbol candidates into an editable musical structure. The core conclusion is that structure decoding is not a thin post-processing step. For polyphonic notation, it requires global reasoning about voices, ticks, durations, and measure boundaries. Our topology formulation and probability-guided tree search make this reasoning explicit: the model provides local evidence, and the solver searches and verifies to select a globally consistent structure.

In experiments on an independent LilyPond test set, probability-guided search substantially improves both tick accuracy and voice assignment over rule-based greedy decoding and a stronger linear-equations baseline. This supports a practical view of OMR: when annotation budgets are limited, combining learned predictors with rule-based exploration and evaluation can raise the ceiling of recognition quality, while also producing structured data that can be fed back to train stronger models.

As a practical resource for future work, we are considering releasing the \texttt{lilypond-topology} dataset used in this paper, including the LilyPond source files, engraved images, and the resulting topology-cluster data. Looking ahead, we also plan to build a more complete benchmark that evaluates a full OMR pipeline from images to a music-language representation, rather than only isolated sub-tasks.

We also run an IMSLP\footnote{\href{https://imslp.org/}{https://imslp.org/}} data mining project\footnote{\href{https://github.com/k-l-lambda/imslp-mining}{https://github.com/k-l-lambda/imslp-mining}} that aims to convert open sheet-music images into a usable symbolic music dataset. Our goal is to make this corpus practical for downstream tasks, including symbolic music generation, algorithmic performance, and automatic transcription from performances back to notation.

Several directions remain open. A stronger learned evaluator could reduce brittleness and improve ranking among near-correct hypotheses. Efficiency is also important for unusually large measures, where the search space becomes extreme. Tighter handling of upstream uncertainty in visually sensitive semantics would make the full pipeline more robust. Finally, beyond data collection, we also view RL-based finetuning of multimodal LLMs using these annotation agent rollouts as a promising direction.

\appendix

\section*{Appendices}

\section{Implementation Notes}
\label{sec:impl}

This appendix collects the engineering details that support reproducibility of the BeadSolver system: the exact input representation fed to the model, the architecture and training hyperparameters, the search budget formulas, and the interface used to feed annotated corrections back into the training pipeline.

\subsection{Input Representation}

\paragraph{Element types.}
Each measure is represented as a flat sequence of elements. Five element types are used: \texttt{PAD} (padding, index~0), \texttt{BOS} (beginning of sequence, index~1), \texttt{EOS} (end of sequence, index~2), \texttt{CHORD} (index~3), and \texttt{REST} (index~4). In a decoded sequence, the model predicts a successor chain that begins at \texttt{BOS} and ends at \texttt{EOS}; chords and rests form the body of the structure.

\paragraph{Geometric fields.}
Each element carries five geometric scalars in staff-space units (one unit = one staff space): $x$ (horizontal position), $\mathit{pivotX}$ (horizontal anchor, used when the visual centre of mass is not centred on the event position, e.g.\ for chords with mixed stem directions), $y_1$ and $y_2$ (upper and lower staff-vertical extent), and $\mathit{staff}$ (integer staff index within the system, zero-based). During training, mild data augmentation is applied: $x$ and $\mathit{pivotX}$ receive Perlin noise, $y_1$ and $y_2$ receive Gaussian noise ($\sigma = 0.12$), and the whole $x$-axis is optionally rescaled by a log-normal factor to simulate horizontal spacing variation.

\paragraph{Local attribute feature vector.}
In addition to geometry, each element receives a 16-dimensional real-valued feature vector that encodes soft evidence about local rhythmic and notational attributes, as estimated by the visual models. These are \emph{not} hard decisions: the values are log-likelihood-style scores that convey prior knowledge without committing the model to a specific interpretation. See \tabref{tab:feature_dims} for the 16 components.

\begin{table}[h]
\centering
\caption[Local Attribute Feature Vector]{Local Attribute Feature Vector. Composition of the 16-dimensional local attribute feature vector.}
\label{tab:feature_dims}
\begin{tabular}{lrl}
\toprule
Field & Dims & Candidates \\
\midrule
\texttt{division}       & 7 & $\geq\!0,\;\geq\!1,\;\ldots,\;\geq\!6$ \; (duration-class lower bounds) \\
\texttt{dots}           & 2 & $\geq\!1$ dot,\; $\geq\!2$ dots \\
\texttt{beam}           & 3 & Open, Continue, Close \\
\texttt{stemDirection}  & 2 & Up, Down \\
\texttt{grace}          & 1 & grace-note flag \\
\texttt{tremoloCatcher} & 1 & tremolo-receiving event flag \\
\bottomrule
\end{tabular}
\end{table}

\paragraph{Time-signature context.}
An additional scalar \texttt{time8th} encodes the expected measure duration in eighth-note ticks (capped at~16). It is derived from the time signature and passed as a global per-measure scalar, giving the model a coarse prior on total measure duration without hard-wiring a specific search constraint.

\subsection{BeadPicker Architecture and Training}

BeadPicker is a transformer encoder that reads the full candidate sequence in a single forward pass and produces predictions for all output heads jointly. See \tabref{tab:arch} for architecture hyperparameters and \tabref{tab:training} for training configuration.

\begin{table}[h]
\centering
\caption[BeadPicker Architecture]{BeadPicker Architecture. Architecture hyperparameters.}
\label{tab:arch}
\begin{tabular}{ll}
\toprule
Parameter & Value \\
\midrule
$d_\text{model}$                  & 128 \\
$d_\text{inner}$ (FFN)            & 512 \\
Attention heads $n_\text{head}$   & 8 \\
Encoder layers $n_\text{layers}$  & 16 \\
$d_k = d_v$                       & 16 \\
Local feature dimension           & 16 \\
Feature activation                & tanh \\
Positional encoding               & sinusoidal~\cite{vaswani2017attention} (cycle 1\,000) \\
\texttt{usePivotX}                & \checkmark \\
\texttt{with\_time8th}            & \checkmark \\
Rectifier version                 & v3 \\
Dropout                           & 0.1 \\
Trainable parameters              & 3.24\,M \\
\bottomrule
\end{tabular}
\end{table}

\begin{table}[h]
\centering
\caption[BeadPicker Training]{BeadPicker Training. Training hyperparameters.}
\label{tab:training}
\begin{tabular}{ll}
\toprule
Parameter & Value \\
\midrule
Batch size          & 1 \\
Optimizer           & Adam ($\beta_1{=}0.9$,\; $\beta_2{=}0.98$,\; $\varepsilon{=}10^{-9}$) \\
LR schedule         & InvSqrt ($d_\text{model}{=}128$,\; lr\_mul${=}0.2$,\; warmup${=}8{,}000$ steps) \\
Epoch size          & 6\,000 steps \\
Total epochs        & 6\,400 \\
Training samples    & 356\,431 (97\% split by group) \\
GPU                 & $2\times$ RTX3090 \\
\bottomrule
\end{tabular}
\end{table}

The model has ten output heads: \texttt{successor} (next-node selection), \texttt{tick} (absolute tick position), \texttt{division} (9 categories), \texttt{dots} (3 categories), \texttt{beam} (4 categories), \texttt{stemDirection} (3 categories), \texttt{grace}, \texttt{timeWarped}, \texttt{fullMeasure}, and \texttt{fake} (binary). The \texttt{successor} head carries the largest loss weight, since it governs chain structure and is the primary search objective. All other heads use individually tuned loss weights that down-weight uncertain or rare attributes.

\subsection{Search Budget}
\label{sec:budget}

BeadSolver maintains a priority queue over partial successor assignments and expands up to a \emph{quota} of nodes at each position. The quota is:
\[
Q(n,\,f) \;=\; \min\!\Bigl(
  \bigl\lceil (n{+}1)\,f\,\ln(n{+}2) \bigr\rceil,\;\;
  \Bigl\lceil Q_{\max} \cdot \min\!\Bigl(1,\;\Bigl(\tfrac{24}{n{+}1}\Bigr)^{\!2}\Bigr) \Bigr\rceil
\Bigr),
\]
where $n$ is the number of candidate events in the measure, $f$ is the quota factor, and $Q_{\max} = 1\,000$. The logarithmic first term scales search effort with measure complexity; the quadratic second term caps expansion for short measures.

Single-pass BeadSolver uses $f = 40$. BeadSolver-Multipass runs three passes with escalating factors ($f = 10$, $40$, $80$) and returns the best solution found across all passes. The \texttt{estimateMeasure} pre-pass (version~3) is not a separate model: it uses the same BeadPicker and performs one glimpse forward pass per cluster under a minimal prefix (only \textsc{BOS} fixed) to populate per-element predispositions, including \texttt{fullMeasure} and tick/duration-related fields, and to re-estimate measure duration when the time signature is doubtful. This prevents whole-measure rests from being drawn into voice assignment and substantially reduces \texttt{irregularTick} errors, at the cost of a minor increase in average tick-twist (see \tabref{tab:prepass}).

\subsection{Annotation Feedback Interface}

Measures that remain difficult after automated regulation are flagged with a status code: \texttt{Solved} (automated solution accepted), \texttt{Issue} (requires review or manual annotation), or \texttt{Fatal} (irrecoverable structural inconsistency). Corrected measures are written back as patch records with a \texttt{patched} flag.

Training sample weights reflect annotation provenance: unannotated baseline samples receive weight~1; OMR-flow annotation samples (machine-predicted, not patched) receive weight~4; human-corrected (patched) samples receive weight~100. These weights are applied during weighted-sampling training (\texttt{torch.multinomial}) so that hard, human-verified cases exert disproportionate influence on the loss despite representing under 1\% of the dataset by count.

The quality evaluation function \texttt{evaluateMeasure} checks each regulated measure against a set of structural invariants: voice monotonicity, beam consistency, tick integrality, time-warp regularity, duration balance, and absence of fake or grace-dominant events. It returns a scalar quality score $q \in [0,1]$ and a set of Boolean error flags (\texttt{beamBroken}, \texttt{tickOverlapped}, \texttt{irregularTick}, etc.) that serve both as training-data filters and as human-readable diagnostics in the annotation interface. A measure is marked erroneous when it contains corrupted voice events, non-monotone voice order, overlapping ticks inside a voice, excessive tick twist, negative staff tick rates, too many unresolved events, missing time-signature context, over-range durations, or implausibly compressed time-warped events.

Among non-erroneous measures, Starry further distinguishes \emph{fine} and \emph{not fine} cases. A measure may be structurally valid but still not fine if it has duration overflow beyond the expected measure length, excessive tick twist, fractional or irregular time warps, irregular tick granularity, surplus voice duration, broken beam state, or grace notes incorrectly included in ordinary voice sequences. These conditions do not always make the measure unusable, but they indicate that the solution is not clean enough to be treated as a high-confidence automatic result. For error-free measures:
\[
q \;=\; (1 - \ell_{\mathrm{space}})\,(1 - \ell_{\mathrm{dur}})\,(1 - \ell_{\mathrm{warp}})\,(1 - \zeta^2),
\]
where $\ell_{\mathrm{space}} = \tanh(|\Delta t_{\mathrm{space}}|)$ penalises unused voice time, $\ell_{\mathrm{dur}} = \max(0,1-r_{\mathrm{dur}})^2$ penalises duration-rate deviation, $\ell_{\mathrm{warp}} = \tanh(N_{\mathrm{irregular}})$ penalises non-standard warp ratios, and $\zeta$ is the maximum tick-twist ratio. Any measure with a structural error flag receives $q = 0$.

\paragraph{Measure-fingerprint solution cache.}
Starry also maintains a solution cache keyed by a measure-level feature fingerprint. The fingerprint summarises the information that is relevant to regulation: candidate events, local rhythmic attributes, staff grouping, measure context, and other structural features that affect the resulting voice and tick assignment. Its design deliberately abstracts away layout-dependent variation and notation details that do not change the rhythmic regulation problem, such as different engravings of the same musical content or expressive markings outside the event topology. When a measure with the same fingerprint is encountered again, the cached regulation solution can be reused directly instead of solving the measure from scratch.

The design principle is cumulative reuse. Human-corrected measures, high-confidence automatic solutions, and previously computed regulation outputs are all stored in the same solution space, so improvements made during annotation can benefit future recognitions. The cache is conservative: it is tied to the measure's structural fingerprint rather than to page identity, image location, or accidental engraving details, so reuse is allowed only when the regulation problem is effectively the same. This makes the annotation workflow incremental: each solved measure becomes training data for the model and, at the same time, a reusable symbolic solution for the engineering pipeline.

Table~\ref{tab:measure-evaluator} reports these evaluator-side diagnostics for the main experiment results on the test set.
\begin{table*}[t]
\centering
\small
\setlength{\tabcolsep}{4pt}
\begin{tabular}{lrrrrrrr}
\toprule
 & \textbf{Greedy} & \textbf{RuleBased} & \textbf{Equations} & \textbf{BeadSlvr} & \textbf{BS-nopre} & \textbf{Multipass} & \textit{Gold.} \\
\midrule
EvalPerf\%  & 46.62 & 52.23 & 64.66 & 76.66 & 78.30 & \textbf{79.69} & 80.78 \\
EvalFine\%  & 60.75 & 64.43 & 79.65 & 90.30 & 92.98 & \textbf{95.11} & 97.87 \\
EvalErr\%   & 30.79 & 33.63 & 12.54 &  7.22 &  4.21 & \textbf{1.37} & \textit{0.55} \\
VoiceRug\%  &  0.00 &  0.00 &  0.38 &  0.00 &  0.00 &  0.00 &  0.00 \\
CmplxTW\%   &  0.00 &  0.00 &  3.87 &  0.00 &  0.00 &  0.00 &  0.11 \\
FracWrp\%   &  0.00 &  0.00 &  6.58 &  1.54 &  0.81 &  2.09 &  1.18 \\
IrrTick\%   &  5.87 &  0.49 &  3.18 &  0.75 &  1.41 &  0.68 &  0.34 \\
TickOvlp\%  &  0.32 & 21.99 &  4.98 &  1.82 &  0.11 &  0.09 &  0.39 \\
BeamBrk\%   & 19.62 & 15.65 &  6.96 &  2.13 &  2.41 &  2.07 &  0.00 \\
TW$\geq$1\% &  0.09 &  0.00 &  2.28 &  0.08 &  0.08 &  1.03 &  0.15 \\
\midrule
SpaceAvg    & 1.7248 & 0.6368 & 0.2004 & 0.0985 & 0.4356 & 0.2315 & 0.0756 \\
SurpAvg     &    0.1 &   88.9 &   16.4 &    4.4 &    0.0 &    0.3 &    0.0 \\
TWAvg       & 0.1099 & 0.0716 & 0.1961 & \textbf{0.0485} & 0.0574 & 0.1057 & 0.0498 \\
DurRate     & 1.1736 & 0.9860 & 0.9822 & 0.9990 & 1.0354 & 1.0002 & 0.9989 \\
QS avg      & 0.6341 & 0.6370 & 0.8169 & 0.9052 & 0.9326 & \textbf{0.9567} & 0.9616 \\
NullEvs     &      0 &  12875 &   3538 &   2734 &    362 &    139 &     68 \\
FakeEvs     &      6 &   1879 &   3541 &   2211 &    371 &    541 &    518 \\
WarpEvs     &      0 &      0 &  11272 &  11329 &   9744 &  11515 &  11717 \\
IrrWrpN     &      0 &      0 &    448 &      0 &      0 &      0 &     52 \\
GraceN      &   1504 &   1488 &   1504 &    588 &    606 &    573 &    633 \\
\bottomrule
\end{tabular}
\caption[\texttt{evaluateMeasure} Diagnostics]{\texttt{evaluateMeasure} diagnostics across all 5,317 measures per experiment. \textit{Gold.}\ is the hand-annotated ground truth of the \texttt{lilypond-topology} test set, included as a quality reference for the annotations themselves; note that its non-zero \textit{EvalErr} (0.55\%) reflects edge cases where the rule-based \texttt{evaluateMeasure} checker flags valid annotations as erroneous---a limitation of the evaluation heuristics rather than annotation mistakes; \textit{RuleBased} and \textit{BS-nopre} (BeadSolver-Multipass without pre-pass) are ablation variants; the remaining four columns (Greedy, Equations, BeadSlvr, Multipass) are from the main-results comparison. All metrics are computed on the experiment output alone, without reference to the golden ground truth. \textbf{Top block}: tier and diagnostic flags (\% of measures). \textit{EvalPerf}: all fields valid, no irregular ticks, no beam breaks, no voice ruggedness. \textit{EvalFine}: passes a relaxed quality check. \textit{EvalErr}: fails the fine check. \textit{VoiceRug} (uneven voice fill), \textit{CmplxTW} (complicated time-warp), \textit{FracWrp} (fractional warp denominator), \textit{IrrTick} (irregular tick positions), \textit{TickOvlp} (overlapping tick ranges), \textit{BeamBrk} (broken beam group), \textit{TW$\geq$1} (tick-twist $\geq 1$). \textbf{Bottom block}: numeric averages and event counts. \textit{SpaceAvg}: space-time product. \textit{SurpAvg}: surplus time (ticks). \textit{TWAvg}: tick-twist average. \textit{DurRate}: note-duration sum / measure duration. \textit{QS avg}: composite quality score. \textit{NullEvs}/\textit{FakeEvs}/\textit{WarpEvs}: null-tick, fake, and time-warped event counts. \textit{IrrWrpN}: irregular warp groups. \textit{GraceN}: grace notes. Source: \texttt{goldenCompare.md} (2026-04-22), lines 29--49.}
\label{tab:measure-evaluator}
\end{table*}

The structured output of regulation can be exported to MusicXML, LilyPond source, or the system-internal spartito JSON format, all of which preserve voice identity, tick positions, and corrected rhythmic attributes.

\subsection{Output Formats}

The structured results of Starry's regulation stage are serialised into several downstream formats, each serving a different use case.

\paragraph{MusicXML 4.0.}
The primary interchange format is MusicXML 4.0 (\texttt{.xml}). The encoder maps Starry's internal voice and tick representation to MusicXML \texttt{<part>}, \texttt{<measure>}, \texttt{<note>}, and \texttt{<direction>} elements. Notes carry voice, staff, beam, tie, and slur attributes; tuplets are expressed via \texttt{<time-modification>} and enclosing \texttt{<tuplet>} notations; dynamics are encoded as direction words. The internal time-division constant is $\text{DIVISIONS} = 480$ ticks per quarter note, matching standard MIDI resolution. This format is the primary path for export to notation editors such as MuseScore, Finale, and Sibelius.

\paragraph{LilyPond source.}
LilyPond (\texttt{.ly}) output targets typeset-quality engraving. The encoder produces relative-pitch notation and supports the full articulation vocabulary: ornaments, ottava brackets, pedal marks, hairpin dynamics, slurs, and tuplets. An optional \texttt{\textbackslash midi\{\}} block can be included for playback. LilyPond output is also used to round-trip topology training samples from symbolic music back to a renderable form.

\paragraph{MIDI.}
MIDI export (\texttt{.mid}) produces a Type~0 file with $\text{ticksPerBeat} = 480$. Each staff is mapped to a separate channel. MIDI is used for quick playback verification and does not preserve engraving annotations (beams, slurs, etc.).

\paragraph{Spartito JSON.}
Spartito (\texttt{.spartito.json}) is the primary internal representation. It is a structured JSON object that records all semantic annotations produced by the OMR pipeline: voice assignments, tick positions, duration attributes, staff layout, and quality flags. The \texttt{Spartito} class includes methods for voice-staff extraction (\texttt{makeVoiceStaves}), re-regulation (\texttt{regulate}), and quality tracking. Spartito files are the format written back by human and agent annotators when correcting measures, and are the canonical input to the annotation feedback loop described in Appendix~F.

\section{Algorithm Summary}

This appendix provides a self-contained walkthrough of the BeadSolver topology-recognition loop. The goal is to make the design intent clear: why the loop is structured the way it is, what each sub-step decides, and how the evaluation function keeps the search honest. Readers who want implementation details can cross-reference Appendix~A; those interested in the experimental setup should see Section~5.

\subsection{Problem Statement}

Before topology recognition, the page-level semantic output is reduced to measure instances. A \emph{measure column} is an interval unit in a system, obtained by segmenting the system with linear semantic \texttt{barMeasure} symbols. Each measure instance contains the candidate events inside one measure column, their measure-local geometry, contextual terms, barline information, staff positions, estimated duration, and local musical basics such as time and key signature. This reduction fixes the spatial support and musical context of the problem; BeadSolver then solves only the remaining event-topology problem.

Let a measure instance have staff groups $G_1,\ldots,G_m$ derived from the staff layout (for example, considering an orchestral score with multiple instruments). The instance is decomposed into one event cluster per non-empty staff group. For a group $G_k$, the corresponding cluster contains only events on staves in $G_k$, with coordinates expressed in the group's local staff frame. This prevents independent instruments from being linked into a single voice graph while preserving cross-staff interaction within the same group.

Within each event cluster, the algorithm receives:
\begin{itemize}
  \item $n$ candidate event elements $e_1,\ldots,e_n$ of type \textsc{Chord} or \textsc{Rest}, bracketed by a \textsc{BOS} sentinel $e_0$ and an \textsc{EOS} sentinel $e_{n+1}$;
  \item each element carries geometric coordinates $(x, y_1, y_2, \mathit{pivotX})$, staff index, and a \emph{predisposition} computed by a single forward pass of BeadPicker: predicted tick $\hat{t}_i$, division distribution $\mathbf{d}_i$, dots distribution $\boldsymbol{\delta}_i$, successor probability vector $\mathbf{s}$, and scalar attributes (grace, timeWarped, fullMeasure, fake\,).
\end{itemize}

The algorithm must assign to every non-sentinel element:
\begin{enumerate}
  \item an integer \emph{order} $o_i$ (consecutive integers mark the same voice; a gap of more than one starts a new voice),
  \item a tick offset $t_i \geq 0$,
  \item a division class $d_i \in \{0,\ldots,8\}$ (whole through 256th) and dot count $\delta_i \in \{0,1,2\}$,
  \item scalar attributes (beam state, stem direction, grace, timeWarp, fullMeasure).
\end{enumerate}
The ordering and duration assignments together determine the voice structure and the tick layout; scalar attributes are resolved deterministically from predispositions after search.

\subsection{Three-Phase Action Sequence}

Each search step handles exactly one event through three sub-decisions executed in sequence.

\begin{description}
  \item[\textbf{Pass.}] Choose the next element to enter the chain (or \textsc{EOS} to close the current voice). The posterior distribution over candidates is the successor vector returned by BeadPicker. Selecting \textsc{EOS} when residue remains triggers a new voice: the \textsc{BOS} sentinel is re-entered and BeadPicker is re-queried with the updated prefix.

  \item[\textbf{Division.}] Assign a division class to the newly chained element. The posterior is the element's predisposition \texttt{divisionVector}.

  \item[\textbf{Dots.}] Select augmentation dots (0, 1, or 2). The posterior is \texttt{dotsVector}. Completing the dots step triggers one call to the evaluation function; if the partial solution is already complete or fatally erroneous the node is sealed.
\end{description}

This factorisation mirrors the model output structure and keeps the branching factor small at each decision: $O(n)$ for Pass, 9 for Division, 3 for Dots.

\subsection{Branch Selection and Pretentiousness Pruning}

At each node the branch with the highest \emph{adjusted probability} is explored next:
\[
  \hat{p}_i = \frac{p_i}{\;\mathit{visits}(c_i) + 1\;},
\]
where $p_i$ is the node's raw posterior probability for choice $i$ and $\mathit{visits}(c_i)$ is the access count of child $c_i$ (zero if not yet created). This rule favours high-probability branches while ensuring every branch is eventually visited.

Each time a branch with raw probability $p$ is taken, the path's accumulated \emph{pretentiousness} increases by $\min(100, -\ln p)$. When the total exceeds $100 \times \lambda$ (where $\lambda$ is the \texttt{ptFactor} tolerance, typically $1.0$--$1.6$), the node is pruned: it returns the current partial evaluation and marks itself as exhausted ($\mathit{accessCount} \leftarrow \infty$). Pretentiousness thus acts as a cumulative log-probability penalty that eliminates paths the model considers collectively implausible, analogous to a beam-search threshold but applied lazily during tree expansion.

\begin{myalgo}{\textsc{Deduce}$(v,\;\text{ctx})$~~---~~one path from node $v$ to a leaf}
\small
\noindent\kw{Input}: node $v$ (type $\in$ \{Pass, Division, Dots\}), context ctx = (picker, ptFactor)\\
\noindent\kw{Output}: ClusterEvaluation\\[4pt]
$v.\mathit{count} \leftarrow v.\mathit{count} + 1$\\
$ni \leftarrow \arg\max_i\;\bigl[p_i \;/\; (\mathit{visits}(c_i)+1)\bigr]$\\
\kw{if} $ni = \varnothing$: \quad $v.\mathit{count}\leftarrow\infty$;\enspace \kw{return} \textsc{Evaluate}(\ldots)\\
$v.\mathit{pt} \leftarrow v.\mathit{pt} + \min(100,\,-\ln p_{ni})$\\
\kw{if} $v.\mathit{pt} > 100\cdot\lambda$: \quad $v.\mathit{count}\leftarrow\infty$;\enspace \kw{return} \textsc{Evaluate}(\ldots)\quad// prune\\[4pt]
\kw{switch} $v.\mathit{type}$:\\
\IN\kw{case} Pass:\\
\IIN \kw{if} $e_{ni}=$ EOS:\\
\IIIN $\mathit{eval} \leftarrow$ \textsc{Evaluate}(cluster, tip, pt)\\
\IIIN \kw{if} no residue \kw{or} fatal:\enspace $v.\mathit{count}\leftarrow\infty$;\enspace \kw{return} $\mathit{eval}$\\
\IIIN // start new voice: re-enter BOS, re-query model\\
\IIIN create $c_{ni}$ as Pass node;\enspace $\mathbf{p} \leftarrow \textsc{PredictCluster}(\text{tip}+1)$\\
\IIN \kw{else}:\\
\IIIN $e_{ni}.\mathit{order} \leftarrow \text{tip}$\\
\IIIN create $c_{ni}$ as Division node;\enspace $\mathbf{p} \leftarrow e_{ni}.\mathit{divisionVector}$\\
\IN\kw{case} Division:\enspace $e_\mathit{cur}.\mathit{div} \leftarrow ni$;\enspace create $c_{ni}$ as Dots node;\enspace $\mathbf{p}\leftarrow e_\mathit{cur}.\mathit{dotsVector}$\\
\IN\kw{case} Dots:\\
\IIN $e_\mathit{cur}.\mathit{dots} \leftarrow ni$;\enspace $\mathit{eval} \leftarrow$ \textsc{Evaluate}(cluster, tip, pt)\\
\IIN \kw{if} no residue \kw{or} fatal:\enspace $v.\mathit{count}\leftarrow\infty$;\enspace \kw{return} $\mathit{eval}$\quad// leaf\\
\IIN create $c_{ni}$ as Pass node;\enspace $\mathbf{p} \leftarrow \textsc{PredictCluster}(\text{tip})$\\[4pt]
$\mathit{result} \leftarrow$ \textsc{Deduce}$(c_{ni},\,\text{ctx})$\\
\kw{if} $\mathit{selfEval} \neq \varnothing$ \kw{and} $\mathit{result}.\mathit{fatal}$:\enspace rollback orders $>$ tip;\enspace \kw{return} $\mathit{selfEval}$\\
\kw{return} $\mathit{result}$
\end{myalgo}

\subsection{Cluster Evaluation}

\textsc{Evaluate} is called at every leaf or pruning point. It receives the partial order assignment, computes ticks cumulatively, and returns a scalar loss.

\paragraph{Tick propagation.}
Events are iterated in order. The current voice-local tick advances by the estimated element duration after each event; when a new voice starts the tick resets to zero. If the model's predicted tick $\hat{t}_i$ is available as an integer, it is used as a lower bound for $t_i$ (allowing parallelism to be encoded via predisposition). Time-warped events (e.g.\ triplets) have their duration scaled by $2/3$.

\paragraph{Tick twist $\tau$.}
A well-engraved measure should be spatially monotone: later notes appear further right. The \emph{twist} metric quantifies violations in the $(x, t)$ plane. For each consecutive pair in $x$-sorted order, the angle $\theta$ between the $({\Delta x}/{x_\text{span}},\,{\Delta t}/{t_\text{span}})$ vector and the diagonal is measured:
\[
  \tau \;=\; \max_{\text{pairs}}\;\Bigl(\tfrac{4}{\pi}\,\arctan\tfrac{\Delta t / t_{\text{span}}}{\Delta x / x_{\text{span}}} - 1\Bigr)^{\!2}.
\]
When all adjacent events progress monotonically the angle equals $45°$, giving $\tau=0$. A reversal (later tick to the left of earlier tick) drives $\tau \to 1$; $\tau \geq 1$ is a fatal error.

\paragraph{Combined loss.}
Let $T = 1920$ (whole-note duration in ticks), $r$ = residue count, $v$ = voice count, $s$ = mean unused voice time, $\pi$ = pretentiousness:
\[
  \mathcal{L} \;=\; \frac{\mathrm{RMSE}(t,\hat{t})}{T} \;+\; \tau \;+\; 0.2\,r \;+\; 0.002\,v \;+\; 0.4\,\frac{s}{T} \;+\; 0.02\,\pi.
\]
The numeric weights (e.g., $0.2, 0.002, 0.4, 0.02$) were chosen by manual, empirical tuning to balance these error terms in practice.
Each term captures a distinct failure mode: tick deviation from the model's estimate; spatial monotonicity violation; unassigned events; overly split voices; temporal gaps within a voice; and implausibility of the chosen path.

\subsection{Cluster-Level Search}

\textsc{SolveCluster} drives the search loop. It initialises a root \textsc{Pass} node, then repeatedly descends from the root until the quota is exhausted, the tree is fully explored, or a sufficiently low loss is reached.

\begin{myalgo}{\textsc{SolveCluster}$(C,\;\text{picker},\;Q,\;s_\text{stop},\;\lambda)$}
\small
\noindent\kw{Input}: cluster $C$, BeadPicker, quota $Q$, stop-loss $s_\text{stop}$, ptFactor $\lambda$\\
\noindent\kw{Output}: best ClusterEvaluation found; cluster state updated in place\\[4pt]
$e_0.\mathit{order} \leftarrow 0$;\quad all other orders $\leftarrow \varnothing$\\
$\mathbf{s}_0 \leftarrow \textsc{PredictCluster}(C,\,\text{tip}=1)$\quad// single forward pass to seed root\\
$\mathit{root} \leftarrow$ new Pass node with $\mathbf{p}=\mathbf{s}_0$, $\mathit{pt}=0$\\
$\mathcal{L}^* \leftarrow \infty$;\quad $S^* \leftarrow \varnothing$\\[4pt]
\kw{while} $Q > 0$ \kw{and} $\mathit{root}.\mathit{count} < \infty$:\\
\IN reset: $e_0.\mathit{order}\leftarrow 0$, all others $\leftarrow \varnothing$\quad// restore start state before each path\\
\IN $\mathit{eval} \leftarrow \textsc{Deduce}(\mathit{root},\,\text{ctx})$;\quad $Q \leftarrow Q - 1$\\
\IN \kw{if} $\mathit{eval}.\mathcal{L} < \mathcal{L}^*$:\\
\IIN $\mathcal{L}^* \leftarrow \mathit{eval}.\mathcal{L}$;\quad $S^* \leftarrow \textsc{SaveState}(C)$\\
\IIN \kw{if} $\mathcal{L}^* \leq s_\text{stop}$:\enspace \kw{break}\quad// solution good enough\\[4pt]
\textsc{RestoreState}$(C,\,S^*)$\\[4pt]
// Post-process residue elements: snap tick to nearest fixed event or grid\\
// Assign beam, stemDirection, grace, timeWarp from predispositions\\
// Build $H$ (horizontal adjacency matrix for layout engine)
\end{myalgo}

Each call to \textsc{Deduce} may itself call \textsc{PredictCluster} once at each new Pass node it creates. The quota $Q$ is therefore a budget on the total number of model inference calls, not on the number of tree nodes. The root node's $\mathit{accessCount}$ counts how many times the root has been visited; when it reaches $\infty$ (all branches sealed), the tree is fully explored and the loop terminates regardless of remaining quota.

\subsection{Measure-Level Assembly}

\textsc{SolveMeasure} is the top-level entry point. It splits the measure into clusters (one per staff group), calls \textsc{SolveCluster} on each, and assembles the voice structure.

The per-cluster quota is $Q = \min(\lceil n\cdot f\rceil,\,Q_\text{max})$ where $n$ is the cluster event count, $f$ is \texttt{quotaFactor}, and $Q_\text{max}=1000$.

After all clusters are solved, voices are extracted by iterating events in order-ascending sequence: consecutive events (order gap exactly 1) belong to the same voice; a gap larger than 1 begins a new voice. Events flagged \texttt{grace} or \texttt{fullMeasure} are excluded from voice sequences but receive ticks from the spatially nearest in-voice event.

The horizontal adjacency matrix $H$ ($H_{ij}=1$ iff event $j$ immediately precedes $i$ in its voice) is constructed and returned as part of the \texttt{RegulationSolution}, which is applied to the measure to populate all fields needed for rendering and export.

\paragraph{Post-regulation refinement.}
After the main voice and tick structure has been selected, Starry performs a lightweight post-regulation pass to attach notational details that are awkward to decide inside the main topology search. Grace notes are assigned to nearby regular events according to their horizontal position and staff. Their ticks are placed immediately before the target event, compressed into the available time interval so that they do not disturb the main metric structure. The target event records the associated grace-note identifiers, allowing export and playback code to preserve the ornament relationship.

Tremolo notation is handled similarly as a post-processing relation. Events marked as tremolo receivers are paired with the nearest compatible preceding pitched event, using rhythmic compatibility and horizontal order as the main cues. The pair is then annotated as a tremolo relation, and the receiver can inherit or infer a tick from the paired event when necessary.

The same post-regulation pass also propagates tick positions to contextual terms such as clefs, key signatures, time signatures, and other measure-local annotations. These symbols are not voice events, but they must still be placed at musically meaningful tick positions for rendering and export. Starry therefore walks each staff from right to left and assigns each contextual term the nearest following event tick. This keeps the final representation suitable for MusicXML, LilyPond, MIDI, and internal editing.

\section{Failure Cases}

This appendix documents four representative measures where successive regulation strategies fail, illustrating the performance boundary of each method. For each entry the left panel shows the original sheet music; the centre panel shows the failing strategy's output; the right panel shows the next stronger strategy (or human annotation) that succeeds. In the topology panels, distinct colours identify distinct voices; the narrow timeline strip at the bottom encodes each voice's tick allocation within the measure. Measures were selected to form a difficulty ladder: each row's failure case is too hard for all previous strategies and is solved by the next.

\begin{enumerate}[label=\textbf{Row~\arabic*.}, leftmargin=*, noitemsep]
  \item \textbf{Greedy $\to$ Equations (rule-based).}\enspace Bach, \textit{BWV}~791, m.\,6.
    A greedy, myopic voice assignment produces six fragmented voices and a quality score of zero.
    The linear-equation solver re-links the fragments correctly, yielding three voices with $q = 0.959$.

  \item \textbf{Equations $\to$ BeadSolver (single-pass).}\enspace Chopin, \textit{\'Etude} Op.\,10 No.\,3, m.\,34.
    The constraint system cannot resolve voice crossings in denser polyphony and again produces an error solution.
    BeadSolver's probability-guided search recovers the correct four-voice layout ($q = 0.950$).

  \item \textbf{BeadSolver $\to$ BeadSolver-Multipass.}\enspace Chopin, \textit{\'Etude} Op.\,10 No.\,3, m.\,57.
    Single-pass BeadSolver fixes a suboptimal partial assignment early and cannot escape it.
    Multipass search with escalating quota factors ($f = 10, 40, 80$) allows global backtracking and recovers four voices ($q = 0.930$).

  \item \textbf{BeadSolver-Multipass $\to$ Human annotation.}\enspace Debussy, \textit{Suite Bergamasque} --- Clair de lune, m.\,14.
    The multipass solver reaches a plausible partial solution ($q = 0.778$) but leaves bubbles in the timeline due to the missing of time warps attributes.
    Human correction finds the correct regulation and reaches $q = 0.215$ --- lower raw score because there is an exceptional time warp fraction: 3/2, which break our assumption about time warp of tuplets.
\end{enumerate}

\begin{figure}[H]
\centering
\small
\setlength\tabcolsep{3pt}
\renewcommand{\arraystretch}{1.0}
\begin{tabular}{@{} m{0.17\textwidth} m{0.21\textwidth} m{0.28\textwidth} m{0.28\textwidth} @{}}
\toprule
\textbf{Strategy} &
\centering\textbf{Sheet Music} &
\centering\textbf{Failure Solution} &
\centering\arraybackslash\textbf{Ground Truth} \\
\midrule\addlinespace[3pt]
\raggedright Greedy $\to$ Equations\par\upshape\smallskip\mbox{}\newline
  {\scriptsize Bach, \textit{BWV}~791, m.\,6} &
\centering\includegraphics[width=1.15\linewidth]{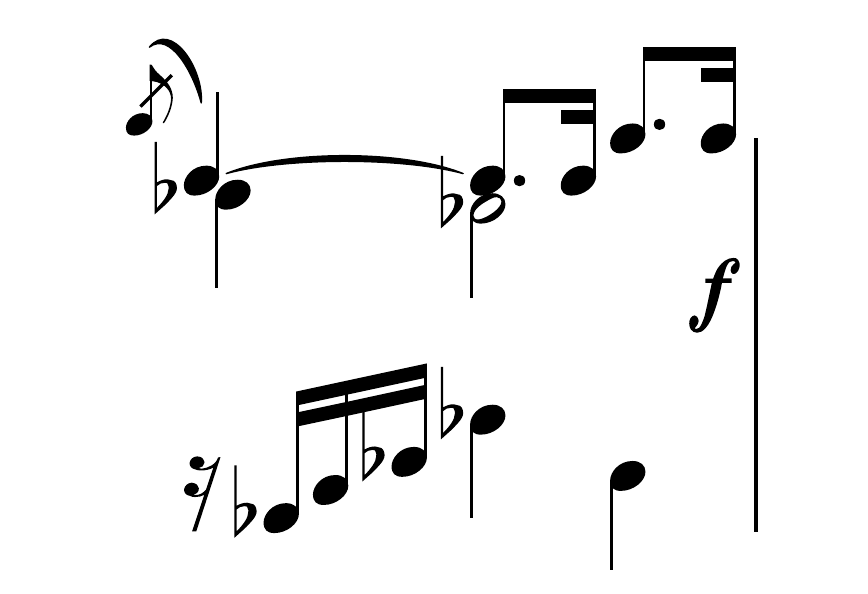} &
\centering\begin{tabular}[b]{@{}c@{}}
  \includegraphics[height=36mm]{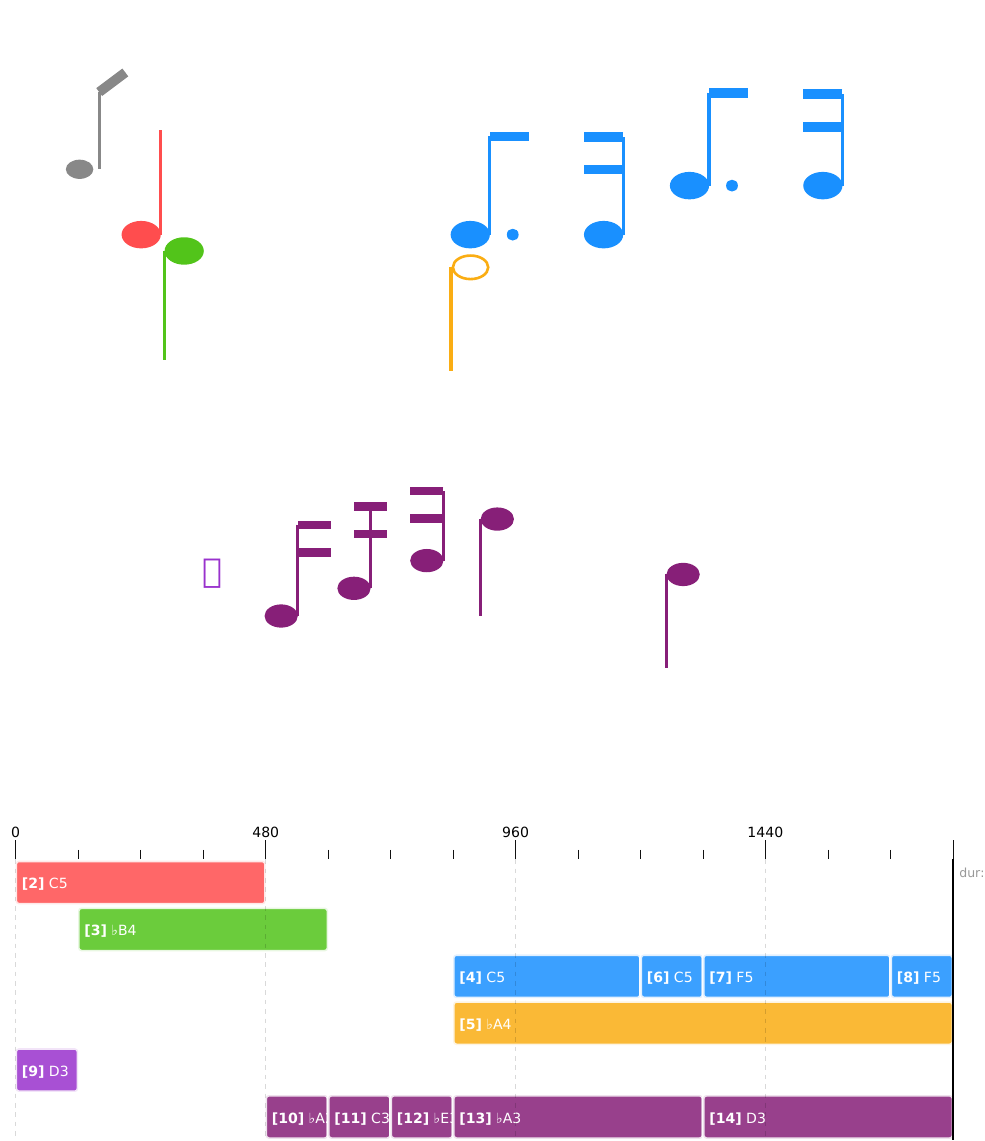}\\[-1pt]
  {\footnotesize\itshape Greedy\upshape\quad 6v\;$q{=}0.000$}
\end{tabular} &
\centering\arraybackslash\begin{tabular}[b]{@{}c@{}}
  \includegraphics[height=36mm]{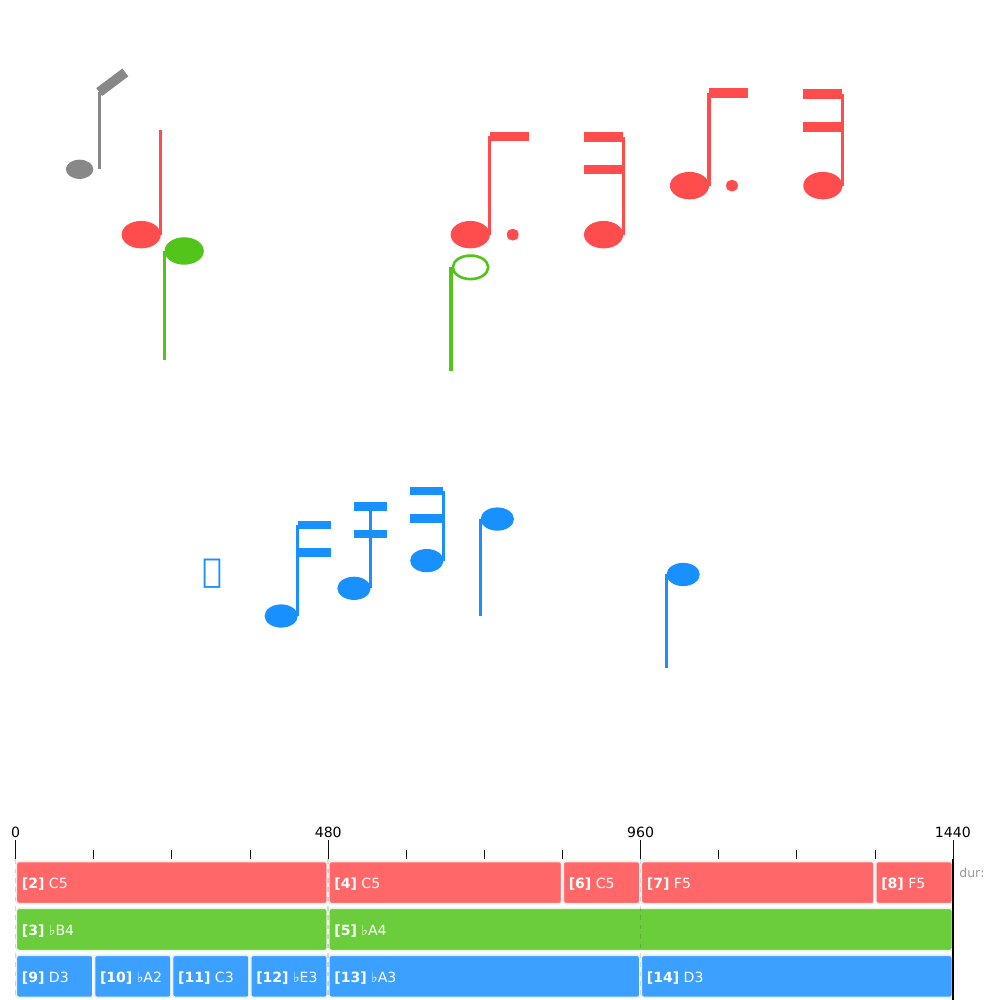}\\[-1pt]
  {\footnotesize\itshape Equations\upshape\quad 3v\;$q{=}0.959$}
\end{tabular} \\[4pt]
\midrule\addlinespace[3pt]
\raggedright Equations $\to$ BeadSolver\par\upshape\smallskip\mbox{}\newline
  {\scriptsize Chopin, \textit{\'Etude} Op.\,10 No.\,3, m.\,34} &
\centering\includegraphics[width=\linewidth]{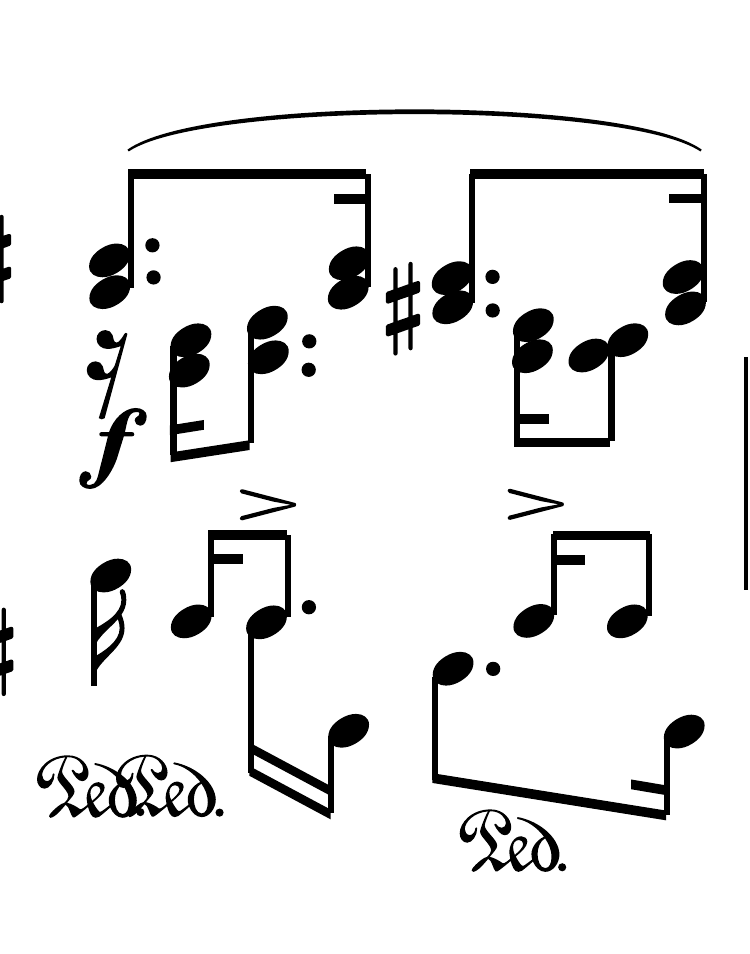} &
\centering\begin{tabular}[b]{@{}c@{}}
  \includegraphics[height=44mm]{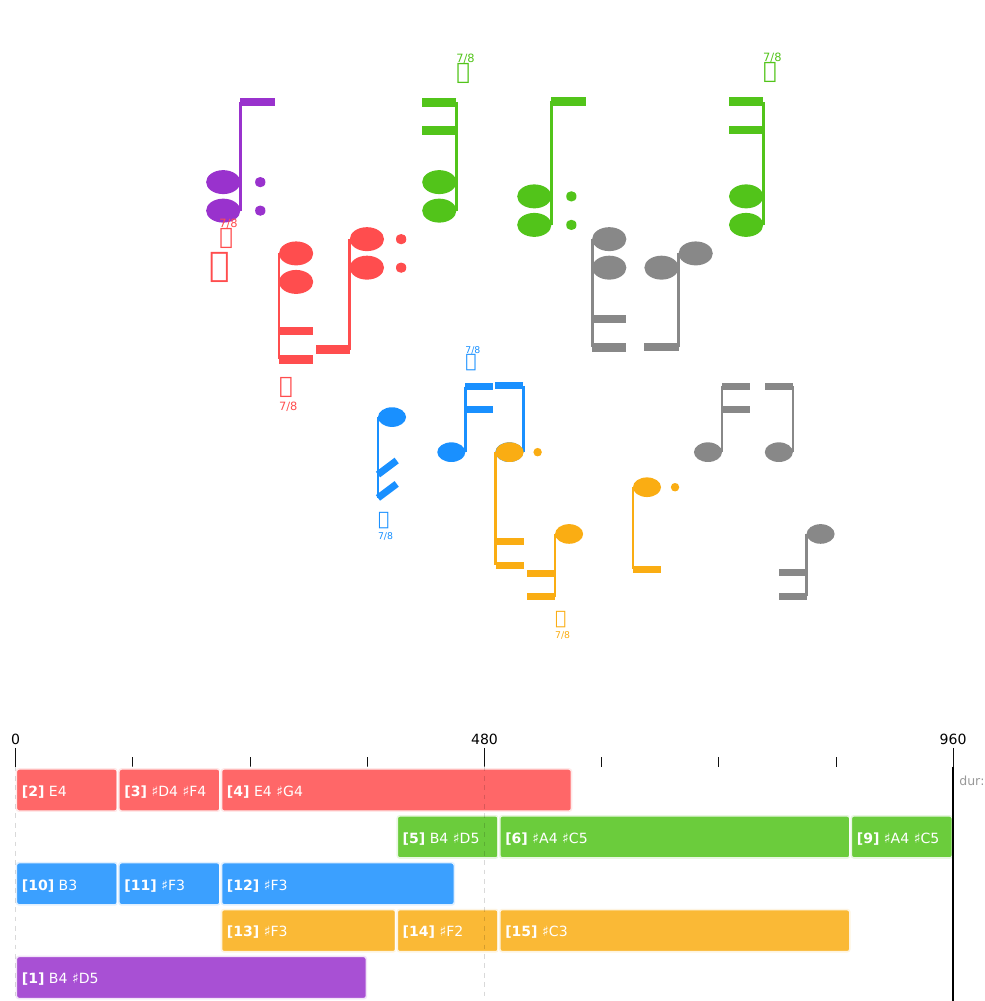}\\[-1pt]
  {\footnotesize\itshape Equations\upshape\quad 5v\;$q{=}0.000$}
\end{tabular} &
\centering\arraybackslash\begin{tabular}[b]{@{}c@{}}
  \includegraphics[height=44mm]{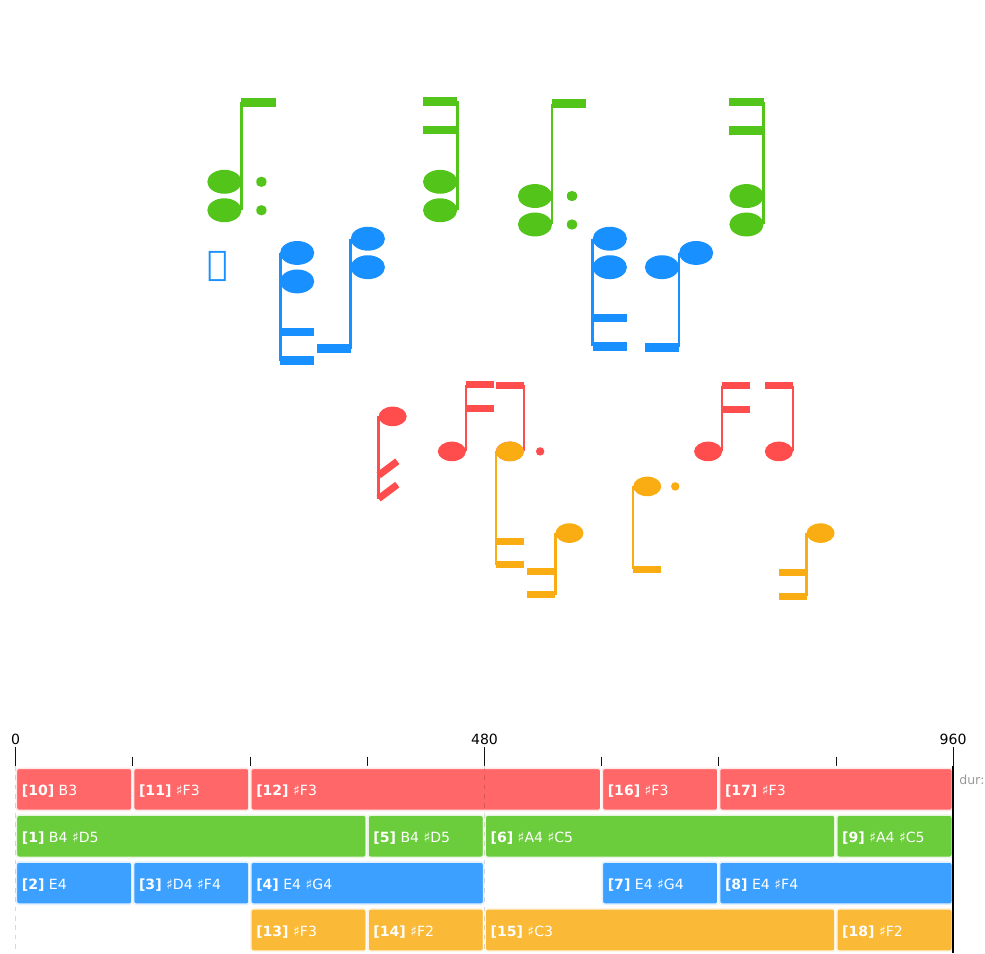}\\[-1pt]
  {\footnotesize\itshape BeadSolver\upshape\quad 4v\;$q{=}0.950$}
\end{tabular} \\[4pt]
\midrule\addlinespace[3pt]
\raggedright BeadSolver $\to$ Multipass\par\upshape\smallskip\mbox{}\newline
  {\scriptsize Chopin, \textit{\'Etude} Op.\,10 No.\,3, m.\,57} &
\centering\includegraphics[width=\linewidth]{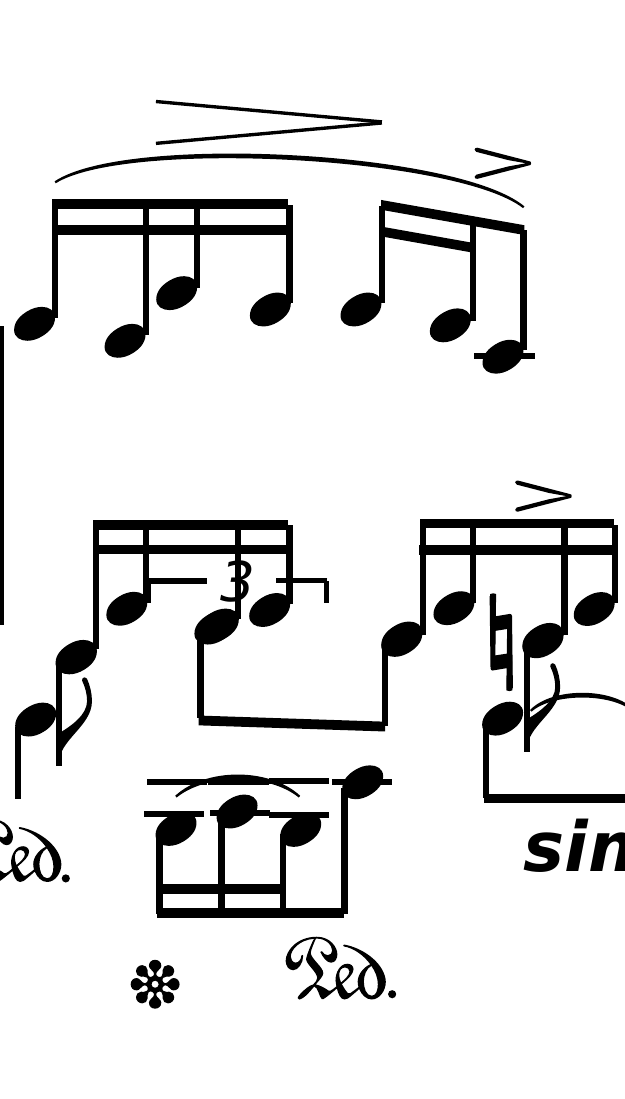} &
\centering\begin{tabular}[b]{@{}c@{}}
  \includegraphics[height=48mm]{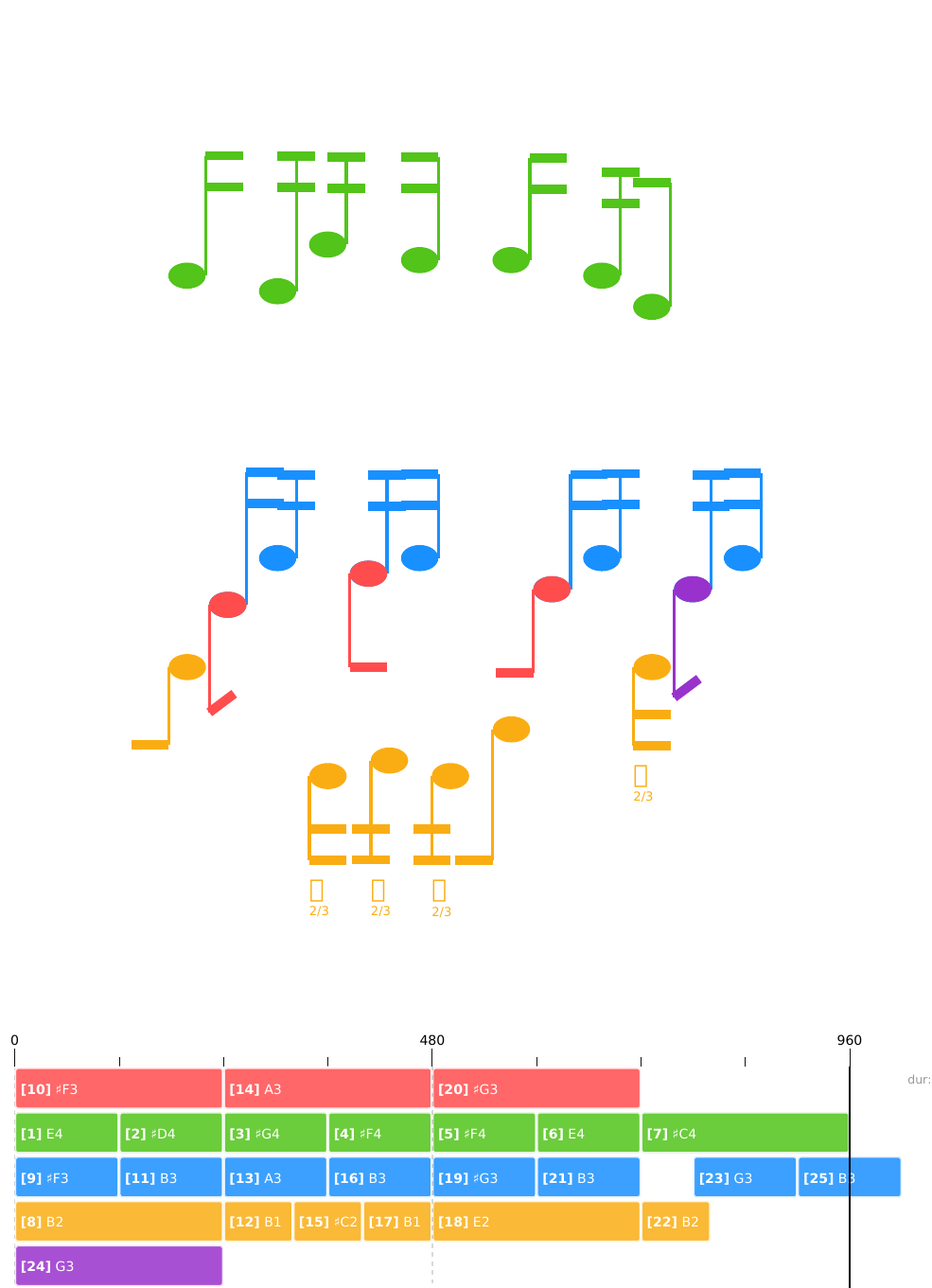}\\[-1pt]
  {\footnotesize\itshape BeadSolver\upshape\quad 5v\;$q{=}0.000$}
\end{tabular} &
\centering\arraybackslash\begin{tabular}[b]{@{}c@{}}
  \includegraphics[height=48mm]{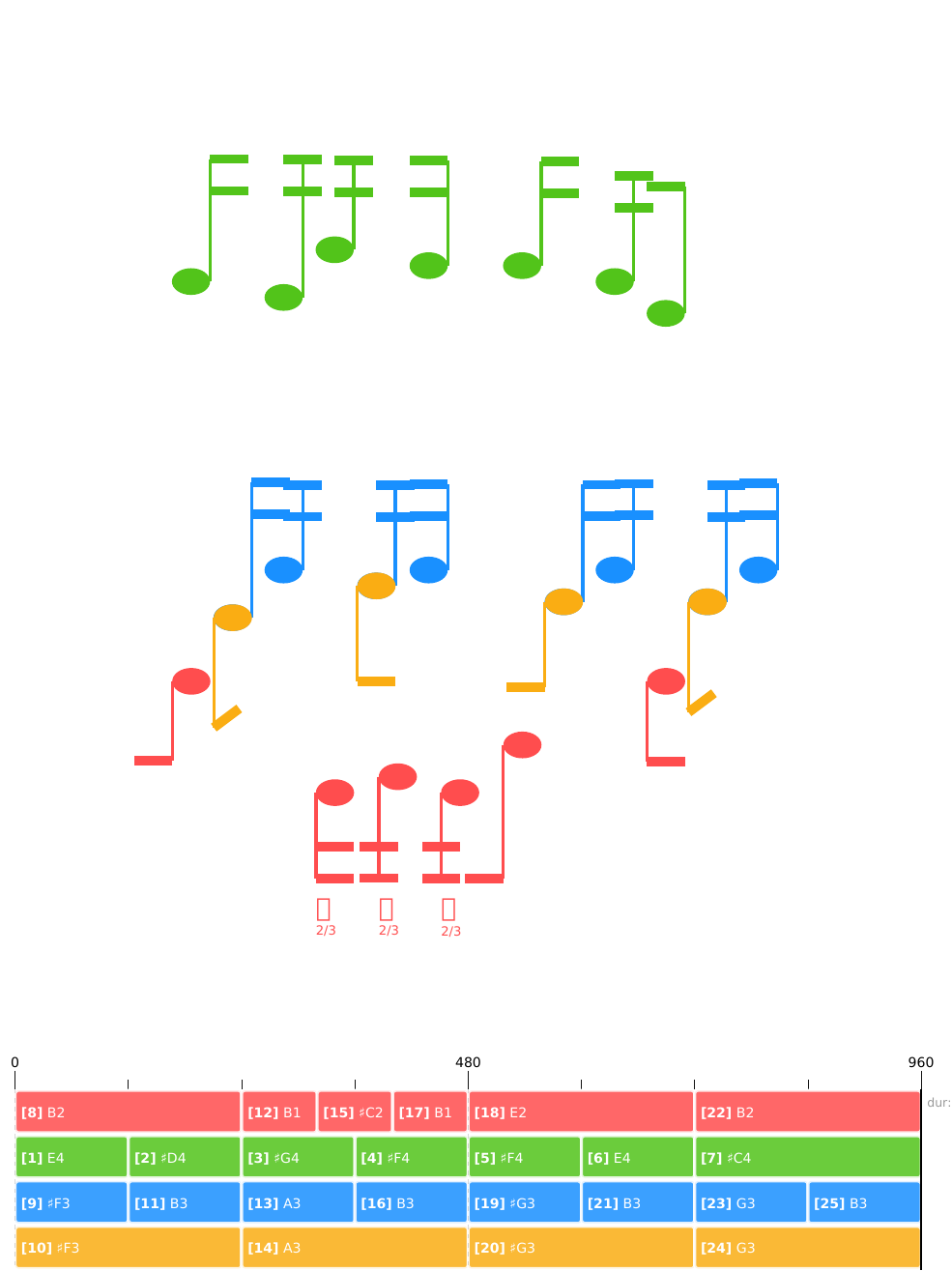}\\[-1pt]
  {\footnotesize\itshape Multipass\upshape\quad 4v\;$q{=}0.930$}
\end{tabular} \\[4pt]
\midrule\addlinespace[3pt]
\raggedright Multipass $\to$ Ground truth\par\upshape\smallskip\mbox{}\newline
  {\scriptsize Debussy, \textit{Suite Berg.} --- Clair de lune, m.\,14} &
\centering\includegraphics[width=\linewidth]{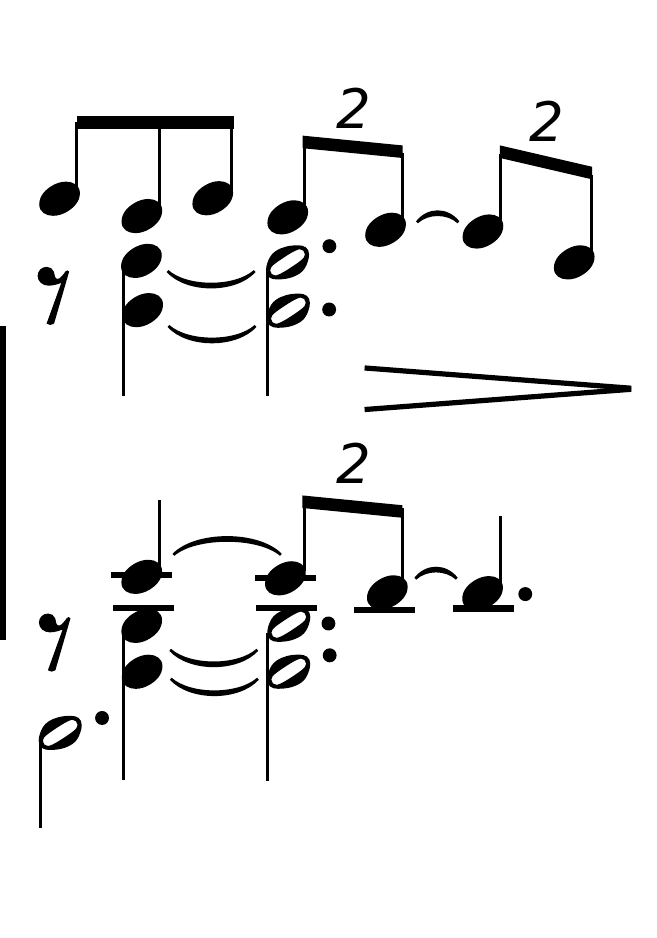} &
\centering\begin{tabular}[b]{@{}c@{}}
  \includegraphics[height=46mm]{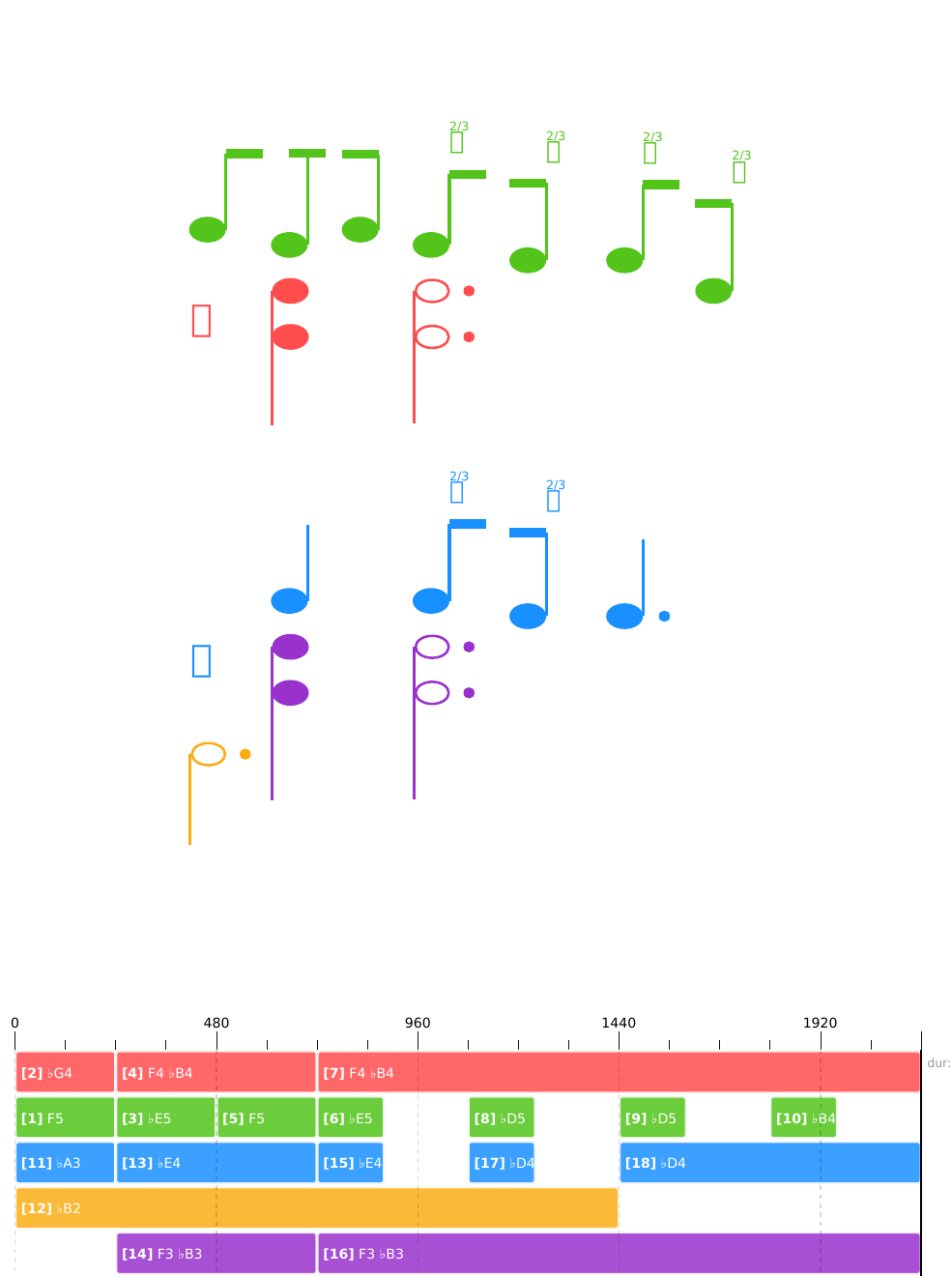}\\[-1pt]
  {\footnotesize\itshape Multipass\upshape\quad 5v\;$q{=}0.778$}
\end{tabular} &
\centering\arraybackslash\begin{tabular}[b]{@{}c@{}}
  \includegraphics[height=46mm]{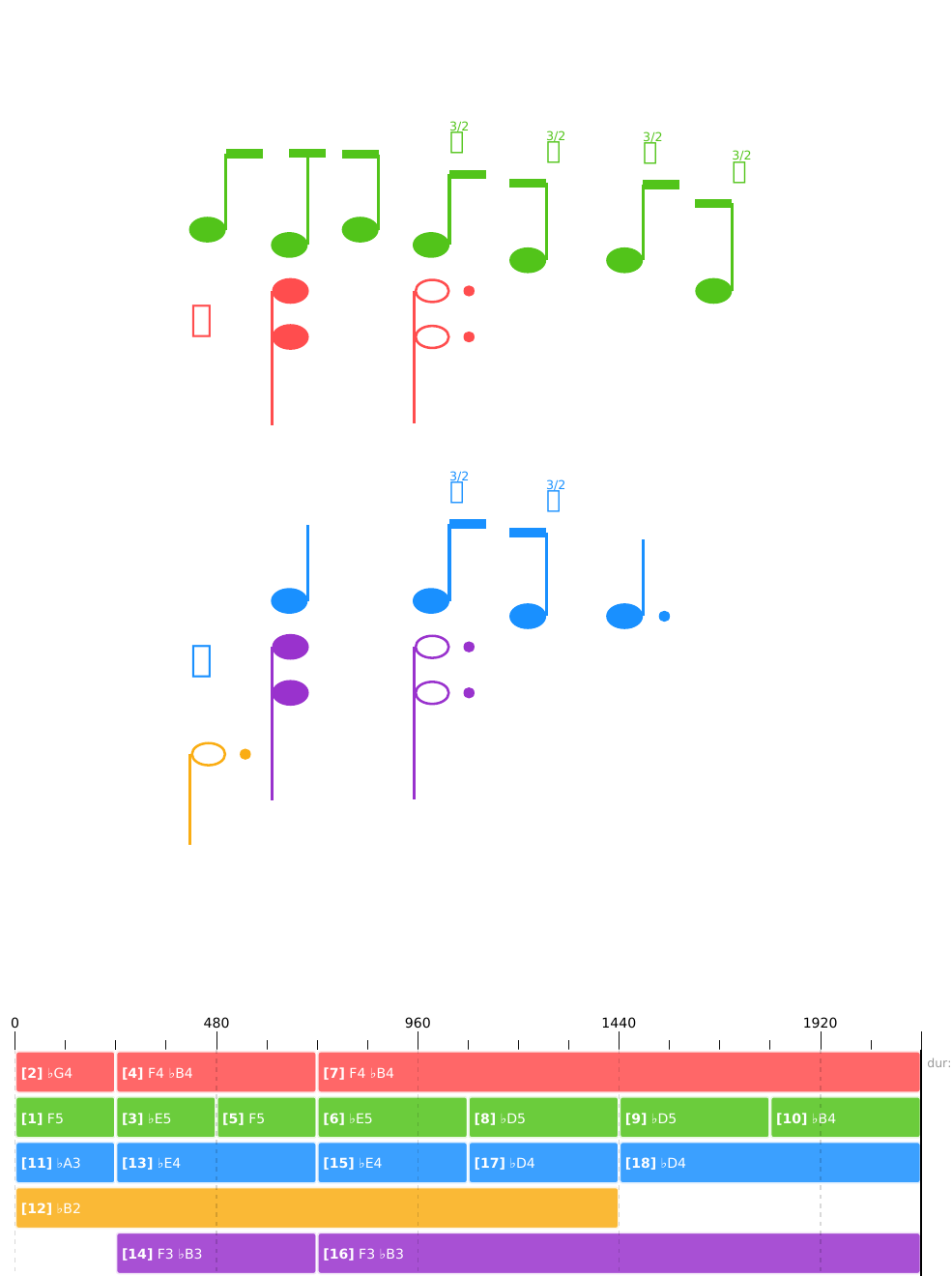}\\[-1pt]
  {\footnotesize\itshape Human\upshape\quad 5v\;$q{=}0.215$}
\end{tabular} \\[2pt]
\bottomrule
\end{tabular}
\caption{Representative failure cases illustrating the performance boundary of each regulation strategy. Each row shows the same measure under the failing strategy (centre) and the succeeding strategy or human annotation (right). Topology panels colour-code voices; the timeline strip beneath each topology view shows per-voice tick allocation. Rows are ordered by increasing difficulty.}
\label{fig:failure-cases}
\end{figure}

\section{Paraff Language and Generation Constraints}

\textbf{Paraff} is a domain-specific language (DSL) for symbolic music, designed to be simultaneously (i)~losslessly convertible from/to LilyPond, (ii)~unambiguously parseable by a formal grammar, and (iii)~friendly to autoregressive language models through a compact, space-separated token vocabulary of fewer than 256 tokens.
In this paper, Paraff serves as the intermediate representation for the topology-data generation pipeline described in Section~4.4.2: scores are generated by an autoregressive model trained on Paraff sequences, then rendered and converted to event-topology training samples for the BeadSolver.
In this conversion, geometric coordinates are taken from engraving rather than synthesized: Paraff is translated to LilyPond, engraved to SVG, and then Lotus parses the SVG to recover token geometry (including $x$ positions) in staff-space units. Layout diversity is introduced by randomizing engraving conditions (e.g., staff size, page dimensions, raggedness, and compact layout presets), which changes the resulting spacing while keeping it physically plausible.

\subsection*{Token Vocabulary and Groups}

All Paraff tokens are pure alphanumeric strings.
Tokens are partitioned into \emph{groups} (analogous to parts of speech), each covering a semantically coherent class of notation elements.
See \tabref{tab:paraff-vocab} for the principal groups.

\begin{table}[H]
\centering
\caption[Paraff Token Groups]{Paraff Token Groups. Principal token groups.}
\label{tab:paraff-vocab}
\small
\begin{tabular}{@{}lll@{}}
\toprule
\textbf{Group (symbol)} & \textbf{Example tokens} & \textbf{Meaning} \\
\midrule
Special       & \texttt{PAD BOM EOM VB}               & Padding, measure boundaries, voice separator \\
Staff (S)     & \texttt{S1 S2 S3}                     & Staff assignment (top to bottom) \\
Clef (C)      & \texttt{Cg Cf}                        & Treble / bass clef \\
Key (K)       & \texttt{K0 K1\ldots K6 K\_1\ldots K\_6} & Key signature ($\pm$1--6 sharps/flats) \\
Time sig.\ numerator (TN) & \texttt{TN2 TN3 TN4 TN6 TN8\ldots} & Beats per bar \\
Time sig.\ denominator (TD) & \texttt{TD2 TD4 TD8}           & Beat unit \\
Pitch (a)     & \texttt{a b c d e f g}                & Scale degree (LilyPond relative mode) \\
Accidental (A)& \texttt{As Af Ass Aff}                & Sharp, flat, double-sharp, double-flat \\
Octave (Osp/Osb) & \texttt{Osup Osub}                & Octave up / down from relative default \\
Duration (Dn) & \texttt{D1 D2 D4 D8 D16 D32 D64}     & Note value ($2^{-n}$ of a whole note) \\
Dot           & \texttt{Dot}                          & Augmentation dot (stacks; each adds $\times\tfrac{1}{2}$) \\
Beam (B)      & \texttt{Bl Br}                        & Beam group left / right endpoint \\
Rest (R)      & \texttt{Rest RSpace}                  & Audible rest / invisible spacer \\
Time warp (W) & \texttt{W2 W3\ldots W16 W Wx}         & Tuplet numerator; \texttt{W} continues group \\
Expressive (E)& \texttt{EslurL EslurR Etie Earp}      & Slur, tie, arpeggio markers \\
\bottomrule
\end{tabular}
\end{table}

\subsection*{Sentence Grammar}

A Paraff sentence encodes exactly one measure:
\begin{center}
\texttt{BOM} \;\textit{voice$_1$}\; [\,\texttt{VB} \;\textit{voice$_2$}\; [\,\texttt{VB} \;\textit{voice$_3$} \,$\cdots$]\,] \;\texttt{EOM}
\end{center}
Each \emph{voice} is a sequence of \emph{terms}.
A \emph{context term} is a single token from \{S, C, K, TN, TD\} that sets a persistent local context for subsequent events.
An \emph{event term} has the form:
\begin{center}
$\underbrace{\textit{pitch}^+}_{\text{chord}}$
\quad
$\underbrace{[\textit{timewarp}]\;\textit{Dn}\;[\textit{Dot}^*]}_{\text{duration}}$
\quad
$\underbrace{[\textit{B/R/E}\,]^*}_{\text{post-events}}$
\end{center}

\paragraph{Hello World --- 4/4, C major, whole note C.}
\begin{verbatim}
BOM K0 TN4 TD4 S1 Cg c D1 EOM
\end{verbatim}

\begin{figure}[H]
\centering
\includegraphics[width=0.7\linewidth]{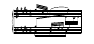}
\caption{Rendered output of the multi-voice cross-staff Paraff example. Three voices share and switch between treble and bass staves; brackets denote beamed groups.}
\label{fig:paraff-complex}
\end{figure}

\paragraph{Multi-voice cross-staff passage (3/8, E major).}
\begin{verbatim}
BOM K4 TN3 TD8
  S1 Cg f As Osup D32 Bl d As D32 b D32 g As D32 Br
  S2 f As Osub D32 Bl d As D32 b D32 g As D32 Br
  S1 d As Osup D32 Bl S2 g As D32 Osub S1 b D32 d As D32 Br VB
  S2 Cf b D8 S1 d As Osup D8 EslurL f As D8 EslurR VB
  S2 Cf b Osub d As D8 Bl b D8 b D8 Br EOM
\end{verbatim}
This example uses three voices, two staves (treble/bass), and cross-staff note assignment via inline \texttt{S1}/\texttt{S2} context switches within a single voice.
Figure~\ref{fig:paraff-complex} shows the rendered output.

\subsection*{Transition Matrix and Constrained Sampling}

To prevent syntactically invalid sequences during autoregressive generation, Paraff applies a hard \emph{logit mask} at each decoding step, derived from a $G \times G$ binary \emph{token-group transition matrix} $T$ (where $G$ is the number of groups).
Entry $T[i][j]=1$ permits a token from group $j$ to follow a token from group $i$; $T[i][j]=0$ forbids it.

Concretely, before sampling token $t_{n+1}$, the model logits $\mathbf{l}\in\mathbb{R}^{|\mathcal{V}|}$ are modified:
\[
\tilde{l}_k = \begin{cases} l_k & \text{if } T\!\left[\mathrm{grp}(t_n),\,\mathrm{grp}(k)\right]=1, \\ -\infty & \text{otherwise.} \end{cases}
\]
Sampling then draws from the renormalised softmax over valid continuations only.
Some illustrative rules: $\mathrm{K}\!\to\!\mathrm{TN}$ (1, key precedes time-sig numerator); $\mathrm{TN}\!\to\!\mathrm{TD}$ (1, numerator before denominator); $\mathrm{TD}\!\to\!\mathrm{S}$ (1, time sig followed by staff selector); $\mathrm{TN}\!\to\!\mathrm{TN}$ (0, two consecutive numerators forbidden).

This shifts grammatical correctness from a post-hoc filter into the sampling process itself, guaranteeing that every generated sequence is parseable --- an important property when Paraff output is piped into downstream topology-extraction tools.

\subsection*{Role in Topology Training Data}

Every valid Paraff sentence carries, by construction, the fields needed to assemble an event-topology supervision sample:
\begin{itemize}[nosep]
  \item \textbf{Voice assignment}: \texttt{VB} boundaries partition events into monophonic voice sequences.
  \item \textbf{Tick positions}: voice-internal order and duration tokens (\texttt{Dn}/\texttt{Dot}/\texttt{W*}) determine each event's absolute tick.
  \item \textbf{Staff membership}: inline \texttt{S1}/\texttt{S2} context tokens record cross-staff assignments.
  \item \textbf{Structural attributes}: beam markers (\texttt{Bl}/\texttt{Br}), tuplet markers (\texttt{W*}), and expressive markers (\texttt{E*}) supply the ancillary fields needed by BeadSolver's evaluation function.
\end{itemize}
A Paraff sentence can therefore be deterministically decoded into the $(x, \mathrm{voices}, \mathrm{tick}, \mathrm{div})$ tuple consumed by the BeadSolver model without additional annotation.
At scale, an autoregressive Paraff generator produces structurally diverse measures that collectively cover the multi-voice and cross-staff patterns most challenging for automated regulation --- precisely the structural domain BeadSolver is designed to handle.

\begin{figure}[h]
\centering
\includegraphics[height=0.8\textheight,keepaspectratio]{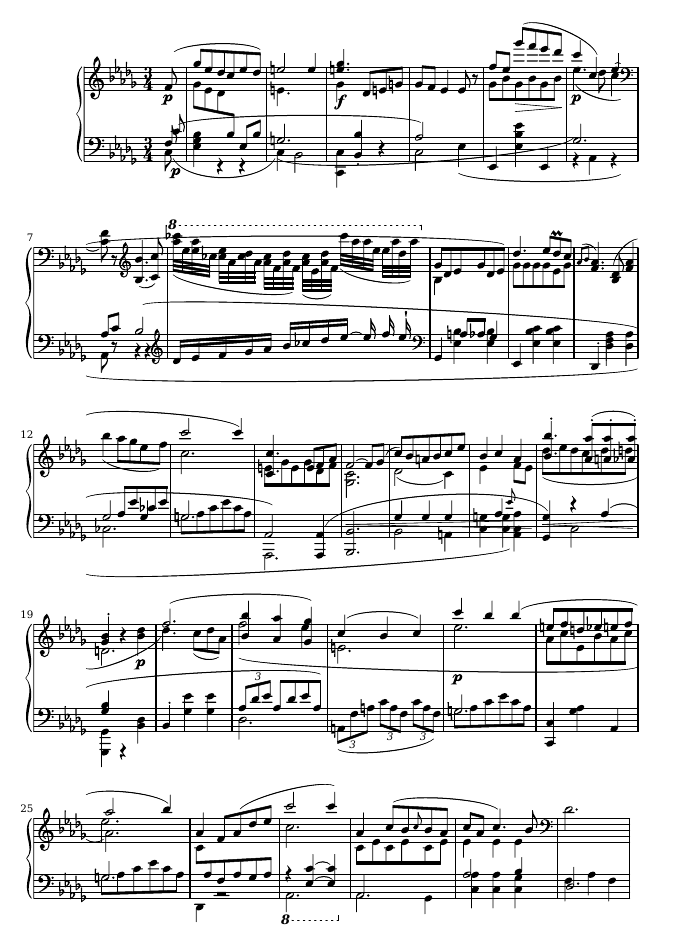}
\caption{A sample score generated by a learned Paraff generation model via constrained autoregressive sampling, with temperature = 1. Engraved by Lilypond 2.22.}
\label{fig:paraff-sample}
\end{figure}

\medskip
\noindent Figure~\ref{fig:paraff-sample} shows a score excerpt sampled from the Paraff generative model.
During sampling, prompt tokens condition the model toward complex structural tendencies.
The result, while somewhat chaos, demonstrates that pitch and rhythmic distributions broadly reflect those found in real engraved scores, suggesting that the Paraff generator captures musically plausible structure at the measure level.

\medskip
\noindent The Paraff implementation is open-sourced at \url{https://github.com/findlab-org/paraff}.

\section{Symbol Semantic Type Inventory}

See \tabref{tab:semtypes} for the label definitions used by the semantic model. In Starry, semantic labels fall into two basic geometric kinds: point labels and vertical-line labels. In the training annotations, both are marked in red on the score image and used directly as supervision targets for the semantic model. The table gives the full inventory of these labels together with a representative glyph for each type. For readability, the labels are grouped by musical role; the \emph{Contextual} group covers helper tokens that carry pitch-position or rendering metadata rather than independent note events.

\begin{longtable}{@{} l @{\hspace{6pt}} l @{\hspace{6pt}} c @{\hspace{8pt}} p{0.46\textwidth} @{}}
\caption[Semantic Type Inventory]{Semantic Type Inventory. Full semantic type inventory with example glyphs.}\label{tab:semtypes}\\
\toprule
\textbf{Type} & \textbf{Category} & \textbf{Case} & \textbf{Description} \\
\midrule
\endfirsthead
\multicolumn{4}{@{}l}{\small\itshape (continued)} \\
\toprule
\textbf{Type} & \textbf{Category} & \textbf{Case} & \textbf{Description} \\
\midrule
\endhead
\midrule
\multicolumn{4}{r@{}}{\small\itshape (continued on next page)} \\
\endfoot
\bottomrule
\endlastfoot
\texttt{ClefG} & Clef & \includegraphics[scale=0.945]{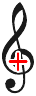} & G clef (treble clef) \\
\texttt{ClefF} & Clef & \includegraphics[scale=0.945]{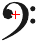} & F clef (bass clef) \\
\texttt{ClefC} & Clef & \includegraphics[scale=0.945]{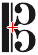} & C clef (alto / tenor clef) \\
\midrule
\texttt{NoteheadS0} & Notehead & \includegraphics[scale=0.945]{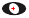} & Whole-note head (open, wide oval) \\
\texttt{NoteheadS1} & Notehead & \includegraphics[scale=0.945]{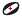} & Half-note head (open oval) \\
\texttt{NoteheadS2} & Notehead & \includegraphics[scale=0.945]{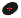} & Filled notehead (quarter, eighth, \ldots{}) \\
\midrule
\texttt{vline\_Stem} & Stem & \includegraphics[scale=0.945]{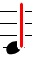} & Vertical stem line \\
\midrule
\texttt{Flag3} & Flag & \includegraphics[scale=0.945]{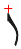} & Eighth-note (and upper) flag on stem \\
\midrule
\texttt{BeamLeft} & Beam & \includegraphics[scale=0.945]{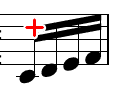} & Beam group, left endpoint \\
\texttt{BeamContinue} & Beam & \includegraphics[scale=0.945]{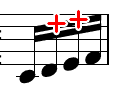} & Beam group, interior segment \\
\texttt{BeamRight} & Beam & \includegraphics[scale=0.945]{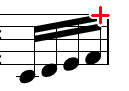} & Beam group, right endpoint \\
\midrule
\texttt{TremoloLeft} & Tremolo & \includegraphics[scale=0.945]{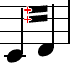} & Tremolo stroke, left endpoint \\
\texttt{TremoloMiddle} & Tremolo & \includegraphics[scale=0.945]{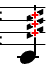} & Tremolo stroke, interior \\
\texttt{TremoloRight} & Tremolo & \includegraphics[scale=0.945]{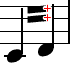} & Tremolo stroke, right endpoint \\
\midrule
\texttt{Dot} & Duration & \includegraphics[scale=0.945]{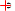} & Augmentation dot \\
\midrule
\texttt{Rest0} & Rest & \includegraphics[scale=0.945]{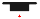} & Breve rest (double whole) \\
\texttt{Rest1} & Rest & \includegraphics[scale=0.945]{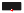} & Whole rest \\
\texttt{Rest2} & Rest & \includegraphics[scale=0.945]{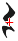} & Half rest \\
\texttt{Rest3} & Rest & \includegraphics[scale=0.945]{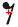} & Quarter rest \\
\texttt{Rest4} & Rest & \includegraphics[scale=0.945]{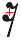} & Eighth rest \\
\texttt{Rest5} & Rest & \includegraphics[scale=0.945]{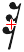} & Sixteenth rest \\
\texttt{Rest6} & Rest & \includegraphics[scale=0.945]{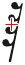} & 32nd rest \\
\midrule
\texttt{AccNatural} & Accidental & \includegraphics[scale=0.945]{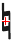} & Natural sign \\
\texttt{AccSharp} & Accidental & \includegraphics[scale=0.945]{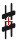} & Sharp \\
\texttt{AccDoublesharp} & Accidental & \includegraphics[scale=0.945]{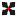} & Double sharp \\
\texttt{AccFlat} & Accidental & \includegraphics[scale=0.945]{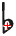} & Flat \\
\texttt{AccFlatflat} & Accidental & \includegraphics[scale=0.945]{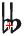} & Double flat \\
\midrule
\texttt{vline\_VoltaLeft} & Volta & \includegraphics[scale=0.945]{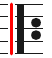} & Volta bracket, left vertical stroke (start-repeat \texttt{".|:"}) \\
\texttt{vline\_VoltaRight} & Volta & \includegraphics[scale=0.945]{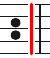} & Volta bracket, right vertical stroke (end-repeat \texttt{":|."}) \\
\midrule
\texttt{vline\_BarMeasure} & Barline & \includegraphics[scale=0.945]{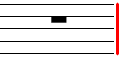} & Barline vertical line segment \\
\texttt{vline\_BarTerminal} & Barline & \includegraphics[scale=0.945]{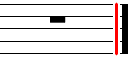} & Terminal barline vertical line \\
\texttt{vline\_BarSegment} & Barline & \includegraphics[scale=0.945]{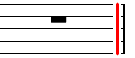} & Segmented (repeat) barline vertical line \\
\midrule
\texttt{SlurBegin} & Slur / Tie & \includegraphics[scale=0.945]{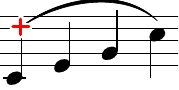} & Slur or tie, begin anchor \\
\texttt{SlurEnd} & Slur / Tie & \includegraphics[scale=0.945]{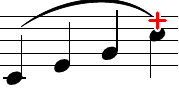} & Slur or tie, end anchor \\
\midrule
\texttt{TimesigC44} & Time Sig. & \includegraphics[scale=0.945]{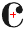} & Common time (4/4) \\
\texttt{TimesigC22} & Time Sig. & \includegraphics[scale=0.945]{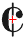} & Cut time / alla breve (2/2) \\
\texttt{TimesigZero} & Time Sig. & \includegraphics[scale=0.945]{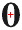} & Time-signature digit 0 \\
\texttt{TimesigOne} & Time Sig. & \includegraphics[scale=0.945]{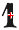} & Time-signature digit 1 \\
\texttt{TimesigTwo} & Time Sig. & \includegraphics[scale=0.945]{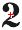} & Time-signature digit 2 \\
\texttt{TimesigThree} & Time Sig. & \includegraphics[scale=0.945]{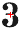} & Time-signature digit 3 \\
\texttt{TimesigFour} & Time Sig. & \includegraphics[scale=0.945]{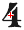} & Time-signature digit 4 \\
\texttt{TimesigFive} & Time Sig. & \includegraphics[scale=0.945]{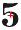} & Time-signature digit 5 \\
\texttt{TimesigSix} & Time Sig. & \includegraphics[scale=0.945]{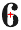} & Time-signature digit 6 \\
\texttt{TimesigSeven} & Time Sig. & \includegraphics[scale=0.945]{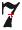} & Time-signature digit 7 \\
\texttt{TimesigEight} & Time Sig. & \includegraphics[scale=0.945]{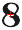} & Time-signature digit 8 \\
\texttt{TimesigNine} & Time Sig. & \includegraphics[scale=0.945]{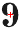} & Time-signature digit 9 \\
\midrule
\texttt{OctaveShift8} & Ottava & \includegraphics[scale=0.945]{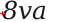} & Ottava "8" digit glyph \\
\texttt{OctaveShift0} & Ottava & \includegraphics[scale=0.945]{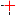} & Ottava termination marker \\
\midrule
\texttt{f} & Dynamic & \includegraphics[scale=0.945]{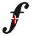} & dynamic letter: forte \\
\texttt{p} & Dynamic & \includegraphics[scale=0.945]{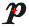} & dynamic letter: piano \\
\texttt{m} & Dynamic & \includegraphics[scale=0.945]{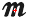} & dynamic letter: mezzo (component of \textit{mf}, \textit{mp}) \\
\texttt{n} & Dynamic & \includegraphics[scale=0.945]{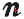} & dynamic letter: niente \\
\texttt{r} & Dynamic & \includegraphics[scale=0.945]{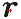} & dynamic letter: rinforzando (\textit{rf}, \textit{rfz}) \\
\texttt{s} & Dynamic & \includegraphics[scale=0.945]{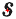} & dynamic letter: sforzando prefix (\textit{sf}, \textit{sfz}, \textit{sff}) \\
\texttt{z} & Dynamic & \includegraphics[scale=0.945]{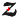} & dynamic letter: sforzando suffix (\textit{fz}, \textit{sfz}) \\
\midrule
\texttt{CrescendoBegin} & Hairpin & \includegraphics[scale=0.945]{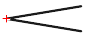} & Crescendo hairpin, open end \\
\texttt{CrescendoEnd} & Hairpin & \includegraphics[scale=0.945]{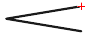} & Crescendo hairpin, closed end \\
\texttt{DecrescendoBegin} & Hairpin & \includegraphics[scale=0.945]{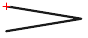} & Decrescendo hairpin, closed end \\
\texttt{DecrescendoEnd} & Hairpin & \includegraphics[scale=0.945]{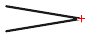} & Decrescendo hairpin, open end \\
\midrule
\texttt{ScriptFermata} & Script & \includegraphics[scale=0.945]{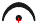} & Fermata \\
\texttt{ScriptShortFermata} & Script & \includegraphics[scale=0.945]{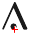} & Short fermata \\
\texttt{ScriptSforzato} & Script & \includegraphics[scale=0.945]{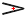} & Accent (sforzato) \\
\texttt{ScriptStaccato} & Script & \includegraphics[scale=0.945]{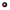} & Staccato dot \\
\texttt{ScriptStaccatissimo} & Script & \includegraphics[scale=0.945]{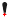} & Staccatissimo wedge \\
\texttt{ScriptTurn} & Script & \includegraphics[scale=0.945]{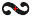} & Turn ornament \\
\texttt{ScriptTrill} & Script & \includegraphics[scale=0.945]{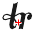} & Trill \\
\texttt{ScriptSegno} & Script & \includegraphics[scale=0.945]{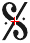} & Segno \\
\texttt{ScriptCoda} & Script & \includegraphics[scale=0.945]{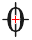} & Coda \\
\texttt{ScriptArpeggio} & Script & \includegraphics[scale=0.945]{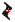} & Arpeggio (wavy line) \\
\texttt{ScriptPrall} & Script & \includegraphics[scale=0.945]{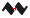} & Prall / upper mordent \\
\texttt{ScriptMordent} & Script & \includegraphics[scale=0.945]{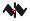} & Mordent / lower mordent \\
\texttt{ScriptMarcato} & Script & \includegraphics[scale=0.945]{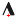} & Marcato \\
\texttt{ScriptTenuto} & Script & \includegraphics[scale=0.945]{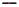} & Tenuto line \\
\texttt{ScriptPortato} & Script & \includegraphics[scale=0.945]{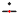} & Portato (tenuto + staccato) \\
\midrule
\texttt{PedalStar} & Pedal & \includegraphics[scale=0.945]{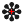} & Pedal release (*) \\
\texttt{PedalPed} & Pedal & \includegraphics[scale=0.945]{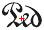} & Piano pedal (Ped) \\
\midrule
\texttt{KeyAcc} & Contextual & \includegraphics[scale=0.945]{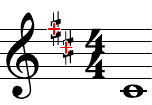} & Key-signature accidental \\
\texttt{TempoNotehead} & Contextual & \includegraphics[scale=0.945]{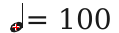} & Notehead in tempo indication \\
\texttt{GraceNotehead} & Contextual & \includegraphics[scale=0.945]{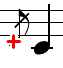} & Grace note head \\
\texttt{SignLined} & Contextual & \includegraphics[scale=0.945]{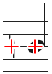} & Pitch-position context: notehead or accidental lies \textbf{on} a staff line (staff-space y is integer) \\
\texttt{SignInterval} & Contextual & \includegraphics[scale=0.945]{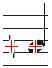} & Pitch-position context: notehead or accidental lies \textbf{in a space} between staff lines (staff-space y is half-integer) \\
\end{longtable}

\section{Agent Annotation Rollout Example}

This appendix reproduces the complete agent-assisted annotation session for piece \texttt{IMSLP\,87102-302241}, measure~274 (model: \texttt{moonshotai/kimi-k2.5}). When the automated BeadSolver regulation scores below threshold, the measure is issued to a spartito annotation agent equipped with an image-read tool, an \texttt{evaluate\_fix} MCP tool (applies a proposed fix and returns quality metrics), and standard file tools. The agent iterates until the measure passes quality checks, then commits the fix to the annotation store.

\subsection*{System Prompt}

{\small
You are a music notation expert annotating regulation issue measures from an OMR (Optical Music Recognition) pipeline. Think deeply and analyze each measure carefully before proposing fixes.

\paragraph{Background}
Regulation assigns tick positions and durations to detected music events. You review measures where regulation failed and output structured fixes. By the time annotation begins, regulation has already run and each event should have: \texttt{tick} (absolute position in the measure, 0 = start), \texttt{division} + \texttt{dots} (note value determining duration), and \texttt{voices} (grouping into monophonic sequences). Your task is to verify and fix these assignments where the algorithm failed.

\paragraph{Duration Formula}
\texttt{duration = 1920 * 2\^{}(-division) * (2 - 2\^{}(-dots))}\\
Whole=1920, Half=960, Quarter=480, Eighth=240, Sixteenth=120; Dotted quarter=720, Dotted half=1440, Double-dotted quarter=840.

\paragraph{Key Concepts}

\textbf{Division, Dots, Tick}
\begin{itemize}[nosep]
  \item \textbf{division}: 0=whole, 1=half, 2=quarter, 3=eighth, 4=sixteenth, 5=32nd, 6=64th
  \item \textbf{dots}: 0=none, 1=dotted, 2=double-dotted
  \item \textbf{tick}: Absolute position within a measure. 0 = beginning, max = measure duration
  \item \textbf{WHOLE\_DURATION} = 1920 ticks per whole note
\end{itemize}

\textbf{Voices}
\begin{itemize}[nosep]
  \item \texttt{voices: number[][]} --- array of monophonic event ID sequences
  \item Each voice is a \texttt{number[]} of event \textbf{IDs} in chronological order
  \item Events in the same voice must not overlap in time
  \item Events not in any voice are ``fake events'' (not sounding)
  \item \textbf{Partial voices}: Not every voice fills the full measure. A voice may cover only part. This is normal --- \texttt{spaceTime > 0} indicates unused time.
  \item \textbf{Cross-staff voices}: Events on different staves belong to different voices.
\end{itemize}

\textbf{Time Warp (Tuplets) --- Verify Carefully}
\begin{itemize}[nosep]
  \item \texttt{timeWarp: \{ numerator, denominator \}} --- tuplet ratio. Effective duration = \texttt{baseDuration * numerator / denominator}.
  \item Formula: $N$ notes in the time of $M$ $\to$ each note gets \texttt{timeWarp: \{ numerator: M, denominator: N \}}.
  \item Common ratios: triplet \texttt{\{2,3\}}, quintuplet \texttt{\{4,5\}}, sextuplet \texttt{\{4,6\}}, octuplet in compound meter \texttt{\{3,4\}}.
  \item Constraints: \texttt{numerator/denominator} must be $>0.5$ ($\leq0.5$ triggers error). Only \texttt{2/3} is ``regular''; other ratios reduce \texttt{qualityScore}.
  \item Within a voice, a tuplet group's total tick sum must be divisible by its denominator, otherwise \texttt{fractionalWarp=true} $\to$ \texttt{fine=false}.
\end{itemize}

\textbf{CRITICAL} --- Do NOT preserve \texttt{timeWarp} from the original regulation blindly. If you can make durations sum correctly WITHOUT \texttt{timeWarp}, always prefer \texttt{null}. Only use \texttt{timeWarp} when the image clearly shows tuplet notation.

\textbf{Quality Metrics}
\begin{description}[nosep,leftmargin=1em]
  \item[\texttt{fine}] Acceptable quality (no fatal errors, \texttt{tickTwist<0.3}, no fractional warp, no irregular tick, no surplus time, no beam broken, no grace in voice)
  \item[\texttt{error}] Fatal problems (\texttt{tickTwist>=1.0}, tick overlap, voice rugged, corrupted events, null events$>$2, overranged, bad timewarp ratio)
  \item[\texttt{perfect}] Ideal regulation. Requires \texttt{fine=true} PLUS \texttt{tickTwist<0.2}, \texttt{spaceTime=0}, no irregular warps, no grace dominant.
  \item[\texttt{qualityScore}] 0--1 composite score. 0 = error, 1 = perfect. Factors: \texttt{spaceTime} loss, duration rate, irregular warps, \texttt{tickTwist}$^2$. Patched measures get 1.0.
  \item[\texttt{tickTwist}] Time-position non-linearity. $<0.2$ = good, $<0.3$ = fine, $\geq1.0$ = error (fatal).
  \item[\texttt{spaceTime}] Unused time in voices (gaps), in whole-note units. Allowed for fine, but must be 0 for perfect.
  \item[\texttt{surplusTime}] Total time exceeding measure duration across all voices. Must be 0.
  \item[\texttt{beamBroken}] Beam Open/Continue/Close sequence is invalid within a voice.
  \item[\texttt{voiceRugged}] Same event ID appears in multiple voices. Fatal error.
  \item[\texttt{tickOverlapped}] Events within a voice overlap in time. Fatal error.
\end{description}

\textbf{Feature Confidence (ML Classifier) --- Trust Over Regulation}
\begin{itemize}[nosep]
  \item \textbf{feature.divisions}: Array of 7 floats (indices 0--6 = whole through 64th). The index with highest value is ML's best guess. Compare with assigned \texttt{event.division} --- if they disagree, the higher-confidence value is usually correct.
  \item \textbf{feature.dots}: \texttt{[dot1\_conf, dot2\_conf]}. If \texttt{feature.dots[1] > 0.1} but \texttt{event.dots = 0}, a dot was likely missed.
  \item \textbf{feature.grace}: Float confidence score. Always verify against the background image.
\end{itemize}

\textbf{Event ID vs Array Index (CRITICAL)}: All fields use \textbf{event ID values} (\texttt{event.id}, typically 1-based), NOT array indices. \texttt{voices} arrays contain event IDs.

\paragraph{Recognition Data Issues (Upstream Errors)}
These are NOT regulation failures --- they are upstream misclassifications that propagate:
\begin{itemize}[nosep]
  \item \textbf{False grace notes}: Events incorrectly tagged with \texttt{grace="grace"}. Fix: set \texttt{grace: false} in the event's solution entry.
  \item \textbf{Wrong division/dots}: Note value misrecognized. Fix: set correct \texttt{division} and \texttt{dots}.
  \item \textbf{Missing dots}: ML sometimes fails to detect augmentation dots on half notes. Check \texttt{feature.dots} confidence.
  \item \textbf{Phantom/duplicate events}: Two events at nearly the same $x$. Keep the more complete one in voices; leave the duplicate out.
  \item \textbf{Missing events}: Events visible in the image but not detected. Cannot fix --- mark as \texttt{status=-1}.
\end{itemize}

\paragraph{Common Agent Mistakes (AVOID THESE)}
\begin{enumerate}[nosep]
  \item \textbf{Excessive voice splitting (most common)}: Creating 3--4 voices when 1--2 would suffice. If events are sequential (non-overlapping) on the same staff, they are ONE voice. Stem direction changes alone do NOT justify a new voice.
  \item \textbf{Blindly copying timeWarp from regulation}: Unless you see actual tuplet brackets in the image, set \texttt{timeWarp: null}.
  \item \textbf{Wrong division leading to cascading tick errors}: Always verify each event's division against \texttt{feature.divisions} AND the image before computing ticks.
  \item \textbf{Not verifying feature.divisions confidence}: The ML classifier's confidence array is often more reliable than the regulation's assigned division.
  \item \textbf{Mixing staves in one voice}: Events with different \texttt{event.staff} values must never be in the same voice array.
\end{enumerate}

\paragraph{Output Format}
Output ONLY a JSON block with fixes. Each fix is a \texttt{RegulationSolution} plus \texttt{measureIndex} and \texttt{status}.

Fix fields: \texttt{measureIndex} (index in spartito); \texttt{events} (array with required \texttt{id}, \texttt{tick}, \texttt{tickGroup}, \texttt{timeWarp}, and optional \texttt{division}/\texttt{dots}/\texttt{beam}/\texttt{grace} overrides); \texttt{voices} (array of event-ID arrays); \texttt{duration} (ticks); \texttt{status} (0=Solved, 1=Issue, $-1$=Discard).

\texttt{status=0} only if each voice's durations sum to $\leq$ measure duration and no tick overlap exists within any voice.

\paragraph{Evaluation Tool}
You have access to the \texttt{evaluate\_fix} tool (via MCP). Workflow: (1) analyze measure data and image; (2) propose a fix; (3) call \texttt{evaluate\_fix} to check quality metrics; (4) if \texttt{fine=false}, adjust and re-evaluate; (5) once satisfied, include the fix in the final JSON output. Always call \texttt{evaluate\_fix} at least once per measure before finalizing.
}

\subsection*{Initial State}

\begin{figure}[H]
\centering
\begin{minipage}[b]{0.22\linewidth}
  \centering
  \includegraphics[width=\linewidth]{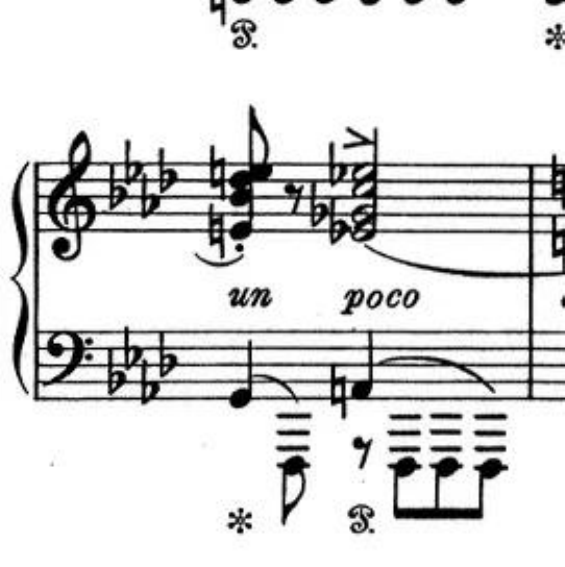}
  \small\textit{Measure image (background)}
\end{minipage}\hfill
\begin{minipage}[b]{0.74\linewidth}
  \centering
  \includegraphics[width=\linewidth]{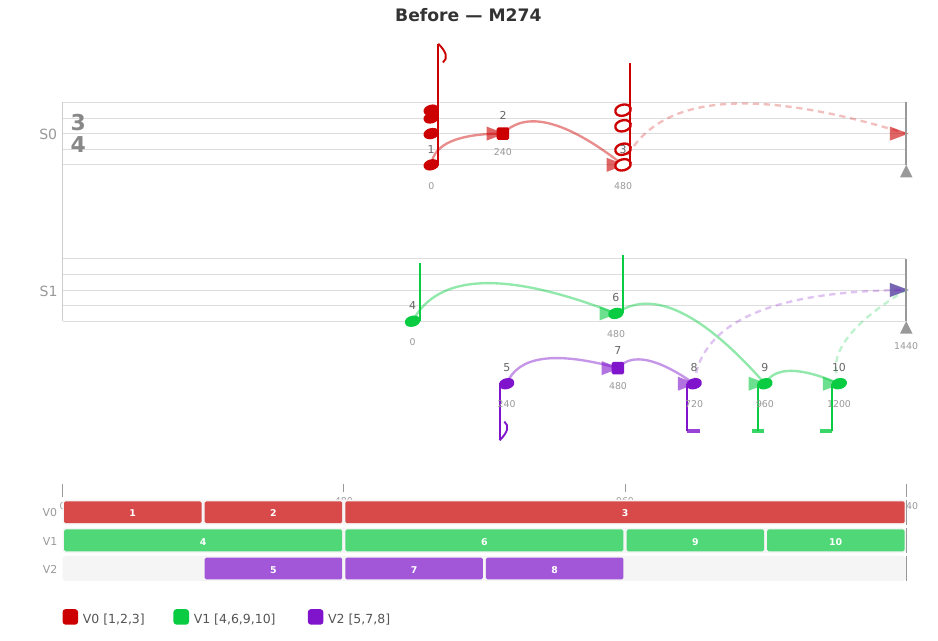}
  \small\textit{Before --- regulation output, \texttt{beamBroken=true}}
\end{minipage}
\caption{Measure~274 initial state. Left: composite stave image read by the agent. Right: topology visualisation of the current regulation (three voices; coloured chains show voice assignments; tick timeline at bottom). Events 8, 9, 10 form a beamed group but are split across Voice~0 and Voice~1, causing \texttt{beamBroken=true}.}
\label{fig:appF-before}
\end{figure}

Measure~274 is a 3/4 bar (duration~=~1440~ticks; whole~=~1920) with two staves and 10 detected events.
Staff~0 (treble, ev.\,1--3): eighth-note chord $\to$ eighth rest $\to$ half-note chord, filling 1440 ticks exactly.
Staff~1 (bass, ev.\,4--10): quarter (ev.\,4, tick~0), eighth stem-down (ev.\,5, tick~240), quarter stem-up (ev.\,6, tick~480), eighth rest (ev.\,7, tick~480), then beamed eighths ev.\,8--10 (ticks 720/960/1200).
Events~6 and~7 are \emph{simultaneous} (both at tick~480), requiring voice separation on staff~1.
The original regulation assigned voices $[4,6,9,10]$ and $[5,7,8]$, but distributed beam events 8--10 across both voices, breaking the beam sequence.

\subsection*{Agent Turn 1: Image Read}

\noindent\textbf{[Assistant]} \quad \texttt{Read("/tmp/spartito-annotate-LHcs8D/m274.webp")}

\medskip
\noindent\textit{[Tool returns the measure image shown in Figure~\ref{fig:appF-before} (left). The topology diagram in Figure~\ref{fig:appF-before} (right) is also shown to the agent at this step.]}

\medskip
\noindent\textit{[Thinking]}
\begin{grayleftbar}
{\footnotesize
Let me analyze the measure image and the event data carefully.

Looking at the image:
\begin{itemize}[nosep]
  \item This is a 3/4 measure (duration = 1440 ticks)
  \item Two staves (staffMask = 3 means staff 0 and staff 1)
  \item Staff 0 (top staff): There's a chord of 4 notes (likely 8th notes), then a rest (8th), then a half note chord
  \item Staff 1 (bottom staff): There's a quarter note, then an 8th note, then a quarter note with a rest on top of it (same x position), then a group of three beamed 8th notes
\end{itemize}

Let me check the current regulation:
\begin{itemize}[nosep]
  \item Staff 0: events 1, 2, 3
  \begin{itemize}[nosep]
    \item Event 1: division=3 (eighth), stem up, chord at tick=0
    \item Event 2: division=3 (eighth), rest, tick=240
    \item Event 3: division=1 (half), stem up, chord at tick=480
    \item Duration: $240 + 240 + 960 = 1440$ \checkmark
  \end{itemize}
  \item Staff 1: events 4, 5, 6, 7, 8, 9, 10
  \begin{itemize}[nosep]
    \item Event 4: division=2 (quarter), stem up, tick=0
    \item Event 5: division=3 (eighth), stem down, tick=240
    \item Event 6: division=2 (quarter), stem up, tick=480
    \item Event 7: division=3 (eighth), rest, tick=480 (same x as event 6)
    \item Event 8: division=3 (eighth), stem down, beam=Open, tick=720
    \item Event 9: division=3 (eighth), stem down, beam=Continue, tick=960
    \item Event 10: division=3 (eighth), stem down, beam=Close, tick=1200
  \end{itemize}
\end{itemize}

Current issues:
\begin{enumerate}[nosep]
  \item beamBroken = true --- the beam group 8--9--10 has events 5 and 7 between them, but they're in different voices
  \item The voices are currently: [1,2,3], [4,6,9,10], [5,7,8] --- three voices
  \item spaceTime = 0.375 (1440 * 0.375 = 540 ticks of unused time)
\end{enumerate}

Examining the feature data reveals a critical mismatch: event 6, which should be a quarter note, has its highest division confidence at index 2 (quarter note) with a value of 5.39, while event 7, an 8th rest, shows strongest confidence for an 8th note.

Events 6 and 7 are very close ($x = 19.56$ vs $19.63$), confirming they're simultaneous or nearly so, meaning they occur in different voices.

My voice distribution looks like this:
\begin{itemize}[nosep]
  \item Voice 1: Staff 0 events (1, 2, 3)
  \item Voice 2: Staff 1 stem-up events (4, 6)
  \item Voice 3: Staff 1 stem-down/beamed events (5, 7, 8, 9, 10)
\end{itemize}

The tricky part is managing tick positioning to prevent beam breaks. Event 5 at tick 240 and event 7 at tick 480 seem to disrupt voice 3's flow, causing the beam broken error.

I'll consolidate the voices into a single sequence for staff 1, prioritizing tick order and ensuring the beamed group remains uninterrupted. The proposed single voice for staff 1 will run from tick 0 to 1440, capturing all events in their natural progression: starting with event 4, then event 5, event 6, event 7, and finally the beamed group (events 8, 9, 10).
}
\end{grayleftbar}

\medskip
\noindent\textbf{[Assistant]} \quad Let me analyze the measure image and event data.

\subsection*{Agent Turn 2: First \texttt{evaluate\_fix} Call}

The agent proposes consolidating all staff-1 events into a single voice:

\begin{verbatim}
voices: [[1, 2, 3], [4, 5, 6, 7, 8, 9, 10]]
\end{verbatim}

\noindent\textbf{[Tool result]}
\begin{framed}
\begin{verbatim}
BEFORE (m274): fine=false, error=false, tickTwist=0.046
  qualityScore=0.874, spaceTime=0.375, surplusTime=0, beamBroken=true
  Events: 10 total, 10 valid, 0 fake, 0 null

AFTER  (m274): fine=false, error=true, tickTwist=0.046
  qualityScore=0.000, spaceTime=0, surplusTime=720, beamBroken=false
  Events: 10 total, 10 valid, 0 fake, 0 null
  tickOverlapped=true

d(tickTwist)=+0.000 -> WORSE (new error)
\end{verbatim}
\end{framed}

\begin{figure}[H]
\centering
\includegraphics[width=0.82\linewidth]{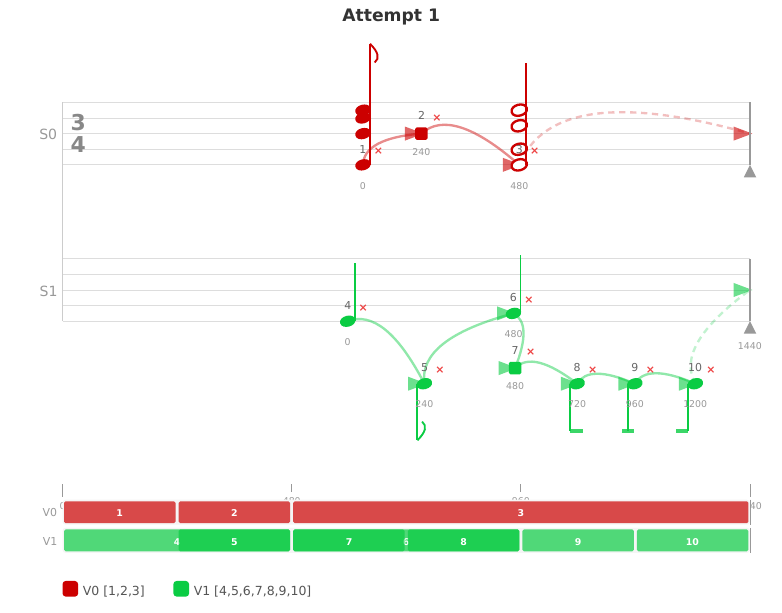}
\caption{Attempt~1 topology. Merging all staff-1 events into one voice places ev.\,6 and ev.\,7 (both tick~480) in the same voice, producing \texttt{tickOverlapped=true}.}
\label{fig:appF-attempt1}
\end{figure}

\subsection*{Agent Turn 3: Analysis and Rethink}

\noindent\textbf{[Assistant]}

I see the issue --- events~6 and~7 are both at tick~480, which causes overlap in a single voice. Looking at the image more carefully:

\begin{itemize}[nosep]
  \item Event~6 (quarter note, stem up) and event~7 (8th rest) appear to be at the same $x$ position, meaning they occur simultaneously but in different voices.
  \item The quarter note is in one voice, the 8th rest is in another voice.
\end{itemize}

So the original voice separation was correct in separating these simultaneous events. The issue is that the beamed group (8, 9, 10) needs to be in a continuous voice without gaps.

Let me reconsider. The original had voice~2 as $[4, 6, 9, 10]$ and voice~3 as $[5, 7, 8]$.

\medskip
Tracing original voice~2: $[4, 6, 9, 10]$
\begin{itemize}[nosep]
  \item 4 at tick~0 (quarter = 480) $\to$ ends at 480
  \item 6 at tick~480 (quarter = 480) $\to$ ends at 960
  \item 9 at tick~960 (8th = 240) $\to$ ends at 1200
  \item 10 at tick~1200 (8th = 240) $\to$ ends at 1440
\end{itemize}

Voice~3: $[5, 7, 8]$
\begin{itemize}[nosep]
  \item 5 at tick~240 (8th = 240) $\to$ ends at 480
  \item 7 at tick~480 (8th rest = 240) $\to$ ends at 720
  \item 8 at tick~720 (8th = 240) $\to$ ends at 960
\end{itemize}

The problem is that the beam is 8--9--10. For the beam to not be broken, all three events need to be in the same voice in sequence. The original has 8 in voice~3, and 9, 10 in voice~2. So when rendering voice~2 we see $4, 6, 9, 10$ --- the beam starts at~9 (\texttt{Continue}), but where's the \texttt{Open}? It's in voice~3 (event~8). Hence, broken beam.

\medskip
To fix: put 8, 9, 10 all in the same voice, consecutively.

\medskip
Let me try: voice~2 $= [4, 6]$, voice~3 $= [5, 7, 8, 9, 10]$:
\begin{itemize}[nosep]
  \item Voice~2: $0{\to}480{\to}960$ (gap 960--1440)
  \item Voice~3: $240{\to}480{\to}720{\to}960{\to}1200{\to}1440$ (fills the measure!)
\end{itemize}

Both voices have gaps (partial voices), but the beam group 8, 9, 10 is now consecutive in voice~3. This is the structure I'll submit.

\subsection*{Agent Turn 4: Second \texttt{evaluate\_fix} Call}

\begin{verbatim}
voices: [[1, 2, 3], [4, 6], [5, 7, 8, 9, 10]]
events: [
  {id:1, tick:0},   {id:2, tick:240},  {id:3, tick:480},
  {id:4, tick:0},   {id:6, tick:480},
  {id:5, tick:240}, {id:7, tick:480},
  {id:8, tick:720,  beam:"Open"},
  {id:9, tick:960,  beam:"Continue"},
  {id:10, tick:1200, beam:"Close"}
]
duration: 1440
\end{verbatim}

\noindent\textbf{[Tool result]}
\begin{framed}
\begin{verbatim}
BEFORE (m274): fine=false, error=false, tickTwist=0.046
  qualityScore=0.874, spaceTime=0.375, surplusTime=0, beamBroken=true
  Events: 10 total, 10 valid, 0 fake, 0 null

AFTER  (m274): fine=true, error=false, tickTwist=0.046
  qualityScore=0.874, spaceTime=0.375, surplusTime=0, beamBroken=false
  Events: 10 total, 10 valid, 0 fake, 0 null

d(tickTwist)=+0.000 -> FIXED!
\end{verbatim}
\end{framed}

\begin{figure}[H]
\centering
\includegraphics[width=0.82\linewidth]{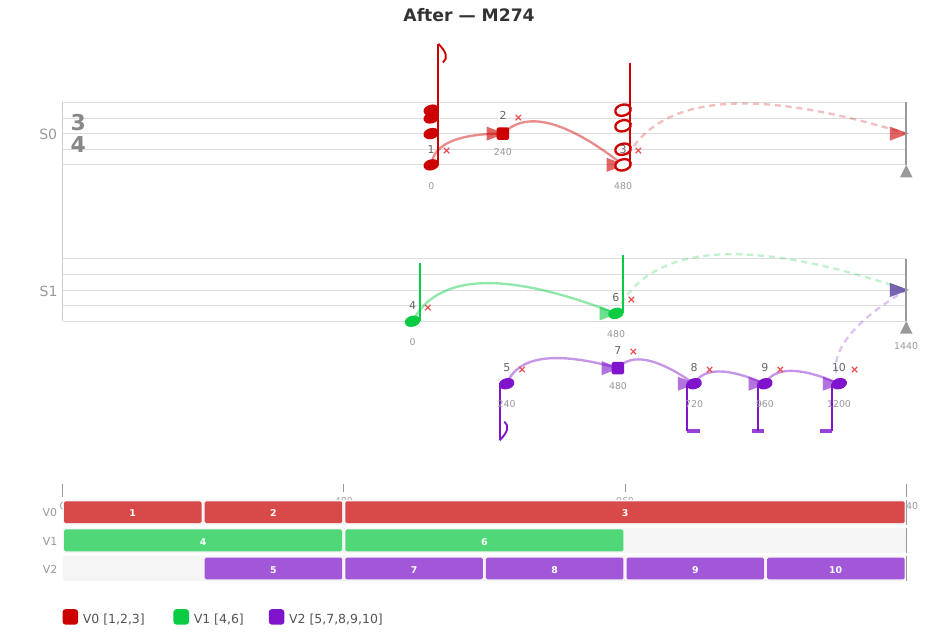}
\caption{Fix summary topology. Voice~0 (red): staff-0 events [1,2,3]. Voice~1 (green): staff-1 stem-up partial [4,6]. Voice~2 (purple): staff-1 stem-down with full beam group [5,7,8,9,10]. The beam Open$\to$Continue$\to$Close sequence is now intact within a single voice.}
\label{fig:appF-fix}
\end{figure}

\subsection*{Agent Turn 5: Final Output}

\noindent\textbf{[Assistant]}

\medskip
\noindent\textit{[Thinking]}
\begin{grayleftbar}
{\footnotesize
The fix works. Now the measure has \texttt{fine=true}, \texttt{error=false}, and \texttt{beamBroken=false}. The \texttt{spaceTime} is still 0.375, but that's acceptable for \texttt{fine=true}.

Let me output the final JSON.
}
\end{grayleftbar}

\noindent\textbf{[Assistant]} \quad The fix resolves the beam break by keeping the beamed group (8, 9, 10) together in voice~2, while voice~1 handles the simultaneous quarter note at tick~480.

\begin{verbatim}
{"fixes": [{
  "measureIndex": 274,
  "voices": [[1,2,3], [4,6], [5,7,8,9,10]],
  "duration": 1440,
  "status": 0
}]}
\end{verbatim}

\noindent\textbf{Fix (status=0 Solved)} \quad Voices: \texttt{[[1,2,3],[4,6],[5,7,8,9,10]]}

\medskip
\noindent\textit{Session: 220.5\,s, 125\,101 input tokens, 8\,388 output tokens.}

\subsection*{Agent Feedback}

After the annotation session, the system posed three follow-up questions to the agent.

\medskip
\noindent\textbf{[User]}

\begin{quote}
Based on your annotation experience just now, please provide a brief summary:
(1) Which principles in the system prompt were most helpful for your annotation work?
(2) What additional guidelines or tips would you suggest adding to the system prompt that are not currently covered?
(3) What common patterns or pitfalls did you encounter during this annotation session?
\end{quote}

\noindent\textbf{[Assistant]}

\textbf{1.\ Most Helpful Principles:}
\begin{itemize}[nosep]
  \item \textbf{Beamed notes must be consecutive in a single voice} --- This was critical for identifying why the beam was broken (events 8, 9, 10 were split across voices).
  \item \textbf{Cross-referencing images with event data} --- The $x$-positions (events~6 and~7 at ${\approx}19.6$) confirmed they were simultaneous, requiring voice separation.
  \item \textbf{Voice separation for simultaneous events} --- Events at the same tick with different staff positions (note vs.\ rest, or different stem directions) must be in different voices.
\end{itemize}

\textbf{2.\ Suggested Additions:}
\begin{itemize}[nosep]
  \item \textbf{Prioritize beam continuity over \texttt{spaceTime} minimization} --- A measure with a continuous beam but some \texttt{spaceTime} gaps is preferred over a ``perfect'' voice assignment with broken beams.
  \item \textbf{Stem direction as a voice hint} --- Notes with opposite stem directions at the same tick are strong indicators they belong to different voices.
  \item \textbf{Simultaneous note+rest pattern} --- When a note and rest share the same tick position, they likely represent different voices (e.g., one voice has a note while another has a rest).
\end{itemize}

\textbf{3.\ Common Patterns/Pitfalls:}
\begin{itemize}[nosep]
  \item \textbf{Broken beams from split voice assignments} --- The beamed group (8, 9, 10) was split with~8 in voice~2 and 9, 10 in voice~1, breaking the beam even though the timing worked.
  \item \textbf{Misinterpreting simultaneous events} --- Events~6 (quarter note) and~7 (eighth rest) at tick~480 appeared to conflict, but belong to different voices.
  \item \textbf{Over-optimizing for \texttt{spaceTime}} --- The original assignment minimized gaps but broke beams; accepting some \texttt{spaceTime} (gaps at measure start/end) fixed the beam issue.
\end{itemize}

\medskip
\noindent\textit{Feedback session: 26.9\,s, 57\,649 input tokens, 656 output tokens.}

\medskip

\end{document}